\documentclass[conference,letter]{IEEEtran}
\usepackage[dvips]{graphicx,color}
\usepackage{latexsym}
\usepackage{amssymb}
\usepackage{amsmath,bm}
\usepackage{array}

\begin{document}
%
%
%
%

\newcommand{\qed}{\hfill$\square$}
\newcommand{\suchthat}{\mbox{~s.t.~}}
\newcommand{\Markov}{\leftrightarrow}
\newcommand{\markov}{\leftrightarrow}
%
%
\newenvironment{pRoof}{%
 \noindent{\em Proof.\ }}{%
 \hspace*{\fill}\qed \\
 \vspace{2ex}}


\newcommand{\ket}[1]{| #1 \rangle}
\newcommand{\bra}[1]{\langle #1 |}
\newcommand{\bol}[1]{\mathbf{#1}}
\newcommand{\rom}[1]{\mathrm{#1}}
\newcommand{\san}[1]{\mathsf{#1}}
\newcommand{\mymid}{:~}
\newcommand{\argmax}{\mathop{\rm argmax}\limits}
\newcommand{\argmin}{\mathop{\rm argmin}\limits}
%
%
%
%
\newcommand{\bc}{\begin{center}}  %
\newcommand{\ec}{\end{center}}
\newcommand{\befi}{\begin{figure}[h]}  %
\newcommand{\enfi}{\end{figure}}
\newcommand{\bsb}{\begin{shadebox}\begin{center}}   %
\newcommand{\esb}{\end{center}\end{shadebox}}
\newcommand{\bs}{\begin{screen}}     %
\newcommand{\es}{\end{screen}}
\newcommand{\bib}{\begin{itembox}}   %
\newcommand{\eib}{\end{itembox}}
\newcommand{\bit}{\begin{itemize}}   %
\newcommand{\eit}{\end{itemize}}
\newcommand{\defeq}{:=}
\newcommand{\Qed}{\hbox{\rule[-2pt]{3pt}{6pt}}}
\newcommand{\beq}{\begin{equation}}
\newcommand{\eeq}{\end{equation}}
\newcommand{\beqa}{\begin{eqnarray}}
\newcommand{\eeqa}{\end{eqnarray}}
\newcommand{\beqno}{\begin{eqnarray*}}
\newcommand{\eeqno}{\end{eqnarray*}}
\newcommand{\ba}{\begin{array}}
\newcommand{\ea}{\end{array}}
\newcommand{\vc}[1]{\mbox{\boldmath $#1$}}
\newcommand{\lvc}[1]{\mbox{\scriptsize \boldmath $#1$}}
\newcommand{\svc}[1]{\mbox{\scriptsize\boldmath $#1$}}

\newcommand{\wh}{\widehat}
\newcommand{\wt}{\widetilde}
\newcommand{\ts}{\textstyle}
\newcommand{\ds}{\displaystyle}
\newcommand{\scs}{\scriptstyle}
\newcommand{\vep}{\varepsilon}
\newcommand{\rhp}{\rightharpoonup}
\newcommand{\cl}{\circ\!\!\!\!\!-}
\newcommand{\bcs}{\dot{\,}.\dot{\,}}
\newcommand{\eqv}{\Leftrightarrow}
\newcommand{\leqv}{\Longleftrightarrow}
\newtheorem{co}{Corollary} 
\newtheorem{lm}{Lemma} 
\newtheorem{Ex}{Example} 
\newtheorem{Th}{Theorem}
\newtheorem{df}{Definition} 
\newtheorem{pr}{Property} 
\newtheorem{pro}{Proposition} 
\newtheorem{rem}{Remark} 

\newcommand{\lcv}{convex } 

\newcommand{\hugel}{{\arraycolsep 0mm
                    \left\{\ba{l}{\,}\\{\,}\ea\right.\!\!}}
\newcommand{\Hugel}{{\arraycolsep 0mm
                    \left\{\ba{l}{\,}\\{\,}\\{\,}\ea\right.\!\!}}
\newcommand{\HUgel}{{\arraycolsep 0mm
                    \left\{\ba{l}{\,}\\{\,}\\{\,}\vspace{-1mm}
                    \\{\,}\ea\right.\!\!}}
\newcommand{\huger}{{\arraycolsep 0mm
                    \left.\ba{l}{\,}\\{\,}\ea\!\!\right\}}}

\newcommand{\Huger}{{\arraycolsep 0mm
                    \left.\ba{l}{\,}\\{\,}\\{\,}\ea\!\!\right\}}}

\newcommand{\HUger}{{\arraycolsep 0mm
                    \left.\ba{l}{\,}\\{\,}\\{\,}\vspace{-1mm}
                    \\{\,}\ea\!\!\right\}}}

\newcommand{\hugebl}{{\arraycolsep 0mm
                    \left[\ba{l}{\,}\\{\,}\ea\right.\!\!}}
\newcommand{\Hugebl}{{\arraycolsep 0mm
                    \left[\ba{l}{\,}\\{\,}\\{\,}\ea\right.\!\!}}
\newcommand{\HUgebl}{{\arraycolsep 0mm
                    \left[\ba{l}{\,}\\{\,}\\{\,}\vspace{-1mm}
                    \\{\,}\ea\right.\!\!}}
\newcommand{\hugebr}{{\arraycolsep 0mm
                    \left.\ba{l}{\,}\\{\,}\ea\!\!\right]}}
\newcommand{\Hugebr}{{\arraycolsep 0mm
                    \left.\ba{l}{\,}\\{\,}\\{\,}\ea\!\!\right]}}
\newcommand{\HugebrB}{{\arraycolsep 0mm
                    \left.\ba{l}{\,}\\{\,}\vspace*{-1mm}\\{\,}\ea\!\!\right]}}

\newcommand{\HUgebr}{{\arraycolsep 0mm
                    \left.\ba{l}{\,}\\{\,}\\{\,}\vspace{-1mm}
                    \\{\,}\ea\!\!\right]}}

\newcommand{\hugecl}{{\arraycolsep 0mm
                    \left(\ba{l}{\,}\\{\,}\ea\right.\!\!}}
\newcommand{\Hugecl}{{\arraycolsep 0mm
                    \left(\ba{l}{\,}\\{\,}\\{\,}\ea\right.\!\!}}
\newcommand{\HUgecl}{{\arraycolsep 0mm
                    \left(\ba{l}{\,}\\{\,}\\{\,}\vspace{-1mm}
                    \\{\,}\ea\right.\!\!}}
\newcommand{\hugecr}{{\arraycolsep 0mm
                    \left.\ba{l}{\,}\\{\,}\ea\!\!\right)}}
\newcommand{\Hugecr}{{\arraycolsep 0mm
                    \left.\ba{l}{\,}\\{\,}\\{\,}\ea\!\!\right)}}
\newcommand{\HUgecr}{{\arraycolsep 0mm
                    \left.\ba{l}{\,}\\{\,}\\{\,}\vspace{-1mm}
                    \\{\,}\ea\!\!\right)}}

\newcommand{\hugepl}{{\arraycolsep 0mm
                    \left|\ba{l}{\,}\\{\,}\ea\right.\!\!}}
\newcommand{\Hugepl}{{\arraycolsep 0mm
                    \left|\ba{l}{\,}\\{\,}\\{\,}\ea\right.\!\!}}
\newcommand{\hugepr}{{\arraycolsep 0mm
                    \left.\ba{l}{\,}\\{\,}\ea\!\!\right|}}
\newcommand{\Hugepr}{{\arraycolsep 0mm
                    \left.\ba{l}{\,}\\{\,}\\{\,}\ea\!\!\right|}}

\newcommand{\MEq}[1]{\stackrel{
{\rm (#1)}}{=}}

\newcommand{\MLeq}[1]{\stackrel{
{\rm (#1)}}{\leq}}

\newcommand{\ML}[1]{\stackrel{
{\rm (#1)}}{<}}

\newcommand{\MGeq}[1]{\stackrel{
{\rm (#1)}}{\geq}}

\newcommand{\MG}[1]{\stackrel{
{\rm (#1)}}{>}}

\newcommand{\MPreq}[1]{\stackrel{
{\rm (#1)}}{\preceq}}

\newcommand{\MSueq}[1]{\stackrel{
{\rm (#1)}}{\succeq}}

\newcommand{\MSubeq}[1]{\stackrel{
{\rm (#1)}}{\subseteq}}

\newcommand{\MSupeq}[1]{\stackrel{
{\rm (#1)}}{\supseteq}}

\newcommand{\Ch}{{\Gamma}}
\newcommand{\Rw}{{W}}

\newcommand{\Cd}{{\cal R}_{\rm d}(\Ch)}
\newcommand{\CdB}{{\cal R}_{\rm d}^{\prime}(\Ch)}
\newcommand{\CdBB}{{\cal R}_{\rm d}^{\prime\prime}(\Ch)}

\newcommand{\Cdi}{{\cal R}_{\rm d}^{\rm (in)}(\Ch)}
\newcommand{\Cdo}{{\cal R}_{\rm d}^{\rm (out)}(\Ch)}

\newcommand{\tCdi}{\tilde{\cal R}_{\rm d}^{\rm (in)}(\Ch)}
\newcommand{\tCdo}{\tilde{\cal R}_{\rm d}^{\rm (out)}(\Ch)}
\newcommand{\hCdo}{  \hat{\cal R}_{\rm d}^{\rm (out)}(\Ch)}

\newcommand{\Cs}{{\cal R}_{\rm s}(\Ch)}
\newcommand{\CsB}{{\cal R}_{\rm s}^{\prime}(\Ch)}
\newcommand{\CsBB}{{\cal R}_{\rm s}^{\prime\prime}(\Ch)}

\newcommand{\Csi}{{\cal R}_{\rm s}^{\rm (in)}(\Ch)}
\newcommand{\Cso}{{\cal R}_{\rm s}^{\rm (out)}(\Ch)}
\newcommand{\tCsi}{\tilde{\cal R}_{\rm s}^{\rm (in)}(\Ch)}
\newcommand{\tCso}{\tilde{\cal R}_{\rm s}^{\rm (out)}(\Ch)}
\newcommand{\cCsi}{\check{\cal R}_{\rm s}^{\rm (in)}(\Ch)}
\newcommand{\Cds}{{\cal C}_{\rm ds}(\Ch)}
\newcommand{\Cdsi}{{\cal C}_{\rm ds}^{\rm (in)}(\Ch)}
\newcommand{\Cdso}{{\cal C}_{\rm ds}^{\rm (out)}(\Ch)}
\newcommand{\tCdsi}{\tilde{\cal C}_{\rm ds}^{\rm (in)}(\Ch)}
\newcommand{\tCdso}{\tilde{\cal C}_{\rm ds}^{\rm (out)}(\Ch)}
\newcommand{\hCdso}{\hat{\cal C}_{\rm ds}^{\rm (out)}(\Ch)}
\newcommand{\Css}{{\cal C}_{\rm ss}(\Ch)}
\newcommand{\Cssi}{{\cal C}_{\rm ss}^{\rm (in)}(\Ch)}
\newcommand{\Csso}{{\cal C}_{\rm ss}^{\rm (out)}(\Ch)}
\newcommand{\tCssi}{\tilde{\cal C}_{\rm ss}^{\rm (in)}(\Ch)}
\newcommand{\tCsso}{\tilde{\cal C}_{\rm ss}^{\rm (out)}(\Ch)}
\newcommand{\Cde}{{\cal R}_{\rm d1e}(\Ch)}
\newcommand{\Cdei}{{\cal R}_{\rm d1e}^{\rm (in)}(\Ch)}
\newcommand{\Cdeo}{{\cal R}_{\rm d1e}^{\rm (out)}(\Ch)}
\newcommand{\tCdei}{\tilde{\cal R}_{\rm d1e}^{\rm (in)}(\Ch)}
\newcommand{\tCdeo}{\tilde{\cal R}_{\rm d1e}^{\rm (out)}(\Ch)}
\newcommand{\hCdeo}{  \hat{\cal R}_{\rm d1e}^{\rm (out)}(\Ch)} 
\newcommand{\Cse}{{\cal R}_{\rm s1e}(\Ch)}
\newcommand{\Csei}{{\cal R}_{\rm s1e}^{\rm (in)}(\Ch)}
\newcommand{\Cseo}{{\cal R}_{\rm s1e}^{\rm (out)}(\Ch)}
\newcommand{\tCsei}{\tilde{\cal R}_{\rm s1e}^{\rm (in)}(\Ch)}
\newcommand{\tCseo}{\tilde{\cal R}_{\rm s1e}^{\rm (out)}(\Ch)}

\newcommand{\Capa}{C}

\newcommand{\ZeTa}{\zeta(S;Y,Z|U)}
\newcommand{\ZeTaI}{\zeta(S_i;Y_i,Z_i|U_i)}

\newcommand{\CEreg}{\irBr{rate} }
\newcommand{\CEregB}{rate\MarkOh{-equivocation }}

\newcommand{\Cls}{class NL}
\newcommand{\vSpa}{\vspace{0.3mm}}
\newcommand{\Prmt}{\zeta}

\newcommand{\pj}{\omega_n}

\newfont{\bg}{cmr10 scaled \magstep4}
\newcommand{\bigzerol}{\smash{\hbox{\bg 0}}}
\newcommand{\bigzerou}{\smash{\lower1.7ex\hbox{\bg 0}}}
\newcommand{\nbn}{\frac{1}{n}}
\newcommand{\ra}{\rightarrow}
\newcommand{\la}{\leftarrow}
\newcommand{\ldo}{\ldots}
\newcommand{\ep}{\epsilon }
\newcommand{\typi}{A_{\epsilon }^{n}}
\newcommand{\bx}{\hspace*{\fill}$\Box$}
\newcommand{\pa}{\vert}
\newcommand{\ignore}[1]{}

\newcommand{\dBar}{||}
\newcommand{\One}{\rm (i)}
\newcommand{\Two}{\rm (ii)}
\newcommand{\Thr}{\rm (iii)}
\newcommand{\Fou}{\rm (iv)}
\newcommand{\Fiv}{\rm (v)}
\newcommand{\OnE}{1}

%
%
%
%

\newenvironment{jenumerate}
	{\begin{enumerate}\itemsep=-0.25em \parindent=1zw}{\end{enumerate}}
\newenvironment{jdescription}
	{\begin{description}\itemsep=-0.25em \parindent=1zw}{\end{description}}
\newenvironment{jitemize}
	{\begin{itemize}\itemsep=-0.25em \parindent=1zw}{\end{itemize}}
\renewcommand{\labelitemii}{$\cdot$}

\newcommand{\iro}[2]{{\color[named]{#1}#2\normalcolor}}
\newcommand{\irr}[1]{{\color[named]{Black}#1\normalcolor}}

\newcommand{\irg}[1]{{\color[named]{Green}#1\normalcolor}}
\newcommand{\irb}[1]{{\color[named]{Black}#1\normalcolor}}

\newcommand{\irBl}[1]{{\color[named]{Black}#1\normalcolor}}
\newcommand{\irWh}[1]{{\color[named]{White}#1\normalcolor}}

\newcommand{\irY}[1]{{\color[named]{Yellow}#1\normalcolor}}
\newcommand{\irO}[1]{{\color[named]{Orange}#1\normalcolor}}
\newcommand{\irBr}[1]{{\color[named]{Black}#1\normalcolor}}
\newcommand{\IrBr}[1]{{\color[named]{Purple}#1\normalcolor}}
\newcommand{\irBw}[1]{{\color[named]{Brown}#1\normalcolor}}
\newcommand{\irPk}[1]{{\color[named]{Magenta}#1\normalcolor}}
\newcommand{\irCb}[1]{{\color[named]{CadetBlue}#1\normalcolor}}

\newcommand{\irMho}[1]{{\color[named]{Mahogany}#1\normalcolor}}
\newcommand{\irOlg}[1]{{\color[named]{Black}#1\normalcolor}}
\newcommand{\irBg}[1]{{\color[named]{BlueGreen}#1\normalcolor}}
\newcommand{\irCy}[1]{{\color[named]{Cyan}#1\normalcolor}}
\newcommand{\irRyp }[1]{{\color[named]{RoyalPurple}#1\normalcolor}}

\newcommand{\irAqm}[1]{{\color[named]{Aquamarine}#1\normalcolor}}
\newcommand{\irRyb}[1]{{\color[named]{RoyalBule}#1\normalcolor}}
\newcommand{\irNvb}[1]{{\color[named]{NavyBlue}#1\normalcolor}}
\newcommand{\irSkb}[1]{{\color[named]{SkyBlue}#1\normalcolor}}
\newcommand{\irTeb}[1]{{\color[named]{TeaBlue}#1\normalcolor}}
\newcommand{\irSep}[1]{{\color[named]{Sepia}#1\normalcolor}}
\newcommand{\irReo}[1]{{\color[named]{RedOrange}#1\normalcolor}}
\newcommand{\irRur}[1]{{\color[named]{RubineRed}#1\normalcolor}}
\newcommand{\irSa }[1]{{\color[named]{Salmon}#1\normalcolor}}
\newcommand{\irAp}[1]{{\color[named]{Apricot}#1\normalcolor}}


%
\newenvironment{indention}[1]{\par
\addtolength{\leftskip}{#1}\begingroup}{\endgroup\par}
%
\newcommand{\namelistlabel}[1]{\mbox{#1}\hfill} 
\newenvironment{namelist}[1]{%
\begin{list}{}
{\let\makelabel\namelistlabel
\settowidth{\labelwidth}{#1}
\setlength{\leftmargin}{1.1\labelwidth}}
}{%
\end{list}}
%
%
\newcommand{\bfig}{\begin{figure}[t]}
\newcommand{\efig}{\end{figure}}
\setcounter{page}{1}

\newtheorem{theorem}{Theorem}
\newcommand{\Ep}{\mbox{\rm e}}

\newcommand{\Exp}{\exp
}
\newcommand{\idenc}{{\varphi}_n}
\newcommand{\resenc}{
{\varphi}_n}
\newcommand{\ID}{\mbox{\scriptsize ID}}
\newcommand{\TR}{\mbox{\scriptsize TR}}
\newcommand{\Av}{\mbox{\sf E}}

\newcommand{\Vl}{|}
\newcommand{\Ag}{(R,P_{X^n}|W^n)}
\newcommand{\Agv}[1]{({#1},P_{X^n}|W^n)}
\newcommand{\Avw}[1]{({#1}|W^n)}

\newcommand{\Jd}{X^nY^n}
\newcommand{\IdR}{r_n}

\newcommand{\Index}{{n,i}}

\newcommand{\cid}{C_{\mbox{\scriptsize ID}}}
\newcommand{\cida}{C_{\mbox{{\scriptsize ID,a}}}}
\newcommand{\ABC}{\mbox{\scriptsize ABC}}


\newcommand{\pOne}{\alpha}
\newcommand{\pTwo}{\beta}
\newcommand{\pThr}{\lambda}
\newcommand{\prmtB}{}

\arraycolsep 0.5mm
\date{}
%
\title{
New Strong Converse for Asymmetric Broadcast Channels
}

\author{%
\IEEEauthorblockN{Yasutada Oohama}
\IEEEauthorblockA{
  University of Electro-Communications, Tokyo, Japan \\
  Email: oohama@uec.ac.jp} 
}
\newcommand{\Empty}{
\author{%
Yasutada Oohama 
\thanks{
Y. Oohama is with 
University of Electro-Communications,
1-5-1 Chofugaoka Chofu-shi, Tokyo 182-8585, Japan.
}%
\thanks{
}
}
\markboth{
}
{
}

}

\maketitle

\begin{abstract}
We consider the discrete memoryless asymmetric broadcast channels. 
We prove that the error probability of decoding tends to one 
exponentially for rates outside the capacity region and 
derive an explicit lower bound of this exponent function. 
We shall demonstrate that the information spectrum approach 
is quite useful for investigating this problem.
\end{abstract}

\section{Asymmetric Broadcast Channels}

\begin{figure}[t]
\bc
\includegraphics[width=7.0cm]{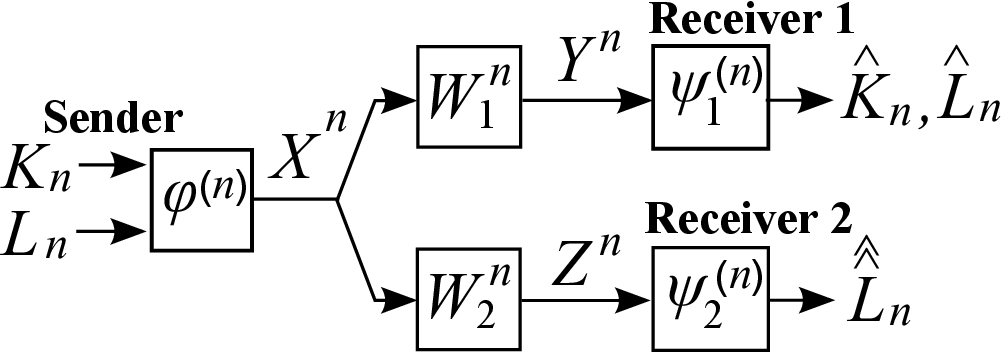}
\caption{Transmission of messages via the BC.}
\label{fig:ABC} 
\ec
\end{figure}

Let ${\cal X}, {\cal Y},$ ${\cal Z}$ be finite sets.
The broadcast channel we study in this paper is defined 
by a discrete memoryless channel specified with the following 
stochastic matrix:
\beq
{W} \defeq \{{W}(y,z| x)\}_{
(x,y,z) 
\in    {\cal X}
\times {\cal Y} 
\times {\cal Z}}.
\eeq
Here the set ${\cal X}$ corresponds to a channel input and 
the sets ${\cal Y}$ and ${\cal Z}$ correspond  to 
two channel outputs. Let $X^n$ be a random variable taking 
values in ${\cal X}^n$. We write an element of ${\cal X}^n$ as   
$x^n=x_{1}x_{2}$$\cdots x_{n}.$ 
Suppose that $X^n$ has a probability distribution on ${\cal X}^n$ 
denoted by 
$p_{X^n}=$ 
$\left\{p_{X^n}(x^n) 
\right\}_{{x^n} \in {\cal X}^n}$.
Similar notations are adopted for other random variables. 
Let $Y^n \in {\cal Y}^n$ and $Z^n \in {\cal Y}^n$  be random 
variables obtained as the channel output by connecting 
$X^n$ to the input of channel. 
We write a conditional distribution of $(Y^n,Z^n)$ 
on given $X^n$ as 
$$
W^n=
\left\{W^n(y^n,z^n|x^n)\right
\}_{(x^n,y^n,z^n)\in {\cal X}^n \times {\cal Y}^n \times {\cal Z}^n}.
$$
Since the channel is memoryless, we have 
\beq
W^n({y}^n,z^n|x^n)=\prod_{t=1}^nW (y_t,z_t|x_t).
\label{eqn:sde0}
\eeq
In this paper we deal with the case where the components 
$W({z},{y}|{x})$ of $W$ satisfy the following 
conditions:
\beq
W({y},{z}|{x})=W_1({y}|{x})W_2({z}|{ x}).
\label{eqn:sde1}
\eeq
In this case the broadcast channel(BC) 
is specified with $(W_1,W_2)$. Under the assumption 
of (\ref{eqn:sde1}), the conditional probability of (\ref{eqn:sde0})
is given by 
\beqno
W^n({y}^n,z^n|x^n)&=&W_1^n({y}^n|x^n)W_2^n(z^n|x^n)
\\
&=&\prod_{t=1}^nW_1(y_t|x_t)W_2(z_t|x_t).
\eeqno 
Transmission of messages via the BC is shown 
in Fig. \ref{fig:ABC}. Let $K_n$ and  $L_n$ be uniformly 
distributed random 
variables taking values in message sets ${\cal K}_n $ and ${\cal L}_n$, 
respectively. 
The random variable $K_n$ is a message sent to the receiver 1.
The random variable $L_n$ is a message sent to the receivers 1 and 2.
A sender transforms $K_n$ and $L_n$ into a transmitted 
sequence $X^n$ using an encoder function $\varphi^{(n)}$ 
and sends it to the receivers 1 and 2. 
In this paper we assume that the encoder function $\varphi^{(n)}$ 
is a stochastic encoder. In this case, $\varphi^{(n)}$ is 
a stochastic matrix given by
$$
\varphi^{(n)}=\{
\varphi^{(n)}(x^n|k,l)\}_{
(k,l,x^n)\in {\cal K}_n\times {\cal L}_n \times {\cal X}^n},
$$ 
where $\varphi^{(n)}(x^n|k,l)$ is a conditional probability 
of $x^n \in {\cal X}^n$ given message pair $(k,l)\in$
${\cal K}_n\times {\cal L}_n$.
The joint probability mass function on 
${\cal K}_n \times {\cal L}_n$ 
$\times {\cal X}^n$ 
$\times {\cal Y}^n$ 
$\times {\cal Z}^n$ 
is given by
\begin{align*}
&\Pr\{(K_n,L_n,X^n,Y^n,Z^n)=(k,l,x^n,y^n, z^n)\}
\nonumber\\
&=
\frac{\varphi^{(n)}(x^n|k,l)}{\pa{\cal K}_n\pa \pa{\cal L}_n\pa}
\prod_{t=1}^n W_1\left(y_t\left|x_t\right.\right)
              W_2\left(z_t\left|y_t\right.\right),
\end{align*}
where $\pa {\cal K}_n \pa$ is a cardinality 
of the set ${\cal K}_n$. The decoding functions 
at the receiver 1 and the receiver 2, respectively, 
are denoted by ${\psi}_1^{(n)}$ and ${\psi}_2^{(n)}$. 
Those functions are formally defined by
$
{\psi}_1^{(n)}: {\cal Y}^{n} \to {\cal K}_n \times {\cal L}_n,
{\psi}_2^{(n)}: {\cal Z}^{n} \to {\cal L}_n.
$
The average error probabilities of decoding at 
the receivers 1 and 2 are defined by 
\begin{align*}
& {\rm P}_{\rm e,1}^{(n)}={\rm P}_{\rm e}^{(n)}(\varphi^{(n)},\psi_1^{(n)})
 \defeq \Pr\{\psi_{1}^{(n)}(Y^n)\neq (K_n,L_n)\},
\\
& {\rm P}_{\rm e,2}^{(n)}={\rm P}_{\rm e}^{(n)}(\varphi^{(n)},\psi_2^{(n)})
\defeq \Pr\{\psi_{2}^{(n)}(Z^n)\neq L_n\}.
\end{align*}
Furthermore, we set
\begin{align*}
& {\rm P}_{\rm e}^{(n)}
={\rm P}_{\rm e}^{(n)}(\varphi^{(n)},\psi_1^{(n)},\psi_2^{(n)})
\\
& \defeq  \Pr\{\psi_{1}^{(n)}(Y^n)\neq (K_n,L_n) \mbox{ or } 
\psi_{2}^{(n)}(Z^n)\neq L_n \}
\end{align*}
It is obvious that we have the following relation.
\beq
{\rm P}_{\rm e}^{(n)}\leq 
{\rm P}_{\rm e,1}^{(n)}+ {\rm P}_{\rm e,2}^{(n)}. 
\eeq
For $k\in {\cal K}_n$ and $l\in {\cal L}_n$, set
\beqno
{\cal D}_1(k,l)&\defeq & \{ y^n: \psi_1^{(n)}(y^n)=(k,l) \},
\\
{\cal D}_2(l)&\defeq & \{ z^n: \psi_2^{(n)}(z^n)=l \}.
\eeqno
The families of sets 
$\{ {\cal D}_1(k,l) \}_{(k,l)\in {\cal K}_n\times {\cal L}_n}$ 
and $\{ {\cal D}_2(l) \}$ ${}_{l \in {\cal L}_n}$ are called 
the decoding regions. Using the decoding region, 
${\rm P}_{\rm e}^{(n)}$ can be written as

\newcommand{\OmitC}{
\beqno
{\rm P}_{\rm e}^{(n)}
&=&\frac{1}{|{\cal K}_n| |{\cal L}_n|} 
\sum_{(k,l)\in {\cal K}_n \times {\cal L}_n }
\sum_{\scs (x^n,y^n,z^n)\in {\cal X}^n \times {\cal Y}^n\times {\cal Z}^n:
       \atop{
       \scs y^n   \in {\cal D}_1^{c}(k,l)\mbox{ or }  
           \scs z^n  \in  {\cal D}_2^{c}(l)
       }
    }\OnE
\\
& &\times 
\varphi^{(n)}(x^n|k,l)W_1^n(y^n|x^n)W_2^n(z^n|x^n).
\eeqno
}
$$
{\rm P}_{\rm e}^{(n)}
=\Pr\{Y^n   \notin {\cal D}_1(K_n,L_n)\mbox{ or }
      Z^n   \notin {\cal D}_2(L_n)\}.
$$
Set 
\begin{align*}
& {\rm P}^{(n)}_{\rm c}=
   {\rm P}^{(n)}_{\rm c}(\varphi^{(n)},\psi_1^{(n)},\psi_2^{(n)})
\\
&\defeq  1-{\rm P}^{(n)}_{\rm e}(\varphi^{(n)},\psi_1^{(n)},\psi_2^{(n)}).
\end{align*}
The quantity ${\rm P}^{(n)}_{\rm c}$ is called the average 
correct probability of decoding. This quantity has the 
following form:
$$
{\rm P}_{\rm c}^{(n)}
=\Pr\{Y^n  \in {\cal D}_1(K_n,L_n),
      Z^n  \in {\cal D}_2(L_n)\}.
$$
\newcommand{\OmitF}{
\beqno
{\rm P}_{\rm c}^{(n)}
&=&\frac{1}{|{\cal K}_n| |{\cal L}_n|} 
\sum_{(k,l)\in {\cal K}_n \times {\cal L}_n }
\sum_{\scs (x^n,y^n,z^n)\in {\cal X}^n \times {\cal Y}^n\times {\cal Z}^n:
       \atop{
       \scs y^n   \in {\cal D}_1(k,l), 
           \scs z^n  \in  {\cal D}_2(l)
       }
    }\OnE
\\
& &\times 
\varphi^{(n)}(x^n|k,l)W_1^n(y^n|x^n)W_2^n(z^n|x^n).
\eeqno
}
For a given $(\varepsilon_1,\varepsilon_2)$ $\in (0,1)^2$, 
a rate pair $(R_1,R_2)$ is $(\varepsilon_1,$ 
$\varepsilon_2)$-{\it achievable} 
if there exists a sequence of triples $\{(\varphi^{(n)},$ 
$\psi_1^{(n)}, \psi_2^{(n)})\}_{n=1}^{\infty}$ 
such that for any $\delta>0$ and for any $n$ with 
$n\geq n_0=n_0(\varepsilon_1,\varepsilon_2,\delta)$ 
\beqa 
& &{\rm P}_{{\rm e,}i}^{(n)}(\varphi^{(n)},\psi_i^{(n)})\leq \varepsilon_i, i=1,2, 
\nonumber\\
& & \nbn \log \pa {\cal K}_n \pa  \geq  \irb{R_1}-\delta,
    \nbn \log \pa {\cal L}_n \pa \geq \irb{R_{2}}-\delta.
\nonumber
\eeqa
The set that consists of all $(\varepsilon_1,\varepsilon_2)$-achievable 
rate pair is denoted by 
${\cal C}_{\rm ABC}(\varepsilon_1,\varepsilon_2| W_1,W_2)$. 
For a given $\varepsilon$ $\in (0,1)$, a pair $(R_1,R_2)$ 
is $\varepsilon$-{\it achievable} if there exists 
a sequence of triples $\{(\varphi^{(n)},$ $\psi_1^{(n)},
\psi_2^{(n)}$ $)\}_{n=1}^{\infty}$ 
such that for any $\delta>0$ and for any $n$ with 
$n\geq n_0=n_0(\varepsilon,\delta)$ 
\beqa 
& &{\rm P}_{{\rm e}}^{(n)}
(\varphi^{(n)},\psi_1^{(n)},\psi_2^{(n)})
\leq \varepsilon, 
\nonumber\\
& &
\nbn \log \pa {\cal K}_n \pa \geq R_1-\delta,
\nbn \log \pa {\cal L}_n \pa \geq R_2-\delta.
\nonumber
\eeqa
The set that consists of all $\varepsilon$-achievable 
rate pair is denoted by 
${\cal C}_{\rm ABC}(\varepsilon|W_1,W_2)$.
It is obvious that for 
$0\leq $ $\varepsilon_1+\varepsilon_2\leq 1$, we have
$$
{\cal C}_{\ABC}(\varepsilon_1,\varepsilon_2|W_1,W_2)
\subseteq {\cal C}_{\ABC}(\varepsilon_1+\varepsilon_2|W_1,W_2).
$$
We set
$$
{\cal C}_{\rm ABC}(W_1,W_2)
\defeq \bigcap_{\varepsilon\in(0,1)}
{\cal C}_{\rm ABC}(\varepsilon|W_1,W_2),
$$
which is called the capacity region of the ABC.
The two maximum error probabilities of decoding 
are defined by as follows:
\beqno
{\rm P}_{{\rm e,\irOlg{m}},1}^{(n)}&=&
   {\rm P}_{{\rm e,\irOlg{m},1}}^{(n)}(\varphi^{(n)},\psi_1^{(n)})
\\
& \defeq & \max_{(k,l) \in {\cal K}_n\times {\cal L}_n}
\Pr\{\psi_1^{(n)}(Y^n)\neq k|K_n=k,L_n=l\},
\\
{\rm P}_{{\rm e,\irOlg{m}},2}^{(n)}&=&
   {\rm P}_{\rm e,\irOlg{m},2}^{(n)}(\varphi^{(n)},\psi_2^{(n)})
\\
& \defeq & \max_{l \in {\cal L}_n}
\Pr\{\psi_2^{(n)}(Z^n)\neq l|L_n=l\}.
\eeqno
Based on those quantities, we define the \irOlg{maximum capacity region} 
as follows. For a given $(\varepsilon_1,\varepsilon_2) \in (0,1)^2$, 
a pair $(\irb{R_1},\irb{R_2})$ 
is $(\varepsilon_1,\varepsilon_2)$-{\it achievable} 
if there exists a sequence of triples 
$\{(\varphi^{(n)},$ $\psi_1^{(n)}, \psi_2^{(n)})\}_{n=1}^{\infty}$ 
such that for any $\delta>0$ and for any $n$ with 
$n\geq n_0=n_0(\varepsilon_1,\varepsilon_2,\delta)$ 
\beqa 
&&{\rm P}_{{\rm e},\irOlg{\rm m},i}^{(n)}
(\varphi^{(n)},\psi_i^{(n)})
\leq \varepsilon_i,i=1,2, 
\nonumber\\
&& \nbn \log \pa {\cal K}_n \pa  \geq  \irb{R_1}-\delta,
\nbn \log \pa {\cal L}_n \pa \geq \irb{R_{2}}-\delta.
\nonumber
\eeqa
The set that consists of all $(\varepsilon_1,\varepsilon_2)$-achievable rate 
pair is denoted by 
${\cal C}_{\irOlg{\rm m}, \ABC}
(\varepsilon_1,\varepsilon_2|\irBr{W_1},\irBr{W_2})$. 
We set 
$$
{\cal C}_{\irOlg{\rm m},\ABC}(\irBr{W_1},\irBr{W_2})
=\bigcap_{(\varepsilon_1,\varepsilon_2)\in (0,1)^2} 
{\cal C}_{\irOlg{\rm m},\ABC}(\varepsilon_1,
\varepsilon_2|\irBr{W_1},\irBr{W_2}),
$$
which is called the \irOlg{maximum} capacity region of the ABC.
It is obvious that 
$$
{\cal C}_{\irOlg{\rm m},\ABC}(\varepsilon_1,\varepsilon_2|\irBr{W_1},\irBr{W_2})
\subseteq  
{\cal C}_{\ABC}(\varepsilon_1,\varepsilon_2|\irBr{W_1},\irBr{W_2}).
$$

We can show that the capacity regions 
${\cal C}_{\rm \ABC }( \varepsilon_1, \varepsilon_2|W_1$ $,W_2)$, 
${\cal C}_{\rm m, \ABC }( \varepsilon_1, \varepsilon_2 |$ $W_1,W_2)$, 
$(\varepsilon_1,\varepsilon_2) \in(0,1)^2$,  
${\cal C}_{\rm \ABC }( \varepsilon |$ $W_1,W_2)$, 
$\varepsilon \in(0,1)$, and ${\cal C}_{\rm \ABC }(W_1,W_2)$
satisfy the following property. 
\begin{pr}\label{pr:pro0a}
$\quad$
\begin{itemize}
\item[a)] 
The regions 
${\cal C}_{\rm \ABC }( \varepsilon_1, \varepsilon_2 |$ $W_1,W_2)$, 
${\cal C}_{\rm m, \ABC }( \varepsilon_1, \varepsilon_2 |$ $W_1,W_2)$, 
$(\varepsilon_1,\varepsilon_2) \in (0,1)^2$,  
${\cal C}_{\rm \ABC }(\varepsilon|W_1,W_2)$, 
$\varepsilon \in(0,1)$, 
and ${\cal C}_{\rm \ABC }($ $W_1,W_2)$
are closed convex sets of 
$\mathbb{R}_{+}^2$ $\defeq \{(R_1,R_2): R_1\geq 0, R_2\geq 0\}.$

\item[b)] 
${\cal C}_{\rm \ABC}(\varepsilon|W_1,W_2)$ has another 
form using $(n,\vep)$-rate region 
${\cal C}_{\rm \ABC}(n,\varepsilon |W_1,W_2)$,
the definition of which is as follows.
We set 
\beqno
& &{\cal C}_{\rm \ABC}(n,\varepsilon |W_1,W_2)
 \defeq \{(R_1,R_2): 
\\
& &\mbox{ There exists }(\varphi^{(n)},\psi_1^{(n)},\psi_2^{(n)})
\mbox{ such that }
\\
& &\frac{1}{n}\log |{\cal K}_n|\geq R_1,  
   \frac{1}{n}\log |{\cal L}_n|\geq R_2,  
\\
& &
{\rm P}_{\rm e}^{(n)}(\varphi^{(n)},\psi_1^{(n)},\psi_2^{(n)})
\leq \varepsilon\}.
\eeqno 
Using ${\cal C}_{\rm \ABC}(n,$ $\varepsilon|W_1,W_2)$, 
${\cal C}_{\rm \ABC}(\varepsilon|W_1,W_2)$ can be 
expressed as 
\begin{align*}
&
{\cal C}_{\rm \ABC}(\varepsilon |W_1,W_2)
\\
&={\rm cl}\left(\bigcup_{m\geq 1}
\bigcap_{n \geq m}{\cal C}_{\rm \ABC}(n,\varepsilon | W_1,W_2)
\right). 
\end{align*}

We define 
${\cal C}_{\rm \ABC}(n,\varepsilon_1,\varepsilon_2 |W_1,W_2)$ and 
${\cal C}_{\rm m, \ABC}(n,\varepsilon_1$ $, \varepsilon_2|W_1,W_2)$
in a manner similar to the definition of 
${\cal C}_{\rm \ABC}(n,$ $\varepsilon|W_1,W_2)$. 
Then we have the following:
\begin{align*}
&
{\cal C}_{\rm \ABC}(\varepsilon_1,\varepsilon_2 |W_1,W_2)
\\
&={\rm cl}\left(\bigcup_{m\geq 1}
\bigcap_{n \geq m}{\cal C}_{\rm \ABC}(n,\varepsilon_1,\varepsilon_2| W_1,W_2)
\right), 
\\
& {\cal C}_{\rm m, \ABC}(\varepsilon_1,\varepsilon_2|W_1,W_2)
\\
&={\rm cl}\left(\bigcup_{m\geq 1}
\bigcap_{n \geq m}{\cal C}_{\rm \ABC}(n,\varepsilon_1,\varepsilon_2| W_1,W_2)
\right). 
\end{align*}
%
\end{itemize}
\end{pr}

To describe previous works on 
${\cal C}_{\rm m,\ABC}(\varepsilon_1,\varepsilon_2|W_1,W_2)$ 
and ${\cal C}_{\ABC}(\varepsilon_1,\varepsilon_2|W_1,W_2)$, 
we introduce an auxiliary random variable $U$ taking 
values in a finite set ${\cal U}$. We assume that the 
joint distribution of $(U,X,Y,Z)$ is 
$$ 
p_{UXYZ}(u,x,y,z)=p_{U}(u)p_{X|U}(x|u)
W_1(y|x)W_2(z|x). 
$$
The above condition is equivalent to 
$Y \leftrightarrow X \leftrightarrow Z$
and 
$U \markov X \leftrightarrow (Y,Z)$. 
Define the set of probability distribution 
$p=p_{UXYZ}$ of $(U,$ $X,$ $Y,$ $Z)$ $\in$ ${\cal U}$
$\times{\cal X}$ $\times{\cal Y}$ $\times{\cal Z}$
by
\beqno
&&{\cal P}(W_1,W_2)
\defeq 
\{p: \pa {\cal U} \pa \leq 
\min\{\pa{\cal X}\pa, \pa{\cal Y}\pa+\pa{\cal Z}\pa \}+1,
\vSpa\\
& &\qquad p_{Y|X}=W_1, p_{Z|X}=W_2,
\vSpa\\ 
& &\qquad U \markov X \markov (Y,Z),
Y \markov X \markov Z \}.
\eeqno
Set 
\beqno
&&{\cal C}(p)
\defeq 
\ba[t]{l}
\{(R_1,R_2): R_1,R_2 \geq 0\,,
\vSpa\\
\ba[t]{rcl}
R_1 & \leq & I_p(X;Y|U),
\\
R_2 & \leq & I_p(U;Z),
\\
R_1+R_2 & \leq & I_p(X;Y)
\},
\ea
\ea
\\
& &{\cal C}(W_1,W_2)
= \bigcup_{p\in {\cal P}(W_1,W_2)}
{\cal C}(p).
\eeqno
We can show that the above functions and sets 
satisfy the following property. 
\begin{pr}\label{pr:pro0}  
$\quad$
\begin{itemize}
\item[a)] 
Set 
\beqno
&&{\cal C}_{\rm ext}(p)
\defeq 
\ba[t]{l}
\{(r_1,r_2,r_3): r_1,r_2,r_3 \geq 0\,,
\vSpa\\
\ba[t]{rcl}
r_1 & \leq & I_p(X;Y|U),
\\ 
r_2 &\leq &I_p(U;Z),
\\
r_3&\leq &I_p(X;Y)
\},
\ea
\ea
\\
& &{\cal C}_{\rm ext}(W_1,W_2)
\defeq \bigcup_{p\in {\cal P}(W_1,W_2)}
{\cal C}_{\rm ext}(p).
\eeqno
Then we have 
\begin{align*}
{\cal C}(W_1,W_2)=&\{(R_1,R_2):(R_1,R_2,R_1+R_2)
\\
&\: \in {\cal C}_{\rm ext}(W_1,W_2)\}.
\end{align*}

\item[b)] The cardinality bound in ${\cal P}(W_1,W_2)$ 
is sufficient to describe ${\cal C}(W_1,W_2)$ 
and ${\cal C}_{\rm ext}(W_1,W_2)$.

\item[c)] 
The region ${\cal C}(W_1,W_2)$ is a closed convex 
subset of $\mathbb{R}_{+}^2$ and the region 
${\cal C}_{\rm ext}(W_1,W_2)$ is a closed convex 
subset of $\mathbb{R}_{+}^3$ 
$\defeq \{(r_1,r_2,r_3): r_i \geq 0,i=1,2,3 \}.$

\item[d)] For any $(W_1,W_2)$, we have 
\beq
\max_{(R_1,R_2)\in {\cal C}(W_1,W_2)}(R_1+R_2)=C(W_1),
\label{eqn:SdEEE}
\eeq
where $C(W_1)$ is the capacity of the channel $W_1$. The maximun 
is attained by $(R_1,R_2)=(C(W_1),0)$. 
This result 
implies that 
the region ${\cal C}(W_1,W_2)$ is always contained in, and may
coincide with, the triangle with vertices 
$(0,0)$, $(C(W_1),0)$, $(0,C(W_1))$.
Furthermore, the point $(C(W_1),0)$ 
always belongs to ${\cal C}(W_1,W_2)$. In general, the upper boundary of 
${\cal C}(W_1,W_2)$ contains a line segment of slope $-1$ going 
through the point $(C(W_1),0)$ but this line segment may reduce 
to the point $(C(W_1),0)$.
\end{itemize}
\end{pr}

Property \ref{pr:pro0} part a) is obvious. Property \ref{pr:pro0} parts 
b) and c) are well known results. Property \ref{pr:pro0} part d) 
is found in \cite{km77}. Proof of those properties are omitted. 

The broadcast channel was posed and investigated by 
Cover \cite{cov72}. The capacity region of 
the ABC was given by K\"orner and Marton \cite{km77}.
They called the ABC the broadcast channels with degraded message sets.
K\"orner and Marton \cite{km77} obtained the following result. 
\begin{Th}[K\"orner and Marton \cite{km77}] \label{th:ddirect}
For each fixed $\varepsilon$ $\in (0,1)$ and 
for any $(W_1,W_2)$, we have
\begin{align*}
& {\cal C}_{\rm m, \ABC}(\varepsilon,\varepsilon|W_1,W_2)
={\cal C}_{\ABC}(W_1,W_2)
\\
&={\cal C}(W_1,W_2).
\end{align*}
\end{Th}

To prove this theorem they used a combinatorial lemma called 
``the blowing up lemma". Their method used to prove the above 
theorem was extended to the method called the entropy and image 
size characterization by Csisz\'ar and K\"orner \cite{ckBook81}, 
where they obtained the following result.  
%
%
\begin{Th}[Csisz\'ar and K\"orner \cite{ckBook81}]
For $\varepsilon \in (0,\frac{1}{2})$, 
\begin{align*}
& {\cal C}_{\ABC}(\varepsilon,\varepsilon|\irBr{W_1},\irBr{W_2})
  ={\cal C}_{\ABC}(\irBr{W_1},\irBr{W_2}).
\end{align*}
\end{Th}

Csisz\'ar and K\"orner \cite{ckBook81} applied the method of entropy and 
image size characterization to other coding problems in multi-user 
information theory to prove strong converse theorems for those problems. 
Universally attainable error exponents for rates inside the capacity 
region ${\cal C}(W_1,W_2)$ was studied by K\"orner and Sgarro \cite{ks80} 
and Kaspi and Merhav \cite{KaMer11}.

To examine an asymptotic behavior of ${\rm P}_{\rm c}^{(n)}$ 
for rates outside the capacity region ${\cal C}(W_1,W_2)$,
we define the following quantity. 
\beqno
& & 
G^{(n)}(R_1,R_2|W_1,W_2)
\\
&&\defeq
\min_{\scs 
(\varphi^{(n)},\psi_1^{(n)},\psi_2^{(n)}):
    \atop{\scs 
         (1/n)\log | {\cal K}_n |\geq R_1,
         \atop{\scs 
         (1/n)\log | {\cal L}_n |\geq R_2
         }
    }
}
\hspace*{-2mm}
\left(-\frac{1}{n}\right)
\log {\rm P}_{\rm c}^{(n)}
(\varphi^{(n)},\psi_1^{(n)},\psi_2^{(n)}).
\eeqno
By time sharing we have that 
\begin{align}
&   G^{(n+m)}\left(\left.
\frac{n R_1+m R_1^{\prime}}{n+m},
\frac{n R_2+m R_2^{\prime}}{n+m}\right|W_1,W_2\right) 
\nonumber\\
&\leq \frac{nG^{(n)}(R_1,R_2|W_1,W_2) 
+mG^{(m)}(R_1^{\prime},R_2^{\prime}|W_1,W_2)}{n+m}.
\label{eqn:aaZ} 
\end{align}
Choosing $R_1=R_1^\prime$ and $R_2=R_2^\prime$ in (\ref{eqn:aaZ}), 
we obtain the following subadditivity property
on $\{G^{(n)}(R_1,R_2|W_1,W_2)$ $\}_{n\geq 1}$: 
\begin{align*}
& G^{(n+m)}(R_1,R_2|{W}_1,W_2) 
\\
&\leq \frac{nG^{(n)}(R_1,R_2|W_1,W_2) 
+mG^{(m)}(R_1,R_2|W_1,W_2)}{n+m},
\end{align*}
from which we have that $G(R_{1},R_2|W_1,W_2)$ exists and 
satisfies the following:  
\beqno
\lim_{n\to\infty}G^{(n)}(R_{1},R_2|W_1,W_2) 
=\inf_{n\geq 1}G^{(n)}(R_1,R_2|W_1,W_2).
\eeqno
Set 
\beqno
&&G(R_1,R_2|W_1,W_2)\defeq
\lim_{n\to\infty}G^{(n)}(R_1,R_2|W_1,W_2),
\\
& &{\cal G}(W_1,W_2)\defeq \{
(R_1,R_2,G): G \geq G(R_1,R_2|W_1,W_2)\}.  
\eeqno
The exponent function $G(R_1,R_2|W_1,W_2)$ is a convex function 
of $(R_1,R_2)$. In fact, from (\ref{eqn:aaZ}), we have that 
for any $\nu \in [0,1]$
\begin{align*}
& G(\nu R_1+\bar{\nu }R_1^{\prime},
     \nu R_2+\bar{\nu }R_2^{\prime}|W_1,W_2)
\\
&\leq 
\nu G(R_1, R_2|W_1,W_2)
+\bar{\nu} G( R_1^{\prime},R_2^{\prime}|W_1,W_2).
\end{align*}
The region ${\cal R}(W_1,W_2)$ is also a closed convex set. 
Our main aim is to 
find an explicit characterization of ${\cal R}(W_1,W_2)$. In 
this paper we derive an explicit outer bound of ${\cal R}$ 
$(W_1,W_2)$ whose section by the plane $G=0$ coincides 
with ${\cal C}(W_1,W_2)$.

%
%
%

\section{Main Results}

In the present section we state our main results 
on the ABC. This section consists of two subsections. 
In the first subsection we state the first main result 
on an explicit lower bound of the optimal exponent 
function $G(R_1,R_2|W_1,W_2)$. 
This exponent function is positive outside the capacity 
region ${\cal C}(W_1,W_2)$ of the ABC. In the second subsection 
we state the second main resul an explicit outer bound of 
the $(n,\varepsilon)$-capacity region 
${\cal C}_{\rm ABC}(n,\varepsilon |W_1,W_2)$ of the ABC.  

\subsection{Explicit Lower bound of $G(R_1,R_2|W_1.W_2)$}

We first explain that the region ${\cal C}_{\rm ext }(W_1,W_2)$ 
can be expressed with two families of supporting hyperplanes. 
We define the set of 
probability distribution $p=p_{UXYZ}$ of $(U,$ $X,$ $Y,$ $Z)$ 
$\in$ ${\cal U}$ $\times{\cal X}$ $\times{\cal Y}$ $\times{\cal Z}$ by
\beqno
\lefteqn{{\cal P}_{\rm sh}(W_1,W_2)}\\
& \defeq & 
\{p=p_{UXYZ}:
\pa {\cal U} \pa \leq \min\{
\pa {\cal X} \pa, \pa {\cal Y} \pa+ \pa {\cal Z} \pa-1\}, 
\vSpa\\
&& \qquad p_{Y|X}=W_1, p_{Z|Y}=W_2,
\vSpa\\
& &\qquad Y\markov X\markov Z, 
 U \markov  X\markov (Y,Z) \}.
\eeqno
We set
\begin{align*}
& C^{(\mu,\gamma)}(W_1,W_2) 
\\
&\defeq 
\max_{p \in {\cal P}_{\rm sh}(W_1,W_2)}
\left\{\gamma \mu I_p(X;Y|U)+\gamma{\prmtB}I_p(U;Z) \right.
\\
& \qquad \qquad \quad 
              +\bar{\gamma}I_p(X;Y)\},
\\
& \underline{\cal C}_{\rm ext, sh}(W_1,W_2)
\defeq  
\bigcap_{\scs \mu\geq 0, \gamma\in [0,1]}
\ba[t]{l} 
\{(r_1,r_2,r_3): 
\\
 \gamma [\mu r_1+ {\prmtB} r_2] +\bar{\gamma}r_3 
\\
\leq C^{(\mu,\gamma)}(W_1,W_2)\},
\ea 
\\
& \underline{\cal C}_{\rm sh}(W_1,W_2)
\defeq 
\bigcap_{\scs 
\mu\geq 0, \gamma \in [0,1]}
\ba[t]{l} 
\{(R_1,R_2): 
\\
(\mu\gamma+\bar{\gamma})R_1+R_2 
\\
\leq C^{(\mu,\gamma)}(W_1,W_2)\},
\ea 
\\
& {\cal C}_{\rm sh}(W_1,W_2)
\defeq  
\bigcap_{\scs (\mu,\gamma)\in 
[0,1]^2}
\ba[t]{l} 
\{(R_1,R_2): 
\\
(\mu\gamma+\bar{\gamma})R_1+ R_2 
\\
\leq C^{(\mu,\gamma)}(W_1,W_2)\}.
\ea 
\end{align*} 

Then we have the following Property. 
\begin{pr}
\label{pr:pro0b}  
$\quad$
\begin{itemize}
\item[a)] 
The cardinality bound 
$
\pa {\cal U} \pa \leq \min\{$ $
\pa {\cal X} \pa, \pa {\cal Y} \pa+ \pa {\cal Z} \pa-1\}
$
appearing in the definition of ${\cal P}_{\rm sh}(W_1,W_2)$
is sufficient to describe 
$C^{(\mu,\gamma)}(W_1,W_2)$. 
\item[b)] For any $W_1$ and $W_2$, we have
\beqno
{\cal C}_{\rm ext}(W_1,W_2)&=&\underline{\cal C}_{\rm ext,sh}(W_1,W_2),
\\
{\cal C}(W_1,W_2)&=&\underline{\cal C}_{\rm sh}(W_1,W_2)
={\cal C}_{\rm sh}(W_1,W_2).
\eeqno
\end{itemize}
\end{pr}

Proof of Property \ref{pr:pro0b} part a) is stated as 
Lemma \ref{lm:CardLm} in Appendix \ref{sub:ApdaAAA}. 
Proof of this lemma is given in this appendix.
Proof of Property \ref{pr:pro0b} part b) 
is given in Appendix \ref{sub:ApdaAAB}.
\newcommand{\ApdaAAA}{
\subsection{
Cardinality Bound on Auxiliary Random Variables
}
\label{sub:ApdaAAA}

In this appendix we prove the cardinality bounds on auxiliary 
random variables appearing in this paper.
We first prove the following lemma.
\begin{lm}\label{lm:CardLm}
The cardinality bound 
$\pa {\cal U}\pa \leq 
\min\{\pa {\cal X} \pa,\pa {\cal Y}\pa+\pa{\cal Z}\pa$ $-1\}
$ appearing in the definition of ${\cal P}_{\rm sh}(W_1,W_2)$
is sufficient to describe $C^{(\mu,\gamma)}(W_1,W_2)$. 
\end{lm}

{\it Proof:} Observe that 
\beqa
\hspace*{-5mm}& &p_{X}(x)
=\sum_{u\in {\cal U}}p_U(u)p_{X|U}(x|u),
\label{eqn:asdfq}
\\
\hspace*{-5mm}& &
\left.
\ba{rcl}
p_{Y}(y)
&=&\ds \sum_{u\in {\cal U}} 
\sum_{x \in {\cal X}}p_U(u)W_1(y|x)p_{X|U}(x|u),
\\
p_{Z}(z)
&=&\ds \sum_{u\in {\cal U}} 
\sum_{x \in {\cal X}}p_U(u)W_2(z|x)p_{X|U}(x|u),
\ea
\right\}
\label{eqn:asdfqq}
\\
\hspace*{-5mm}
& &\gamma [\mu I_p(X;Y|U)+ {\prmtB} I_p(U;Z)] +\bar{\gamma}I_p(X;Y)
\nonumber\\
\hspace*{-5mm}
&&=\sum_{ u \in {\cal U}}p_{U}(u)[ 
\pi_1(p_{X|U}(\cdot|u))+\pi_2(p_{X|U}(\cdot|u))],
\label{eqn:aqqqaq}
\eeqa
where we set
\beqno
& &\pi_1(p_{{X}|U}(\cdot|u))
\defeq \sum_{(x,y)\in{\cal X}\times{\cal Y}}
p_{X|U}(x|u) W_1(y|x)
\\
& &\times \log \left\{
\frac{
\frac{\ds W_1(y|x)}
{\ds \sum_{\tilde{x} \in {\cal X}}
W_1(y|\tilde{x}) p_{X}(\tilde{x})}
}
{\left[\ds 
\frac{\ds \sum_{\tilde{x} \in {\cal X}}
W_1(y|\tilde{x}) p_{X|U}(\tilde{x}|u)}
{p_{Y}(y)
}
\right]^{\gamma \mu }}
\right\},
\\
&&\pi_2(p_{X|U}(\cdot|u))
\defeq \sum_{(x,z)\in{\cal X}\times{\cal Z}}
p_{X|U}(x|u) W_2(z|x)
\\
& &\times \log \left\{
\left[
\ds \frac{ 
\ds \sum_{\tilde{x} \in {\cal X}}
W_2(z|\tilde{x}) p_{X|U}( \tilde{x}|u)}{p_Z(z)}
\right]^{\gamma{\prmtB}}
\right\}.
\eeqno
We first derive a sufficient value of 
$|{\cal U}|$ to express $|{\cal X}|-1$ values 
of (\ref{eqn:asdfq}) and (\ref{eqn:aqqqaq}). 
Note that by (\ref{eqn:asdfqq})
the quantities $p_{Y}(\cdot)$ and $p_{Z}(\cdot)$ 
appearing in the above definitions of $\pi_i(p_{X|U}(\cdot|u)),$ 
$i=1,2$, are regarded as constants under (\ref{eqn:asdfq}). For 
each $u\in {\cal U}$, $\pi_i(p_{{X}|U}(\cdot|u)),$ $i=1,2$ is a 
continuous function of $p_{X|U}(\cdot|u)$. Then by the support 
lemma,
$$
|{\cal U}| \leq |{\cal X}|-1 +1= |{\cal X}| 
$$
is sufficient to express $|{\cal X}|-1$ values 
of (\ref{eqn:asdfq}) and one value of (\ref{eqn:aqqqaq}). 
We next derive a sufficient value of $|{\cal U}|$ to express 
$|{\cal Y}|+$ $|{\cal Z}|-2$ values of (\ref{eqn:asdfqq}) 
and (\ref{eqn:aqqqaq}). 
Note that the quantities $p_{Y}(\cdot)$ and $p_{Z}(\cdot)$ 
appearing in the above definitions of $\pi_i(p_{X|U}(\cdot|u)),$ 
$i=1,2$, are regarded as constants under (\ref{eqn:asdfqq}). For 
each $u\in {\cal U}$, $\pi_i(p_{{X}|U}(\cdot|u)),$ $i=1,2$ is a 
continuous function of $p_{X|U}(\cdot|u)$. Then by the support 
lemma,
$$
|{\cal U}| \leq |{\cal Y}|+|{\cal Z}|-2 +1=
|{\cal Y}|+|{\cal Z}|-1
$$
is sufficient 
to express $|{\cal Y}|+|{\cal Z}|-2$ values of 
(\ref{eqn:asdfqq}) and 
one value of (\ref{eqn:aqqqaq}). 
\hfill \IEEEQED
} 
\newcommand{\ApdaAAB}{
\subsection{
Supporting Hyperplane Expression of the Capacity Region 
}
\label{sub:ApdaAAB}


In this appendix we prove Property \ref{pr:pro0b} part b). 
From Property \ref{pr:pro0} part c), we have the 
following lemma. 
\begin{lm}\label{lm:asgsq} 
Suppose that  
$(\hat{r}_1,
  \hat{r}_2,
  \hat{r}_3)$ 
does not belong to ${\cal C}_{\rm ext}(W_1,W_2)$. 
Then there exist $\epsilon>0$ and 
$\mu^*\geq 0,$ $\gamma^*\in [0,1]$ such that 
for any $(r_1,r_2,r_3)\in {\cal C}_{\rm ext}(W_1,W_2)$
we have  
\begin{align*}
&\gamma^* \mu^*   (r_1-\hat{r}_1)
        +\gamma^*{\prmtB} (r_2-\hat{r}_2)
      +\bar{\gamma}^*   (r_3-\hat{r}_3)
  +\epsilon \leq 0.
\end{align*}
\end{lm} 

Proof of this lemma is omitted here. This lemma will be used 
to prove Property \ref{pr:pro0b} part b). 

{\it Proof of Property \ref{pr:pro0b} part b):} \ We recall 
the following definitions of 
${\cal P}(W_1,W_2)$, ${\cal P}_{\rm sh}(W_1,W_2)$, and ${\cal Q}$: 
\begin{align*}
&{\cal P}(W_1,W_2)
\\
&\defeq \{p=p_{UXYZ}: \pa {\cal U} \pa \leq 
        \min\{ \pa {\cal X} \pa, \pa {\cal Y} \pa+ \pa {\cal Z} \pa\}+1, 
\\
&\quad p_{Y|X}=W_1,p_{Z|X}=W_2,
\\
&\quad Y \markov X \markov Z, U \markov  X \markov (Y,Z) \},
\\
&{\cal P}_{\rm sh}(W_1,W_2)
\\
&\defeq \{p=p_{UXYZ}: \pa {\cal U} \pa \leq 
          \min\{\pa {\cal X} \pa, \pa {\cal Y}\pa+\pa {\cal Z} \pa-1\}, 
\\
&\quad p_{Y|X}=W_1, p_{Z|Y}=W_2,
\\
&\quad Y \markov X\markov Z, U \markov  X\markov (Y,Z) \},
\\
& {\cal Q} \defeq \{q=q_{UXYZ}:\pa {\cal U} \pa 
           \leq \pa {\cal Y}\pa + \pa {\cal Z} \pa-1 \}.
\end{align*}
We first prove 
$\underline{\cal C}_{\rm ext,sh}($ $W_1,W_2)$ 
$\subseteq $ ${\cal C}_{\rm ext}$ $(W_1,W_2)$. 
We assume that 
$(\hat{r}_1,$ 
$\hat{r}_2$, 
$\hat{r}_3) \notin $ 
${\cal C}_{\rm ext}(W_1,W_2)$.
Then by Lemma \ref{lm:asgsq}, there exist 
$\epsilon>0$ and $\mu^*\geq 0, \gamma\in [0,1]$, 
such that for any $(r_1,r_2,r_3) \in {\cal C}_{\rm ext}(W_1,W_2)$
we have 
\beqno
&&     \gamma^* \mu^*(r_1-\hat{r}_1)
      +\gamma^*      (r_2-\hat{r}_2)
 +\bar{\gamma}^*     (r_3-\hat{r}_3) +\epsilon \leq 0.
\eeqno
Then we have
\begin{align}
&\gamma^*(\mu^* \hat{r}_1 + {\prmtB} \hat{r}_2)
  +\bar{\gamma}^* \hat{r}_3 
\nonumber\\
&\geq 
   \max_{\scs (r_1,r_2,r_3)
   \atop{\scs \in {\cal C}(W_1,W_2)}} 
   \left\{ \gamma^*(\mu^* r_1+ {\prmtB} r_2)
          +\bar{\gamma^*} r_3 \right\} 
   +\epsilon
\nonumber\\ 
&\MEq{a} \max_{p\in {\cal P}(W_1,W_2)}
\{ \gamma^*\mu^* I_p(X;Y|U) + \gamma^* {\prmtB} I_p(U;Z)
\nonumber\\ 
& \qquad\qquad\qquad +\bar{\gamma}^* I_p(X;Y) \}+\epsilon
\nonumber\\ 
&\geq  \max_{p\in {\cal P}_{\rm sh}(W_1,W_2)}
\{ \gamma^*\mu^* I_p(X;Y|U) + \gamma^* {\prmtB} I_p(U;Z)
\nonumber\\ 
& \qquad\qquad\qquad +\bar{\gamma}^* I_p(X;Y) \}+\epsilon
\nonumber\\ 
&=C^{(\gamma^*,\mu^*)}(W_1,W_2)+\epsilon.
\label{eqn:sddsd}
\end{align}
Step (a) follows from the definition of ${\cal C}_{\rm ext}(W_1,W_2)$. 
The bound (\ref{eqn:sddsd}) implies that 
$(\hat{r}_1,\hat{r}_2,$$\hat{r}_3)$ 
$\notin \underline{\cal C}_{\rm ext,sh}(W_1,W_2)$.
Thus $\underline{\cal C}_{\rm ext,sh}(W_1,W_2)
\subseteq {\cal C}_{\rm ext}(W_1,W_2)$ is proved. 
We next prove ${\cal C}_{\rm ext}($ $W_1,W_2)$ 
$\subseteq $ $\underline{\cal C}_{\rm ext, sh}$
$(W_1,W_2)$. We assume that $(R_1,$ $R_2) \in$ ${\cal C}_{\rm ext}(W_1,W_2)$.
Then there exists $p\in$ ${\cal P}$ $(W_1,W_2)$ such that
\beq
\left. 
\ba{rcl}
r_1&\leq &I_p(X;Y|U), 
r_2 \leq I_p(U;Z),
\\
r_3&\leq &I_p(X;Y).
\ea
\right\}
\label{eqn:assz0}
\eeq
Then, for $(r_1,r_2,$ $r_3)$ $ \in {\cal C}_{\rm ext}(W_1,W_2)$, 
we have the following chain of inequalities:
\begin{align*}
& \gamma(\mu r_1 +{\prmtB} r_2)+ \bar{\gamma} r_3 
\\
&\MLeq{a} \gamma[ \mu I_p(X;Y|U)+{\prmtB} I_p(U;Z)] 
+\bar{\gamma}I_p(X;Y)
\\
&\leq \max_{p \in {\cal P}(W_1,W_2)}
 \ba[t]{l}
 \{\gamma \mu I_p(X;Y|U)+\gamma{\prmtB} I_p(U;Z)
 \vSpa\\
  \:\:+\bar{\gamma}I_p(X;Y)\}
\ea
\\
&\MEq{b}
\max_{p \in {\cal P}_{\rm sh}(W_1,W_2)}
\ba[t]{l}
\{\gamma[\mu I_p(X;Y|U) + {\prmtB} I_p(U;Z)]
\vSpa\\
\:\:+\bar{\gamma}I_p(X;Y)\}
\ea
\\
&=C^{(\mu,\gamma)}(W_1,W_2).
\end{align*}
Step (a) follows from (\ref{eqn:assz0}).
Step (b) follows from that by 
Property \ref{pr:pro0b} part a), the cardinality bound in 
${\cal P}_{\rm }(W_1,W_2)$ can be reduced to 
that in ${\cal P}_{\rm sh}(W_1,W_2)$. 
Hence we have ${\cal C}_{\rm ext}(W_1,W_2)$ $\subseteq $ 
$\underline{\cal C}_{\rm ext,sh}(W_1,W_2)$.
We thirdly prove 
$\underline{\cal C}_{\rm sh}(W_1,W_2)={\cal C}($ $W_1,W_2)$. 
We have the following chain of equalities:
\begin{align*}
&{\cal C}_{\rm sh}(W_1,W_2)=\{(R_1,R_2): 
\\
&\quad (\mu\gamma+\bar{\gamma})R_1+R_2 
   \leq C^{(\mu,\gamma)}(W_1,W_2)
\\
&\quad \mbox{ for any }(\mu,\gamma)\in [0,1]^2\}
\\
&=\{(R_1,R_2): 
\\
&\quad \mu\gamma R_1+{\prmtB}\gamma R_2+\bar{\gamma}(R_1+R_2) 
\leq C^{(\mu,\gamma)}(W_1,W_2)
\\
&\quad \mbox{ for any }(\mu,\gamma) \in [0,1]^2\}
\\
&=\{(R_1,R_2):(R_1,R_2,R_1+R_2)\in \underline{\cal C}_{\rm ext,sh}(W_1,W_2) \}
\\
&\MEq{a}\{(R_1,R_2):(R_1,R_2,R_1+R_2)\in {\cal C}_{\rm ext}(W_1,W_2)\}
\\
&={\cal C}(W_1,W_2).
\end{align*}
Step (a) follows from 
$\underline{\cal C}_{\rm ext,sh }(W_1,W_2)$
$=$${\cal C}_{\rm ext}(W_1,W_2)$. 
We finally show ${\cal C}_{\rm sh}(W_1,W_2)
={\cal C}(W_1,W_2)$. For any $\mu \geq 0,$ $\gamma\in [0,1]$, we have 
the following chain of inequalities:
\begin{align}
&\max_{(R_1,R_2)\in {\cal C}(W_1,W_2)}
\{(\gamma\mu+\bar{\gamma})R_1 +R_2 
\notag\\
&=\max_{ \scs (R_1,R_2):
   \atop{ \scs (R_1,R_2,R_1+R_2)
       \atop{\scs  \in {\cal C}_{\rm ext}(W_1,W_2)}
        }
   }\{ \gamma\mu R_1+\gamma{\prmtB} R_2+\gamma(R_1+R_2)\}
\notag\\
&\leq
\max_{p \in {\cal P}(W_1,W_2)}
\left\{\gamma \mu I_p(X;Y|U)+\gamma{\prmtB}I_p(U;Z) \right.
\notag\\
& \qquad \qquad \quad 
              +\bar{\gamma}I_p(X;Y)\}
\notag\\
&\MEq{a}
\max_{p \in {\cal P}_{\rm sh}(W_1,W_2)}
\left\{\gamma \mu I_p(X;Y|U) + \gamma {\prmtB} I_p(U;Z) \right.
\notag\\
& \qquad \qquad \quad 
              +\bar{\gamma}I_p(X;Y)\}
\notag\\
&=C^{(\mu,\gamma)}(W_1,W_2).
\label{eqn:SDAzz}
\end{align}
Step (a) follows from Property \ref{pr:pro0b} part a).
We now consider the case where $\mu\geq 1,\gamma\in [0,1]$. 
In this case we have the following:
\begin{align}
&\max_{(R_1,R_2)\in {\cal C}(W_1,W_2)}
\{(\gamma\mu+\bar{\gamma})R_1+R_2\} 
\notag\\
&\leq 
 (\gamma\mu+\bar{\gamma})\max_{(R_1,R_2)\in {\cal C}(W_1,W_2)}
\{R_1+R_2\} 
\notag\\
& \MEq{a} (\gamma\mu+\bar{\gamma})C(W_1).
\label{eqn:SDAzzP}
\end{align}
Step (a) follows from Property \ref{pr:pro0} part d). 
In (\ref{eqn:SDAzzP}), the equality is attained by $(R_1,R_2)=(C(W_1),0)$.
Thus we have that for any $\mu\geq 1,\gamma\in [0,1]$, 
\begin{align}
& \max_{(R_1,R_2)\in {\cal C}(W_1,W_2)}
\{(\gamma\mu+\bar{\gamma})R_1+R_2\}
\nonumber\\
&=(\gamma\mu+\bar{\gamma})C(W_1). 
\label{eqn:aSDDD}
\end{align}
From (\ref{eqn:SDAzz}) and (\ref{eqn:aSDDD}), we have that for any 
$\mu\geq 1$ and any $\gamma\in$ $ [0,1]$,
\beq
C^{(\mu,\gamma)}(W_1,W_2) \geq (\gamma\mu+\bar{\gamma})C(W_1).
\label{eqn:ssaSDDD}
\eeq
Set
\begin{align*}
& \check{\cal C}_{\rm sh}(W_1,W_2)
= 
\bigcap_{\scs \mu\geq 1,\gamma\in [0,1]}
\ba[t]{l} 
\{(R_1,R_2): 
\\
(\mu\gamma+\bar{\gamma})R_1+R_2 
\\
\leq C^{(\mu,\gamma)}(W_1,W_2)\}.
\ea 
\end{align*}
Then we have 
\beqa
&&\check{\cal C}_{\rm sh}(W_1,W_2)\cap{\cal C}_{\rm sh}(W_1,W_2)
\nonumber\\
&&=\underline{\cal C}_{\rm sh}(W_1,W_2)={\cal C}(W_1,W_2).  
\label{eqn:aSffDDD}
\eeqa
On $\check{\cal C}_{\rm sh}(W_1,W_2)$, we have the following: 
\begin{align}
& \check{\cal C}_{\rm sh}(W_1,W_2)
\MSupeq{a} 
\bigcap_{\scs \mu\geq 1,\gamma\in [0,1]}
\ba[t]{l} 
\{(R_1,R_2): 
\\
(\gamma\mu+\bar{\gamma})R_1+R_2 
\\
\leq (\gamma\mu+\bar{\gamma})C(W_1)\}
\ea 
\notag\\
&\MSupeq{b} 
\bigcap_{\scs \mu\geq 1,\gamma\in [0,1]}
\ba[t]{l} 
\{(R_1,R_2): 
R_2 \leq C(W_1)-R_1
\}
\ea 
\notag\\
&=\ba[t]{l} 
\{(R_1,R_2): 
R_1+ R_2 \leq C^{(1,0)}(W_1)
\}
\ea 
\notag\\
&\supseteq  {\cal C}_{\rm sh}(W_1,W_2).  
\label{eqn:SErrr}
\end{align}
Step (a) follows from (\ref{eqn:ssaSDDD}).
Step (b) follows from $\gamma\mu+\bar{\gamma}$ $\geq 1$.
From (\ref{eqn:aSffDDD}), (\ref{eqn:SErrr}), we have 
${\cal C}_{\rm sh}(W_1,W_2)$ $={\cal C}(W_1,W_2).$
\hfill\IEEEQED
}
Set
\begin{align*}
&{\cal Q}
 \defeq 
 \{q=q_{UXYZ}:\pa {\cal U} \pa \leq 
\pa {\cal Y} \pa + \pa {\cal Z} \pa-1 \}.
\end{align*}
Define 
\begin{align*}
& \omega^{(\mu,\gamma,\pOne)}_{q}(x,y,z|u)
\defeq \log \left[\frac{W_1(y|x)W_2(z|x)}
{q_{YZ|XU}(y,z|x,u)}\right]
\\
&\quad +\pOne \gamma\left[
          \mu \log \frac{W_1(y|x)}{q_{Y|U}(y|u)}
    +{\prmtB}\log \frac{q_{Z|U}(z|u)}{q_{Z}(z)}
    \right]
\\
&\quad +\pOne\bar{\gamma}\log \frac{W_1(y|x)}{q_Y(y)}.
\end{align*}
Furthermore, define
\begin{align*}
& \Lambda^{(\mu,\gamma,\pOne,\pTwo)}(q|W_1,W_2)
\defeq 
{\rm E}_q
\left[\exp\left\{\pTwo \omega^{(\mu,\gamma,\pOne)}_{q}(X,Y,Z|U)
\right\}\right],
\\
&  \Omega^{(\mu,\gamma,\pOne,\pTwo)}(q|W_1,W_2)
\defeq\log \Lambda^{(\mu,\gamma,\pOne,\pTwo)}(q|W_1,W_2),
\\
&  {\Omega}^{(\mu,\gamma,\pOne,\pTwo)}(W_1,W_2)
 \defeq \max_{q\in {\cal Q}} 
\Omega^{(\mu,\gamma,\pOne,\pTwo)}(q|W_1,W_2),
\\
& F^{(\mu,\gamma,\pOne,\pTwo)}(R_1,R_2|W_1,W_2)
\\
&\defeq 
\frac{\pTwo\pOne \bigl[ 
 (\gamma\mu +\bar{\gamma})R_1+R_2 \bigr]
 -{\Omega}^{(\mu,\gamma,\pOne,\pTwo)}(W_1,W_2)}
{1+\pTwo[2+\pOne(1+\gamma+\gamma\mu)]},
\\
& {F}(R_1,R_2|W_1,W_2)
\\& \defeq \sup_{\scs 
(\mu,\gamma) \in [0,1]^2,
 \atop{\scs \pOne,\pTwo\geq 0}}
F^{(\mu,\gamma,\pOne,\pTwo)}(R_1,R_2|W_1,W_2),
\\
& \overline{\cal R}(W_1,W_2)
\defeq 
\left\{(R_1,R_2,G): G\geq F(R_1,R_2|W_1,W_2)\right\}.
\end{align*}
We next define a function serving as a lower bound of 
$F(R_1,R_2|W_1,W_2)$. For each 
$p_{UXYZ}\in {\cal P}_{\rm sh}(W_1,W_2)$, define
\begin{align*}
& \tilde{\omega}_{p}^{(\mu,\gamma)}(x,y,z|u)
\\
&\defeq  \gamma\biggl[
           \mu \log \frac{W_1(y|x)}{p_{Y|U}(y|u)}
+{\prmtB}\log \frac{p_{Z|U}(z|u)}{p_{Z}(z)}
     \biggr]+\bar{\gamma}\log \frac{W_1(y|x)}{p_Y(y)},
\\
& \tilde{\Omega}^{(\mu,\gamma,\lambda)}(p)
\defeq 
\log 
{\rm E}_{p}
\left[\exp\left\{\lambda
\tilde{\omega}^{(\mu,\gamma)}_p (X,Y,Z|U)\right\}\right].
\end{align*}
Furthermore, set
\begin{align*}
& \tilde{\Omega}^{(\mu,\gamma,\lambda)}(W_1,W_2)
 \defeq \min_{\scs \atop{\scs 
p \in {{\cal P}_{\rm sh}(W_1,W_2)}}}
\tilde{\Omega}^{(\mu,\gamma,\lambda)}(p),
\\
&\tilde{F}^{(\mu,\gamma,\pOne)}(R_1,R_2|W_1,W_2)
\\
&\defeq
\frac{\lambda \bigl[ 
 (\gamma\mu +\bar{\gamma})R_1+ R_2 \bigr]
 -{\Omega}^{(\mu,\gamma,\lambda)}(W_1,W_2)}
{3+10\lambda},
\\
& \tilde{F}(R_1,R_2|W_1,W_2)
\\
&=\sup_{\scs (\mu,\gamma)\in [0,1]^2,
 \atop{\scs \lambda\geq 0}
}
\tilde{F}^{(\mu,\gamma,\lambda)}(R_1,R_2|W_1,W_2).
\end{align*}

We can show that the above functions and sets 
satisfy the following property. 
\begin{pr}\label{pr:pro1}  
$\quad$
\begin{itemize}
\item[a)]
The cardinality bound 
$\pa {\cal U} \pa \leq 
 \pa {\cal Y} \pa + \pa {\cal Z} \pa-1$
appearing in the definition of ${\cal Q}$ is 
sufficient to describe 
$\Omega^{(\mu,\gamma,\pOne,\pTwo)}(W_1,W_2)$.
Furthermore, the cardinality bound 
$
\pa {\cal U} \pa \leq \min\{$ $
\pa {\cal X} \pa, \pa {\cal Y} \pa+ \pa {\cal Z} \pa-1\}
$
appearing in the definition of ${\cal P}_{\rm sh}(W_1,W_2)$
is sufficient to describe 
$\tilde{\Omega}^{(\mu,\gamma,\lambda)}(W_1,W_2)$. 

\item[b)] 
For any $R_1,R_2>0$, we have  
\beqno
& &F(R_1,R_2|W_1,W_2)\geq \tilde{F}(R_1,R_2|W_1,W_2).
\eeqno

\item[c)]
For any $(\mu,\gamma, \lambda)\in [0,1]^2\times [0,1/2]$ 
and any $q \in {\cal P}_{\rm sh}(W_1,W_2)$, we have 
$$
1 \leq \Lambda^{(\mu,\lambda)}(q)\leq 
|{\cal X}|^2|{\cal Y}||{\cal Z}|.
$$

\item[d)]
Fix any $q \in {\cal P}_{\rm sh}(W_1,W_2)$ and $\mu \in[0,1]$.
Define a probability distribution 
$q^{(\lambda)}=q_{UXYZ}^{(\lambda)}$ by
\begin{align*} 
& q^{(\lambda)}(u,x,y,z)
\\
&\defeq
\frac{
q(u,x,y,z)
\exp\left\{\lambda
\tilde{\omega}^{(\mu,\gamma)}_{q }(x,y,z|u)\right\}
}{{\rm E}_{q}
\left[\exp\left\{\lambda 
\tilde{\omega}^{(\mu,\gamma)}_{q}(X,Y,Z|U)\right\}\right]}.
\end{align*}
\irr{Then, for $\lambda \in [0,1/4]$, $\Omega^{(\mu, \lambda)}(q)$ 
is twice differentiable. Furthermore, for 
$\lambda \in [0,1/4]$, we have }
\beqno
& & \frac{\rm d}{{\rm d}\lambda} 
\tilde{\Omega}^{(\mu,\gamma, \lambda)}(q)
={\rm E}_{q^{(\lambda)}}
\left[ \tilde{\omega}^{(\mu,\gamma)}_{q}(X,Y,Z|U)
\right],
\\
& & \frac{\rm d^2}{{\rm d}\lambda^2} \tilde{\Omega}
^{(\mu,\gamma,\lambda)}(q)
={\rm Var}_{q^{(\lambda)}}
\left[\tilde{\omega}^{(\mu,\gamma)}_{q}(X,Y,Z|U)\right].
\eeqno
The second equality implies that 
$\tilde{\Omega}^{(\mu,\gamma,\lambda)}(q)$ 
is a convex function of $\lambda \in [0,1/4]$. 
\item[e)]
For $(\mu,\gamma,\lambda) \in [0,1]^2\times [0,1/4]$, 
define
\begin{align*}
& \rho^{(\mu,\gamma,\lambda)}(W_1,W_2)
\\
& \defeq
\max_{\scs  \irr{(c,q)\in [0,\lambda]}
      \atop{\scs \irr{\times} {\cal P}_{\rm sh}(W_1,W_2),
            \atop{\scs \tilde{\Omega}^{(\mu,\gamma,\lambda)}(q)
                 \atop{\scs =\tilde{\Omega}^{(\mu,\gamma,\lambda)}({W}_1,W_2)
                  }
             }
      }
}
{\rm Var}_{q^{\irr{(c)}}}
\left[\tilde{\omega}^{(\mu,\gamma)}_{q}
(X,Y,Z|U)\right],
\end{align*}
and set 
\begin{align*}
& \rho=\rho(W_1,W_2)
\defeq \max_{\scs (\mu,\gamma,\lambda)
\atop{\scs 
\in [0,1]^2\times [0,1/4]
}}
\rho^{(\mu,\gamma,\lambda)}(W_1,W_2).
\end{align*}
Then, we have $\rho(W_1,W_2)< \infty$. Furthermore, 
for any $(\mu,\gamma, \lambda) \irr{ \in [0,1]^2 \times [0,1/4] }$, 
we have  
\begin{align*}
& \Omega^{(\mu, \gamma,\lambda)}(W_1,W_2) 
\\
& \leq \lambda C^{(\mu,\gamma)}(W_1,W_2)
+\frac{\lambda^2}{2}\rho(W_1,W_2).
\end{align*}

\item[f)] For every $\tau \in (0,(1/4)\rho(W_1,W_2)]$,
$(R_1, R_2-{\tau}) \notin {\cal C}(W_1,W_2)$
implies 
\begin{align*}
&\tilde{F}(R_1,R_2|W_1,W_2)
\\
& > \frac{\rho(W_1,W_2)}{6} \cdot g^2
  \left(\frac{\tau}{\rho(W_1,W_2)}\right)>0,
\end{align*}
where $g$ is the inverse function of 
$\vartheta(a) \defeq a+(5/3)a^2, a\geq 0$.
\end{itemize}
\end{pr}

Proof of Property \ref{pr:pro0b} part a) is stated as 
Lemma \ref{lm:CardLmA} in Appendix \ref{sub:ApdaAAA}. 
Proof of this lemma is given in this appendix.
Proofs of Property \ref{pr:pro1} 
part b) is given in Appendix \ref{sub:ApdaAACa}. 
Proofs of Property \ref{pr:pro1} parts c), d), e) are 
given in Appendix \ref{sub:ApdaAACb}. 
Proof of Property \ref{pr:pro1} part f) are 
given in Appendix \ref{sub:ApdaAACc}. 
\newcommand{\ApdaAACaa}{

We next prove the following lemma.
\begin{lm}\label{lm:CardLmA}    
The cardinality bound 
$\pa {\cal U} \pa \leq 
 \pa {\cal Y} \pa + \pa {\cal Z} \pa-1$
appearing in the definition of ${\cal Q}$ is 
sufficient to describe 
$\Omega^{(\mu,\gamma,\pOne,\pTwo)}(W_1,W_2)$.
Furthermore, the cardinality bound 
$\pa {\cal U} \pa \leq \min\{$$\pa {\cal X}\pa, 
\pa {\cal Y} \pa+\pa {\cal Z}\pa-1\}$ appearing 
in the definition of ${\cal P}_{\rm sh}(W_1,W_2)$
is sufficient to describe 
$\tilde{\Omega}^{(\mu,\gamma,\lambda)}(W_1,W_2)$. 
\end{lm} 

{\it Proof: }We first bound  
the cardinality $|{\cal U}|$ of ${U}$ in ${\cal Q}$ 
to show that the bound 
$|{\cal U}| \leq |{\cal Y}|$$+|{\cal Z}|-1$
is sufficient to describe 
${\Omega}^{(\mu,\gamma,\pOne,\pTwo)}$ 
$(W_1,W_2)$. Observe that 
\begin{align}
& 
\left.
\ba{l} 
\ds
q_{{Y}}(y)
=\sum_{u\in {\cal U}}q_U(u)
q_{Y|U}(y|u),
\vspace*{1mm}\\
\ds q_{Z}(z)
=\sum_{u\in {\cal U}}q_U(u)
q_{Z|U}(z|u),
\ea
\right\}
\label{eqn:asdf}
\\
& \exp\left\{ 
\Omega^{(\mu,\gamma,\pOne,\pTwo)}(q|W_1,W_2)
\right\}
\nonumber\\
&=\sum_{u\in {\cal U}}q_U(u)
\zeta^{(\mu,\gamma,\pOne,\pTwo)}(q_{XYZ|U}(\cdot|u)),
\label{eqn:aqqqa}
\end{align}
where we set
\begin{align*}
& \zeta^{(\mu,\gamma,\pOne,\pTwo)}(q_{XYZ|U}(\cdot,\cdot,\cdot|u))
\\
&\defeq  \sum_{\scs (x,y,z)\atop{\scs 
\in{\cal X}\times{\cal Y}\times{\cal Z}}}
q_{{XYZ}|U}(x,y,z|u)
\exp\left\{\pTwo 
\omega^{(\mu,\gamma,\pOne)}_{q}(x,y,z|u)\right\}.
\end{align*}
For the quantities $q_{Y}(\cdot)$ and $q_{Z}(\cdot)$ 
contained in the forms of 
$\zeta^{(\mu,\gamma,\pOne)}$ 
$(q_{XYZ|U}(\cdot|u)),$ $u\in {\cal U}$, we regard them as 
constants under (\ref{eqn:asdf}). For each $u \in{\cal U}$, 
$
\zeta^{(\mu,\gamma,\pOne)}$ $(q_{XYZ|U}(\cdot|u))
$
is a continuous function of $q_{XYZ|U}(\cdot,\cdot,\cdot|u)$.
Then by the support lemma, 
$$
|{\cal U}| \leq |{\cal Y}|+|{\cal Z}|-2+1
=|{\cal Y}|+|{\cal Z}|-1 
$$
is sufficient to express $|{\cal Y}|+|{\cal Z}|-2$ 
values of (\ref{eqn:asdf}) 
and one value of (\ref{eqn:aqqqa}). 
We next bound the cardinality $|{\cal U}|$ of $U$ in 
${\cal P}_{\rm sh}(W_1,W_2)$ to show that the bound $|{\cal U}|\leq $
$\min \{|{\cal X}|,|{\cal Y}|+$ $|{\cal Z}|-1 \}$ is sufficient 
to describe $\tilde{\Omega}^{(\mu,\gamma,\lambda)}(W_1,W_2)$. 
Observe that for any $q\in {\cal P}_{\rm sh}(W_1,W_2)$, we have
\begin{align}
&  q_{{X}}(x)=\sum_{u\in {\cal U}}q_U(u)q_{X|U}(x|u),
\label{eqn:Zasdf}
\\
& \exp\left\{ 
\tilde{\Omega}^{(\mu,\gamma,\lambda)}(q|W_1,W_2)
\right\}
\nonumber\\
&=\sum_{u\in {\cal U}}q_U(u)
\tilde{\zeta}^{(\mu,\gamma,\lambda)}(q_{XYZ|U}(\cdot|u)),
\label{eqn:Zaqqqa}
\end{align}
where we set
\begin{align*}
& \tilde{\zeta}^{(\mu,\gamma,\lambda)}(q_{XYZ|U}(\cdot,\cdot,\cdot|u))
\\
&\defeq  \sum_{\scs (x,y,z)\atop{\scs 
\in {\cal X}\times{\cal Y}\times{\cal Z}}}
q_{{XYZ}|U}(x,y,z|u)
\exp\left\{\pTwo 
\omega^{(\mu,\gamma)}_{q}(x,y,z|u)\right\}.
\end{align*}
For the quantities $q_{Y}(\cdot)$ and $q_{Z}(\cdot)$ 
contained in the forms of 
$\zeta^{(\mu,\gamma,\pOne)}$ 
$(q_{XYZ|U}(\cdot|u)),$ $u\in {\cal U}$, we regard them as 
constants under (\ref{eqn:asdf}). For each $u \in{\cal U}$, 
$
\tilde{\zeta}^{(\mu,\gamma,\pOne)}$ $(q_{XYZ|U}(\cdot|u))
$
is a continuous function of $q_{XYZ|U}(\cdot,\cdot,\cdot|u)$.
Then by the support lemma, 
$$
|{\cal U}| \leq |{\cal Y}|+|{\cal Z}|-2+1
=|{\cal Y}|+|{\cal Z}|-1 
$$
is sufficient to express $|{\cal Y}|+|{\cal Z}|-2$ 
values of (\ref{eqn:asdf}) 
and one value of (\ref{eqn:Zaqqqa}). We also have that 
by the support lemma, 
$$
|{\cal U}| \leq |{\cal X}|-1+1=|{\cal X}|
$$
is sufficient to express  $|{\cal X}|-1$ values of 
(\ref{eqn:Zasdf}) and one value of (\ref{eqn:Zaqqqa}). 
\hfill \IEEEQED
}
\newcommand{\ApdaAACa}{
\subsection{
Proof of Property \ref{pr:pro1} part b) 
} 
\label{sub:ApdaAACa}

In this appendix we prove Property \ref{pr:pro1} part b). Fix 
$q=q_{UXYZ}\in{\cal Q}$, arbitrary. For $p_{UXYZ}=(q_{UX},W_1,W_2) 
\in {\cal P}_{\rm sh}(W_1,W_2)$ and for $q_{Z|U}$, define
\begin{align*}
& \hat{\omega}_{p,q_{Z|U}}^{(\mu,\gamma)}(x,y,z|u)
\defeq  \pOne \gamma\biggl[
           \mu \log \frac{W_1(y|x)}{q_{Y|U}(y|u)}
\\
&\qquad \quad+{\prmtB}\log \frac{q_{Z|U}(z|u)}{p_{Z}(z)}
     \biggr]+\pOne\bar{\gamma}\log \frac{W_1(y|x)}{p_Y(y)},
\\
& \hat{\Omega}^{(\mu,\gamma,\pOne)}(p,q_{Z|U})
\defeq 
\log 
{\rm E}_{p}
\left[\exp\left\{\pOne
\omega^{(\mu,\gamma)}_p (X,Y,Z|U)\right\}\right].
\end{align*}

Then we have the following two lemmas.  
\begin{lm}\label{lm:lemmaSS}
For any $(\mu,\gamma)$ $\in [0,1]^2$, $\pOne \geq 0$, 
$\pTwo\in [0,\frac{1}{1+2\pOne}]$ and 
any $q=q_{UXYZ}\in {\cal Q}$, there exists
$p=p_{UXYZ}\in {\cal P}_{\rm sh}(W_1,W_2)$ such that 
\beq
{\Omega}^{(\mu,\gamma,\pOne,\pTwo)}(q|W_1,W_2)
\leq 
\pTwo \hat{\Omega}^{(\mu,\gamma,\pOne)}(p,q_{Z|U}).
\label{eqn:Zsss}
\eeq
Specifically, if $\gamma=0$, we have 
\beq
{\Omega}^{(\mu,0,\pOne,\pTwo)}(q|W_1,W_2)
\leq 
\pTwo \hat{\Omega}^{(\mu,0,\pOne)}(p)
=\pTwo \tilde{\Omega}^{(\mu,0,\pOne)}(p).
\label{eqn:ZsssPP}
\eeq
\end{lm}

\begin{lm}\label{lm:lemmaSSz}
For any $\mu,\gamma,\pOne$ satisfying $(\mu,\gamma)$ 
$\in [0,1] \times (0,1]$, $\pOne \in [0,\frac{1}{{\prmtB}\gamma})$,  
any $p=p_{UXYZ}\in {\cal P}_{\rm sh}(W_1,W_2)$, and any stochastic 
matrix $q_{Z|U}$ induced by $q_{UXYZ}\in {\cal Q}$, we have 
\begin{align}
& \hat{\Omega}^{(\mu,\gamma,\pOne)}(p,q_{Z|U})
\leq (1-{\prmtB}\gamma\pOne)
\tilde{\Omega}^{(\mu,\gamma,
\frac{\pOne}{1-{\prmtB}\gamma\pOne})}(p).
\label{eqn:ZsssZZZ}
\end{align}
\end{lm}

From Lemmas \ref{lm:lemmaSS} and \ref{lm:lemmaSSz} we have 
the following corollary.
\begin{co}\label{co:CoOne}
For any $\mu,\gamma,\pOne,\pTwo$ satisfying
$(\mu,\gamma)$ $\in [0,1]\times (0,1]$,  
$\pOne \in [0,\frac{1}{{\prmtB}\gamma})$,  
$\pTwo\in [0,\frac{1}{1+2\pOne}]$ and any 
$q=q_{UXYZ}\in {\cal Q}$, there exists 
$p=p_{UXYZ}\in {\cal P}_{\rm sh}(W_1,W_2)$
such that 
\begin{align}
& \Omega^{(\mu,\gamma,\pOne,\pTwo)}(q|W_1,W_2)
\leq \pTwo(1-{\prmtB}\gamma\pOne)
\tilde{\Omega}^{(\mu,\gamma,\frac{\pOne}{1-{\prmtB}\gamma\pOne})}
(p).
\label{eqn:ZsssZZZqq}
\end{align}
From (\ref{eqn:ZsssZZZqq}), we have that
for any 
$(\mu,\gamma)$ $\in [0,1]\times (0,1]$,  
$\pOne \in [0,\frac{1}{{\prmtB}\gamma})$,  
$\pTwo\in [0,\frac{1}{1+2\pOne}]$, we have 
\begin{align}
& \Omega^{(\mu,\gamma,\pOne,\pTwo)}(W_1,W_2)
 \notag\\
& \leq \pTwo(1-{\prmtB}\gamma\pOne)
\tilde{\Omega}^{(\mu,\gamma,\frac{\pOne}{1-{\prmtB}\gamma\pOne})}
(W_1,W_2).
\label{eqn:ZsssAZZqq}
\end{align}
If $\gamma=0$, then for any 
$\mu$ $\in [0,1]$,  
$\pOne \geq 0$, $\pTwo \in [0,\frac{1}{1+2\pOne}]$, 
we have 
\begin{align}
& \Omega^{(\mu,0,\pOne,\pTwo)}(W_1,W_2)
\leq \pTwo
\tilde{\Omega}^{(\mu,0,\pOne)}
(W_1,W_2).
\label{eqn:ZsssAZZqqq}
\end{align}
\end{co}

{\it Proof of Lemma \ref{lm:lemmaSS}:} We fix $(\mu,\gamma) \in [0,1]^2$, 
$\pOne \geq 0$, $\pTwo \in [0,1]$ arbitrary. For each 
$q=q_{UXYZ} \in {\cal Q}$, we choose $p=p_{UXYZ} \in {\cal P}_{\rm sh}(W_1,W_2)$ 
so that $p_{UX}=q_{UX}$. Then for any 
$(u,x,y,z)$ 
$\in {\cal U}$
$\times {\cal X}$
$\times {\cal Y}$
$\times {\cal Z}$, 
we have the following: 
\begin{align}
& \frac{W_1(y|x)W_2(z|x)}{q_{YZ|XU}(y,z|x,u)}
 =\frac{p_{UXYZ}(u,x,y,z)}{q_{UXYZ}(u,x,y,z)}.
\label{eqn:Zsppp}
\end{align}
On upper bounds of $\exp \left\{ 
{\Omega}^{(\mu,\gamma,\pOne,\pTwo)}(q|W_1,W_2)
\right\}$, we have the following chain of inequalities:
\begin{align}
& \exp \left\{ 
{\Omega}^{(\mu,\gamma,\pOne,\pTwo)}(q|W_1,W_2)
\right\}
\MEq{a}
{\rm E}_{q}
\HUgebl
\Biggl\{
\frac{p_{UXYZ}(U,X,Y,Z)}{q_{UXYZ}(U,X,Y,Z)}
\nonumber\\
&\quad \times 
\frac{W_1^{\mu\gamma \pOne}(Y|X)}{p^{\mu\gamma \pOne}_{Y|U}(Y|U)}
\left. \frac{q_{Z|U}^{{\prmtB}\gamma \pOne}(Z|U)}
              {p_{Z}^{{\prmtB}\gamma \pOne}(Z)}
\frac{W_1^{\bar{\gamma}\pOne}(Y|X)}{p_Y^{\bar{\gamma}\pOne}(Y)}
\right\}^{\pTwo}
\nonumber\\
&\quad \times 
\left\{\frac{p_{Y|U}^{(1+\mu\gamma)\pOne\frac{\pTwo}{\bar{\pTwo}}}(Y|U)}
            {q_{Y|U}^{(1+\mu\gamma)\pOne\frac{\pTwo}{\bar{\pTwo}}}(Y|U)}
\right\}^{\frac{\mu\gamma\bar{\pTwo}}{1+\mu\gamma}}
\left\{\frac{p_{Z}^{(1+\mu\gamma)\pOne\frac{\pTwo}{\bar{\pTwo}}}(Z)}
{q_{Z}^{(1+\mu\gamma)\pOne\frac{\pTwo}{\bar{\pTwo}}}(Z)}
\right\}^{\frac{{\prmtB}\gamma\bar{\pTwo}}{1+\mu\gamma}}
\nonumber\\
&\quad \times 
\left\{\frac{p_{Y}^{(1+\mu\gamma)\pOne\frac{\pTwo}{\bar{\pTwo}}}(Y)}
            {q_{Y}^{(1+\mu\gamma)\pOne\frac{\pTwo}{\bar{\pTwo}}}(Y)}
\right\}^{\frac{\bar{\gamma}\bar{\pTwo}}{1+\mu\gamma}}
\HUgebr
\nonumber\\
&\MLeq{b}
\exp \left\{ \pTwo
\hat{\Omega}^{(\mu,\gamma,\pOne)}
  (p,q_{Z|U})\right\}
A_1^{\frac{{\mu}\gamma\bar{\pTwo}}{1+\mu\gamma}}
A_2^{\frac{{\prmtB}\gamma\bar{\pTwo}}{1+\mu\gamma}}
A_3^{\frac{\bar{\gamma}\bar{\pTwo}}{1+\mu\gamma}},
\label{eqn:EddxSSS}
\end{align}
where we set
\beqno
A_1 &\defeq& {\rm E}_q\left[
\frac{p_{Y|U}^{(1+\mu\gamma)\pOne \frac{\pTwo}{\bar{\pTwo}}}(Y|U)}
     {q_{Y|U}^{(1+\mu\gamma)\pOne \frac{\pTwo}{\bar{\pTwo}}}(Y|U)} \right],
A_2 \defeq {\rm E}_q\left[
\frac{p_{Z}^{(1+\mu\gamma)\pOne \frac{\pTwo}{\bar{\pTwo}}}(Z)}
     {q_{Z}^{(1+\mu\gamma)\pOne \frac{\pTwo}{\bar{\pTwo}}}(Z)}\right]
\\
A_3 &\defeq & {\rm E}_q\left[
\frac{p_{Y}^{(1+\mu\gamma)\pOne \frac{\pTwo}{\bar{\pTwo}} }(Y)}
     {q_{Y}^{(1+\mu\gamma)\pOne \frac{\pTwo}{\bar{\pTwo}} }(Y)} \right].
\eeqno
Step (a) follows from (\ref{eqn:Zsppp}). Step (b) 
follows from H\"older's inequality. From (\ref{eqn:EddxSSS}), we can 
see that it suffices to show $A_i \leq 1,i=1,2,3.$ to 
complete the proof. Note here that 
$\pTwo \in [0, \frac{1}{1+(1+\mu\gamma)\pOne}]$ implies that 
$(1+\mu\gamma)\pOne\frac{\pTwo}{\bar{\pTwo}}\leq 1$.
Then, we have the following: 
\begin{align*}  
 \pTwo \in \left[0,\frac{1}{1+2\pOne}\right]
 & \Rightarrow 
    \pTwo \in \left[0,\frac{1}{1+(1+\mu\gamma)\pOne}\right]
\\
& \Rightarrow (1+\mu\gamma)\pOne\frac{\pTwo}{\bar{\pTwo}}\leq 1.
\end{align*}
Since we have $(1+\mu\gamma)\pOne\frac{\pTwo}{\bar{\pTwo}}\leq 1$ 
under $\pTwo \in [0,\frac{1}{1+2\pOne}]$, we can apply H\"older's 
inequality to $A_1$ to obtain 
\begin{align*}
&A_1={\rm E}_q \left[
\frac{p_{Y|U}^{(1+\mu\gamma)\pOne\frac{\pTwo}{\bar{\pTwo}}}(Y|U)}
     {q_{Y|U}^{(1+\mu\gamma)\pOne\frac{\pTwo}{\bar{\pTwo}}}(Y|U)} \right]
\\
&\leq \left({\rm E}_q\left[
\frac{p_{Y|U}(Y|U)}{q_{Y|U}(Y|U)} \right]
     \right)^{(1+\mu\gamma)\pOne\frac{\pTwo}{\bar{\pTwo}}}
=1.
\end{align*}
In a similar manner we can prove $A_i\leq 1$ for $i=2,3$.
Hence we have (\ref{eqn:Zsss}) in Lemma \ref{lm:lemmaSS}. 
\hfill \IEEEQED 

{\it Proof of Lemma \ref{lm:lemmaSSz}:}
We fix 
$(\mu,\gamma)$ $\in [0,1]\times (0,1]$,  
$\pOne \in [0,\frac{1}{{\prmtB}\gamma})$,  
$\pTwo\in [0,\frac{1}{1+2\pOne}]$, arbitrary.
For any 
$p=p_{UXYZ}\in {\cal P}_{\rm sh}(W_1,W_2)$, 
and any $q=q_{UXYZ}\in {\cal Q}$, we have the following 
chain of inequalities: 
\begin{align}
& \exp \left\{ 
\hat{\Omega}^{(\mu,\gamma,\pOne)}(p,q_{Z|U}|W_1,W_2)
\right\}
\nonumber\\
&=
{\rm E}_{p}
\left[
\left\{
\frac{W_1^{\mu\gamma \pOne}(Y|X)}
      {p^{\mu\gamma \pOne}_{Y|U}(Y|U)}
\frac{p_{Z|U}^{{\prmtB}\gamma \pOne}(Z|U)}
       {p_{Z}^{{\prmtB}\gamma \pOne}(Z)}
\frac{W_1^{\bar{\gamma}\pOne}(Y|X)}{p_Y^{\bar{\gamma}\pOne}(Y)}
\right\}^{\frac{1-{\prmtB}\gamma\pOne}{1-{\prmtB}\gamma\pOne}}
\right.
\nonumber\\
&\quad\times 
\left\{\frac{q_{Z|U}(Z|U)}
            {p_{Z|U}(Z|U)}\right\}^{{\prmtB}\gamma\pOne}
\Hugebr
\nonumber\\
&\MLeq{a}
\exp \left\{(1-{\prmtB}\gamma\pOne)
\tilde{\Omega}^{(\mu,\gamma,\frac{\pOne}{1-{\prmtB}\gamma\pOne})}
  (p|W_1,W_2)\right\}
\nonumber\\
&\quad \times \left({\rm E}_{p}\left[
\frac{q_{Z|U}(Z|U)}{p_{Z|U}(Z|U)}
\right]\right)^{{\prmtB}\gamma\pOne}
\nonumber\\
&=\exp \left\{(1-{\prmtB}\gamma\pOne)
\tilde{\Omega}^{(\mu,\gamma,\frac{\pOne}{1-{\prmtB}\gamma\pOne})}
  (p|W_1,W_2)\right\}.
\nonumber
\end{align}
Step (a) follows from H\"older's inequality. 
Thus we have (\ref{eqn:ZsssZZZ}) in 
Lemma \ref{lm:lemmaSSz}. 
\hfill \IEEEQED 

{\it Proof of Property \ref{pr:pro1} part b):} 
When $\gamma \neq 0$, we evaluate lower bounds 
of $F(R_1,R_2 |W_1,W_2)$ to obtain 
the following chain of inequalities:
\begin{align}
& F(R_1,R_2|W_1,W_2)
\notag\\
&\MGeq{a}\sup_{\scs  (\mu,\gamma)\in [0,1] \times (0,1],
     \atop{\scs  \pOne \in [0, \frac{1}{{\prmtB}\gamma}),
                \pTwo \in [0,\frac{1}{1+2\pOne}]
          }
             }\frac{\pTwo}
{1+\pTwo[2+\pOne(1+ \gamma+ \gamma \mu)]}
\notag\\
&\quad \times
\Bigl\{\pOne \bigl[ 
 (\gamma\mu +\bar{\gamma})R_1+ R_2 \bigr]
\notag\\ 
& \qquad \qquad -(1-{\prmtB}\gamma\pOne)\tilde{\Omega}
^{(\mu,\gamma,\frac{\pOne}{1-{\prmtB}\gamma\pOne})}
(W_1,W_2)\Bigr\}
\notag\\
&\MGeq{b}\sup_{\scs  (\mu,\gamma)\in [0,1] \times (0,1],
     \atop{\scs  \pOne \in [0, \frac{1}{{\prmtB}\gamma})                
          }
             }
\sup_{\scs \pTwo \in [0,\frac{1}{1+2\pOne}]}
\frac{\pTwo}{1+\pTwo[2+3\pOne]}
\notag\\
&\quad \times
\Bigl\{\pOne \bigl[ 
 (\gamma\mu +\bar{\gamma})R_1+ R_2 \bigr]
\notag\\ 
& \qquad \qquad -(1-{\prmtB}\gamma\pOne)\tilde{\Omega}
^{(\mu,\gamma,\frac{\pOne}{1-{\prmtB}\gamma\pOne})}(W_1,W_2)\Bigr\}
\notag\\
\newcommand{\SDff}{
&\MGeq{b}
\sup_{\scs  (\mu,\gamma)\in [0,1] \times (0,1],
     \atop{\scs  \pOne \in [0, \frac{1}{{\prmtB}\gamma}),
          }
             }
\bigg\{\pOne \bigl[(\gamma \mu +\bar{\gamma})R_1
             +R_2 \bigr]
\notag\\
&\quad -\tilde{\Omega}^{(\mu,\gamma,\pOne}(W_1,W_2)\biggr\}
\times 
\notag\\
}
&\MGeq{c}\sup_{\scs  (\mu,\gamma)\in [0,1] \times (0,1],
     \atop{\scs  \pOne \in [0, \frac{1}{{\prmtB}\gamma})                
          }
             }
\frac{1}{3+5\pOne}
\Bigl\{\pOne \bigl[ (\gamma\mu +\bar{\gamma})R_1
\notag\\
&\quad +R_2 \bigr]
-(1-{\prmtB}\gamma\pOne)
\tilde{\Omega}^{(\mu,\gamma,\frac{\pOne}{1-{\prmtB}\gamma\pOne})}(W_1,W_2)\Bigr\}
\notag\\
&\MEq{d}\sup_{\scs  (\mu,\gamma)\in [0,1] \times (0,1],
     \atop{\scs  \pOne=\frac{\lambda}{1+{\prmtB}\gamma\lambda},
           \lambda\geq 0
          }
             }
\frac{1}{3+5({\prmtB}\gamma+1)\lambda}
\Bigl\{\lambda \bigl[(\gamma\mu +\bar{\gamma})R_1
\notag\\
&\quad + R_2 \bigr]
-\tilde{\Omega}^{(\mu,\gamma,\lambda)}(W_1,W_2)\Bigr\}
\notag\\
&\MGeq{e} \sup_{\scs  (\mu,\gamma)\in [0,1] \times (0,1],
     \atop{\scs  \lambda \geq 0}} 
\tilde{F}^{(\mu,\gamma,\lambda)}(R_1,R_2|W_1,W_2).
\label{eqn:ZZi}
\end{align}
Step (a) follows from the definition of ${F}(R_1,R_2|W_1,W_2)$ and 
$(\ref{eqn:ZsssZZZqq})$ in Corollary \ref{co:CoOne}. 
Step (b) follows from that for any $(\mu,\gamma) 
\in [0,1]\times (0,1]$, $\pOne \geq 0$, 
$\pTwo \in [0,\frac{1}{\pOne+2}]$ 
$$
\frac{\pTwo}
{1+\pTwo[2+\pOne(1+ \gamma+ \gamma\mu )]}
\geq \frac{\pTwo}{1+\pTwo[2+3\pOne]}.
$$
Step (c) follows from that for each $\pOne\geq 0$,
$$
\sup_{\scs \pTwo 
\in [0,\frac{1}{1+2\pOne}]}\frac{\pTwo}{1+\pTwo[2+3\pOne]}
= \frac{1}{3+5\pOne}.
$$
Step (d) follows from that
$$
\lambda=\frac{\pOne}{1-{\prmtB}\gamma\pOne}, 
\pOne\in [0,{\ts \frac{1}{{\prmtB}\gamma}}) 
\Leftrightarrow
\pOne=\frac{\lambda}{1+{\prmtB}\gamma\lambda},\lambda \geq 0. 
$$
Step (e) follows from that
for any $(\mu,\gamma) \in [0,1]\times (0,1]$, $\lambda\geq 0$,
$$
\frac{1}{3+5({\prmtB}\gamma+1)\lambda}
\geq \frac{1}{3+10\lambda}.
$$
When $\gamma=0$, we evaluate lower bounds 
of $F(R_1,R_2 |W_1,W_2)$ to obtain the following 
chain of inequalities:
\begin{align}
& F(R_1,R_2|W_1,W_2)
\MGeq{a}\sup_{\scs  \mu \in [0,1],
     \atop{\scs  \pOne \geq 0}
             }
\sup_{\scs \pTwo \in [0,\frac{1}{1+2\pOne}]}
\frac{\pTwo}{1+\pTwo[2+\pOne]}
\notag\\
&\quad \times \Bigl\{\pOne \bigl[R_1+R_2 \bigr]
-\tilde{\Omega}^{(\mu,0,\pOne)}(W_1,W_2)\Bigr\}
\notag\\
\newcommand{\SDff}{
&\MGeq{b}
\sup_{\scs  (\mu,\gamma)\in [0,1]^2,
     \atop{\scs  \pOne \in [0, \frac{1}{{\prmtB}\gamma}),
          }
             }
\bigg\{\pOne \bigl[(\gamma \mu +\bar{\gamma})R_1
             +(\gamma {\prmtB}+\bar{\gamma})R_2 \bigr]
\notag \\
&\quad -\tilde{\Omega}^{(\mu,\gamma,\pOne}(W_1,W_2)\biggr\}
\times 
\notag\\
}
&\MEq{b}\sup_{\scs  \mu \in [0,1],
     \atop{\scs  \pOne \geq 0}
             }
\frac{1}{3+3\pOne}
\Bigl\{\pOne \bigl[R_1+R_2\bigr]
-\tilde{\Omega}^{(\mu,0,\pOne)}(W_1,W_2)\Bigr\}
\notag\\
&\geq \sup_{\scs  \mu\in [0,1],
      \atop{\scs  \pOne \geq 0}} 
\tilde{F}^{(\mu,0,\pOne)}(R_1,R_2|W_1,W_2).
\label{eqn:ZZii}
\end{align}
Step (a) follows from the definition of 
${F}(R_1,R_2|W_1,W_2)$ and $(\ref{eqn:ZsssAZZqq})$ 
in Corollary \ref{co:CoOne}. 
Step (b) follows from that for each $\pOne\geq 0$,
$$
\sup_{\scs \pTwo \in [0,\frac{1}{1+2\pOne}]}\frac{\pTwo}{1+\pTwo[2+\pOne]}
= \frac{1}{3+3\pOne}.
$$
From (\ref{eqn:ZZi}) and (\ref{eqn:ZZii}), we have
\begin{align*}
& F(R_1,R_2|W_1,W_2)
\\
&\geq \sup_{\scs (\mu,\gamma)\in [0,1]^2,
     \atop{\scs \lambda \geq 0}} 
\tilde{F}^{(\mu,\gamma,\lambda)}(R_1,R_2|W_1,W_2)
\\
&=\tilde{F}(R_1,R_2|W_1,W_2),
\end{align*}
completing the proof. 
\hfill\IEEEQED

}
\newcommand{\ApdaAACb}{
\subsection{
Proofs of Property \ref{pr:pro1} Parts c),d), and e)
} 
\label{sub:ApdaAACb}

In this appendix we prove Property \ref{pr:pro1} 
parts c), d), and e). 

{\it Proof of Property \ref{pr:pro1} part c):} 
\irr{
We first observe the following form of 
$\exp[{\Omega}^{(\mu,\lambda)}(q)]$:
\begin{align}
&\exp[{\Omega}^{(\mu,\lambda)}(q)]
=\sum_{\scs (u,x,y,z)
\atop{\scs \in {\cal U}\times{\cal X}
                  \times{\cal Y}
                  \times{\cal Z}}}
q_{UXYZ}(u,x,y,z)
\notag\\
&\quad
\times 
\left[ \frac{ q_{Y|X}(y|x)}{q_{Y|U}(y|u)}\right]^{\lambda\gamma\mu}
\left[\frac{q_{Z|U}(z|u)}{q_{Z}(z)}\right]^{\lambda \gamma } 
\left[\frac{q_{Y|X}(y|x)}{q_{Y}(y)}\right]^{\lambda \bar{\gamma} } 
\label{eqn:SdZq}
\\
&=\sum_{\scs (u,x,y,z)
\atop{\scs \in {\cal U}\times{\cal X}
                  \times{\cal Y}
                  \times{\cal Z}}}
B_1^{\gamma\mu}
B_2^{\gamma}
B_3^{\bar{\gamma}}
\label{eqn:Sdppx},
\end{align} 
where
\begin{align*}
&B_1 \defeq q_{UXYZ}^{\frac{1}{1+\mu\gamma}} (u,x,y,z)
\left[\frac{ q_{Y|X}(y|x)}{q_{Y|U}(y|u)}\right]^{\lambda},
\\
&B_2 \defeq 
q_{UXYZ}^{\frac{1}{1+\mu\gamma}} (u,x,y,z)
\left[\frac{q_{Z|U}(z|u)}{q_{Z}(z)}\right]^{\lambda},
\\
&B_3 \defeq 
q_{UXYZ}^{\frac{1}{1+\mu\gamma}} (u,x,y,z)
\left[\frac{q_{Y|X}(y|x)}{q_{Y}(y)}\right]^{\lambda}.
\end{align*}
From (\ref{eqn:Sdppx}), and the defintions of $B_i,i=1,2,3$, 
we can see that if $\lambda \in [0,1/2]$, then for any $\mu\in [0,1]$,
\begin{align*}
& \exp[{\Omega}^{(\mu,\lambda)}(q)]
\leq |{\cal U}||{\cal X}||{\cal Y}||{\cal Z}|  
\MLeq{a}|{\cal X}|^2 |{\cal Y}||{\cal Z}|.  
\end{align*}
Step (a) follows from $q \in {\cal P}_{\rm sh}(W_1,W_2)$. 
We next prove that for any $q \in {\cal P}_{\rm sh}(W_1,W_2)$, 
any $\lambda\in [0,1]$, and 
any $\mu\in [0,1]$, $\exp[{\Omega}^{(\mu,\lambda)}(q)] \geq 1$. 
On lower bounds of $\exp[{\Omega}^{(\mu,\lambda)}(q)]$, 
we have the following chain of inequalities:
\begin{align}
&\exp [ {\Omega}^{(\mu,\lambda)}(q)]
\MEq{a} {\rm E}_q
\left[ 
\left\{ 
\frac{q_{Y|X}(Y|X)}{q_{Y|U}(Y|U)}
\right\}^{\lambda\gamma\mu} 
\left\{
\frac{q_{Z|U} (Z|U) } {q_Z(Z)}
\right\}^{\lambda\gamma\prmtB}   
\right.
\notag\\
&\quad \left.\times 
\left\{
\frac{q_{Y|X}(Y|X)}{q_Y(Y)}
\right\}^{\lambda\bar{\gamma}}   
\right]
\notag\\
&\MEq{b}
{\rm E}_q
\left[ 
\left\{ 
\frac{q_{Y|U}(Y|U)}{q_{Y|XU}(Y|X,U)}
\right\}^{-\lambda\mu\gamma} 
\left\{
\frac{q_Z(Z)}{q_{Z|U}(Z|U)} 
\right\}^{-\lambda\gamma}  
\right.
\notag\\
& \quad \times
\left.
\left\{
\frac{q_Y(Y)}{q_{Y|X}(Y|X)} 
\right\}^{-\lambda\bar{\gamma}}  
\right] 
\notag\\
&\MGeq{c}
\left\{ 
{\rm E}_q
\left[ 
\frac{q_{Y|U}(Y|U)}{q_{Y|XU}(Y|X,U)}
\right] 
\right\}^{-\lambda \mu \gamma} 
\left\{
{\rm E}_q
\left[ 
\frac{q_Z(Z)}{q_{Z|U}(Z|U) } 
\right] 
\right\}^{-\lambda \gamma }
\notag\\
&\quad \times 
\left\{
{\rm E}_q
\left[ 
\frac{q_Y(Y)}{q_{Y|X}(Y|X)} 
\right] 
\right\}^{-\lambda \bar{\gamma}}=1. 
\notag
\end{align}
Step (a) follows from (\ref{eqn:SdZq}).  
Step (b) follows from that since $q \in {\cal P}_{\rm sh}(W_1,W_2)$,
we have $U \Markov X \Markov Y$.    
Step (c) follows from the reverse H\"older's inequality. }
\hfill\IEEEQED


{\it Proof of Property \ref{pr:pro1} part d):}  
For simplicity of notations, set
\beqno
& & 
\underline{a} \defeq (u,x,y,z), \underline{A}\defeq (U,X,Y,Z),
\underline{\cal A} \defeq 
 {\cal U} \times 
 {\cal X} \times 
 {\cal Y} \times 
 {\cal Z},
\\
& & \tilde{\omega}^{(\mu,\gamma)}_{q}(x,y,z|u) 
  \defeq \varsigma(\underline{a}),
\tilde{\Omega}^{(\mu,\gamma,\lambda)}(q|W_1,W_2)\defeq \xi(\lambda).
\eeqno
Then we have 
\beq
\tilde{\Omega}^{(\mu,\gamma,\lambda)}(q|W_1,W_2)=
\xi(\lambda)=\log
\left
[\sum_{\underline{a}\in \underline{\cal A} }q(\underline{a})
{\rm e}^{\lambda {\varsigma}(\underline{a})}\right].
\label{eqn:SdfV}
\eeq
The quantity 
$
q^{(\lambda)}(\underline{a}), \underline{a}\in {\cal A}
$
has the following form:
\beq
q^{(\lambda)}(\underline{a})
={\rm e}^{-\xi(\lambda)}
q
(\underline{a})
{\rm e}^{\lambda {\varsigma}(\underline{a})}.
\label{eqn:aazP011}
\eeq
By simple computations we have 
\beqa
& &\xi^{\prime}(\lambda)=
{\rm e}^{-\xi(\lambda)}
\left[\sum_{\underline{a}} q
(\underline{a})
\varsigma(\underline{a})
{\rm e}^{\lambda {\varsigma}(\underline{a})}\right],
\label{eqn:aaz11}
\eeqa
\beqa
&&\xi^{\prime\prime}(\lambda)=
{\rm e}^{-2\xi(\lambda)}
\nonumber\\
&&\quad \times 
\left[\sum_{\underline{a}, 
            \underline{b}\in \underline{\cal A}}
 q
(\underline{a})
 q
(\underline{b})
 \frac{\left\{{\varsigma}(\underline{a})
       -{\varsigma}(\underline{b})\right\}^2}{2}
{\rm e}^{\lambda\left\{{\varsigma}(\underline{a})
         +{\varsigma}(\underline{b})\right\}}\right]
\nonumber\\
\hspace*{-5mm}&&=
\sum_{\underline{a}, \underline{b}\in \underline{\cal A}}
q
^{(\lambda)}(\underline{a})
q
^{(\lambda)}(\underline{b})
\frac{\left\{{\varsigma}(\underline{a})
- {\varsigma}(\underline{b})\right\}^2}{2}
\nonumber\\
\hspace*{-5mm}&&=
\sum_{\underline{a}\in \underline{\cal A}}
q^{(\lambda)}(\underline{a})\varsigma^2(\underline{a})
-\left[
\sum_{\underline{a}\in \underline{\cal A}} 
q^{(\lambda)}(\underline{a})
{\varsigma}(\underline{a})\right]^2.
\label{eqn:aaz011}
\eeqa
\irr{
On upper bound of $\xi^{\prime\prime}(\lambda)$,
we have the following chain of inequalities:
\begin{align}
& \xi^{\prime\prime}(\lambda)
\MLeq{a}\sum_{\underline{a}\in \underline{\cal A}}
q^{(\lambda)}(\underline{a})\varsigma^2(\underline{a})
\MEq{b}
\sum_{\underline{a}\in \underline{\cal A}}
q(\underline{a})\varsigma^2(\underline{a})
{\rm e}^{\lambda{\varsigma}(\underline{a})-\xi(\lambda)}
\nonumber\\
&={\rm e}^{-\xi(\lambda)}
\sum_{\underline{a}\in \underline{\cal A}}
q(\underline{a})\sqrt{{\rm e}^{2\lambda{\varsigma}(\underline{a})}}
\sqrt{\varsigma^4(\underline{a})}
\nonumber\\
& \MLeq{c} 
\sqrt{{\rm e}^{\xi(2\lambda)-2\xi(\lambda)}}
\sqrt{
\sum_{\underline{a}\in \underline{\cal A}}
q(\underline{a})\varsigma^4(\underline{a})}.
\label{eqn:Sdppi}
\end{align}
Step (a) follows from (\ref{eqn:aaz011}).
Step (b) follows from (\ref{eqn:aazP011}).
Step (c) follows from Cauchy-Schwarz inequality and 
(\ref{eqn:SdfV}). 
Since 
$$
\sum_{\underline{a}\in \underline{\cal A}}
q(\underline{a})\varsigma^4(\underline{a})< \infty
$$
and the bound (\ref{eqn:Sdppi}), it sufficies to examine 
the quantity ${\rm e}^{\xi(2\lambda)-2\xi(\lambda)}$.
By Property \ref{pr:pro1} part b), 
this quantity is bounded for $\lambda \in [0,1/4]$.
Hence $\xi^{\prime\prime}(\lambda)$ exists
for $\lambda \in [0,1/4]$. 
By simple analytical argument we know 
that $\xi^{\prime}(\lambda)$ exists for 
$\lambda \in [0,1/4]$. }\hfill\IEEEQED

{\it Proof of Property \ref{pr:pro1} part e):} 
\irr{Fix any $(\mu,\gamma,\lambda) 
\in [0,1]^2 \times [0,1/4]$ 
and any $q \in {\cal P}_{\rm sh}(W_1,W_2)$.}
By the Taylor expansion of 
$\tilde{\Omega}^{(\mu,\gamma,\lambda)}(q)$
with respect to $\lambda$ around $\lambda=0$,
we have that 
\irr{for any $q \in $ ${\cal P}_{\rm sh}(W_1,W_2)$ and 
for some $c \in [0,\lambda]$,} 
\begin{align}
& \tilde{\Omega}^{(\mu,\gamma,\lambda)}(q)
=\xi(\lambda)=\xi(0)+ \xi^{\prime}(0)\lambda
  +\frac{1}{2}\xi^{\prime\prime}(\irr{c})\irr{\lambda^2}
\notag\\
&=\lambda 
{\rm E}_{q}\left[\tilde{\omega}^{(\mu,\gamma)}_{q}(X,Y,Z|U)\right]
\notag\\
&\quad +\frac{\irr{\lambda^2}}{2}{\rm Var}_{\irr{q^{(c)}}}
\left[\tilde{\omega}^{(\mu,\gamma)}_{q}(X,Y,Z|U)\right]
\notag\\
&\irr{\MLeq{a}} 
  \lambda C^{(\mu,\gamma)}(W_1,W_2)
+\frac{\lambda^2}{2}{\rm Var}_{\irr{q^{(c)}}}
\left[\tilde{\omega}^{(\mu)}_{q}(X,Y,Z|U)\right].
\label{eqn:ZsddAA}
\end{align}
\irr{Step (a) follows from $q \in {\cal P}_{\rm sh}(W_1,W_2)$,  
\begin{align*}
&{\rm E}_{q}
\left[
\tilde{\omega}^{(\mu,\gamma)}_{q}(X,Y,Z|U)
\right]
\\
&= \gamma[\mu I_q(X;Y|U)+ I(U;Z)]+\bar{\gamma}I(X;Y),
\end{align*}
and the definition of $C^{(\mu,\gamma)}(W_1,W_2)$. Let 
$(c_{\rm opt}, q_{\rm opt})$ 
$ \in [0, \lambda] \times {\cal P}_{\rm sh}(W_1,W_2)$
be a pair which attains 
$\rho^{(\mu,\gamma,\lambda)}(W_1,W_2)$.
By this definition 
we have that
\begin{align}
&\tilde{\Omega}^{(\mu,\gamma, \lambda)}(q_{\rm opt})
=\tilde{\Omega}^{(\mu,\gamma,\lambda)}(W_1,W_2) 
\label{eqn:asWWWd}
\end{align}
and that for any $c \in [0,\lambda],$
\begin{align}
&{\rm Var}_{ q_{\rm opt}^{(c)}}
\left[\tilde{\omega}_{ q_{\rm opt} }^{(\mu,\gamma)}(X,Y,Z|U)\right]
\notag\\
&  \leq {\rm Var}_{q_{\rm opt}^{(c_{\rm opt})}}
   \left[\tilde{\omega}_{q_{\rm opt}}^{(\mu,\gamma)}(X,Y,Z|U)\right]
=\rho^{(\mu,\gamma,\lambda)}(W_1,W_2).
\label{eqn:asWxxd}
\end{align}
On upper bounds of $\tilde{\Omega}^{(\mu,\gamma,\lambda)}(W_1,W_2)$, 
we have the following chain of inequalities:
}
\begin{align*}
&   \tilde{\Omega}^{(\mu,\gamma,\lambda)}(W_1,W_2)
\MEq{a} 
\tilde{\Omega}^{(\mu,\gamma,\lambda)}(\irr{q_{\rm opt}}) 
\nonumber\\
&\MLeq{b} \lambda C^{(\mu,\gamma)}(W_1,W_2)
+\frac{\lambda^2}{2}
{\rm Var}_{\irr{q_{\rm opt}^{(c)}}}
\left[\tilde{\omega}^{(\mu,\gamma)}_{\irr{q_{\rm opt}}}(X,Y,Z|U)\right]
\\
&\irr{\MLeq{c}} \lambda {C}^{(\mu,\gamma)}(W_1,W_2)
+\frac{\lambda^2}{2}\rho^{(\mu,\gamma,\lambda)}(W_1,W_2)
\\
&\MLeq{d} \lambda C^{(\mu,\gamma)}(W_1,W_2)
+\frac{\lambda^2}{2}\rho(W_1,W_2).
\end{align*}
Step (a) follows from \irr{(\ref{eqn:asWWWd})}. 
Step (b) follows from (\ref{eqn:ZsddAA}).
Step (c) follows from \irr{(\ref{eqn:asWxxd})}. 
Step (d) follows from the definition of 
$\rho(W_1,W_2)$.
\hfill\IEEEQED

}
\newcommand{\ApdaAACc}{
\subsection{Proof of Property \ref{pr:pro1} part f)} 
\label{sub:ApdaAACc}
In this appendix we prove Property \ref{pr:pro1} part f). 
To prove this property we use the following lemma. 
\begin{lm}\label{lm:LemSaS}
Let $\upsilon$ be some positive constant.
When $\tau \in (0, \rho/4]$, the maximum of 
$$
\frac{1}{1+2 \upsilon \lambda}
\left\{
{\tau} \lambda-\frac{\rho}{2}\lambda^2 
\right\}
$$
for $\lambda \in (0,1/4]$ is attained by 
the positive $\lambda_0$ satisfying
\beq
\vartheta(\lambda_0) \defeq \lambda_0+ \upsilon \lambda_0^2
=\frac{\tau}{\rho}\leq \frac{1}{4}.
\label{eqn:AsssD}
\eeq
Let $g(a)$ be the inverse function of $\vartheta(a)$ for $a\geq 0$. 
Then the condition of (\ref{eqn:AsssD}) is equivalent to 
$\lambda_0=g(\frac{\tau}{\rho})$. The maximum is given by 
$$
\frac{1}{1+2 \upsilon \lambda_0}
\left\{
{\tau} \lambda_0-\frac{\rho}{2}\lambda_0^2 
\right\}=\frac{\rho}{2}\lambda^2_0
=\frac{\rho}{2}g^2\left(\frac{\tau}{\rho}\right).
$$
\end{lm}   

By an elementary computation we can prove this lemma. 
We omit the detail.

{\it Proof of Property \ref{pr:pro1} part e):} 
By the hyperplane expression ${\cal C}_{\rm sh}(W_1,W_2)$ of 
${\cal C}(W_1,W_2)$ stated Property \ref{pr:pro0} part b)
we have that when $(R_1,R_2-\tau) \notin {\cal C}(W_1,W_2)$, 
we have 
\begin{align} 
&(\gamma^*{\mu}^*+\bar{\gamma}^*)R_1+R_2
>C^{(\mu^*,\gamma^*)}(W_1,W_2)+{\tau}
\label{eqn:Xddcc}
\end{align} 
for some $(\mu^*,\gamma^*)\in (0,1]^2$.
Then for each positive ${\tau}$, we have 
the following chain of inequalities: 
\begin{align*}
&  \tilde{F}(R_1,R_2|W_1,W_2)
\geq \sup_{\lambda \in (0,1/4]}
\tilde{F}^{(\mu^{*},\gamma^*,\lambda)}
(R_1,R_2|W_1,W_2)
\\
&=\sup_{\lambda \in (0,1/4]}
\frac{1}{3+10\lambda}\biggl\{
      \lambda\left[(\gamma^*{\mu}^*+\bar{\gamma}^*)R_1
        +R_2\right]
\\
&\quad -\Omega^{(\mu^*,\gamma^*,\lambda)}(W_1,W_2)\biggr\}
\\
&\MGeq{a}  
\sup_{\lambda \in (0,1/4]}
\frac{1}{3+10\lambda}
\biggl\{ 
      \lambda\left[(\gamma^*{\mu}^*+\bar{\gamma}^*)R_1+R_2\right]
\\
& \qquad -\lambda C^{(\mu^*,\gamma^*)}(W_1,W_2)
   -\frac{\lambda^2}{2}\rho(W_1,W_2) \biggr\}
\\
&\MG{b} 
\sup_{\lambda \in (0,1/4]} \frac{1}{3}\cdot\frac{1}{1+2(\frac{5}{3})\lambda}
\left\{
{\tau} \lambda-\frac{\rho}{2}\lambda^2 
\right\}
\MEq{c}
\frac{\rho}{6}g^2\left(\frac{\tau}{\rho}\right),
\end{align*}
where $g=\vartheta^{-1}$ is the inverse function of 
$\vartheta(a)=a+(5/3)a^2$.
Step (a) follows from Property \ref{pr:pro1} part d).
Step (b) follows from (\ref{eqn:Xddcc}). Step (c) follows from 
Lemma \ref{lm:LemSaS}. 
\hfill\IEEEQED
}
Our main result is the following. 
\begin{Th}\label{Th:main}
For any $R_1,R_2\geq 0$, any $(W_1,W_2)$, and 
for any $(\varphi^{(n)},$ $\psi_1^{(n)},$ 
$\psi_2^{(n)})$ satisfying 
$(1/n)\log |{\cal K}_n| \geq R_1,$ 
$(1/n)\log |{\cal L}_n| \geq R_2,$
we have 
\beqno
& &{\rm P}_{\rm c}^{(n)}(\varphi^{(n)},
\psi_1^{(n)},\psi_2^{(n)})
\leq 6\exp \left\{-n F(R_1,R_2|W_1,W_2)\right\}.
\eeqno
\end{Th}

It follows from Theorem \ref{Th:main} and 
Property \ref{pr:pro1} part c) that if $(R_1,R_2)$ is 
outside the capacity region, then the error probability 
of decoding goes to one exponentially and its exponent 
is not below $F(R_1,R_2|W_1,W_2)$. It immediately 
follows from Theorem \ref{Th:main} that we have 
the following corollary. 

\begin{co}\label{co:mainCo}
For any BC $(W_1,W_2)$, we have 
\beqa
G(R_1,R_2|W_1,W_2) &\geq& F(R_1,R_2|W_1,W_2),
\label{eqn:mainIeq}
\\
{\cal R}(W_1,W_2) &\subseteq& \overline{\cal R}(W_1,W_2).
\eeqa
\end{co}

Proof of Theorem \ref{Th:main} will be given in the next section. The 
technique to prove strong converse theorems in multi-user information 
theory developed by Csisz\'ar and K\"orner \cite{ckBook81} is not useful 
to prove Theorem \ref{Th:main}. Some novel techniques based on the 
information spectrum method introduced by Han \cite{han} are necessary 
to prove this theorem.

From Theorem \ref{Th:main} and Property \ref{pr:pro1} part d), 
we can obtain an explicit 
outer bound of ${\cal C}_{\rm ABC}(\varepsilon |W_1,W_2)$ 
with an asymptotically vanishing deviation from 
${\cal C}_{\rm ABC}(W_1,W_2)$ $={\cal C}(W_1,W_2)$. 
The strong converse theorem 
immediately follows 
from this corollary. To discribe this outer bound, 
for $\nu>0$, we set 
To discribe this outer bound, 
for $\nu>0$, we set 
\begin{align*}
&  {\cal C}(W_1,W_2)+\nu(0,1)
\\
& \defeq  \{(R_1, R_2+\nu): (R_1,R_2) 
\in {\cal C}(W_1,W_2)\}, 
\end{align*}
which serves as an outer bound of ${\cal C}(W_1,W_2)$. 
We define $\nu_n =\nu_n(\varepsilon,\rho(W_1,W_2))$ by
\beqa
\nu_n &\defeq &
\rho(W_1,W_2) \vartheta\left(
\sqrt{ \frac{6}{n\rho(W_1,W_2)} \log\left(\frac{6}{1-\varepsilon}\right)} 
\right)
\label{eqn:zdd}\\
&\MEq{a}&
\sqrt{ 
\frac{6\rho(W_1,W_2)}{n}
\log\left(\frac{6}{1-\varepsilon}\right)}
+\frac{10}{n} \log \left(\frac{6}{1-\varepsilon}\right).
\nonumber
\eeqa
Step (a) follows from $\vartheta(a)=a+(5/3)a^2$.  
Since $\nu_n \to 0$ as $n\to \infty$, we have the 
smallest positive integer $n_0=n_0(\varepsilon,\rho(W_1,W_2))$
such that $\nu_n \leq (1/4)\rho(W_1,W_2)$ for $n\geq n_0$.
From Theorem \ref{Th:main} and Property \ref{pr:pro1} part e), 
we have the following corollary.
\begin{co}\label{co:StConv}  
For each fixed $\varepsilon$ $ \in (0,1)$, 
we choose the above positive integer $n_0(\varepsilon,\rho(W_1,W_2))$.
Then, for any $n\geq n_0$, 
\beqno
&&{\cal C}_{\ABC}(n,\varepsilon |W_1,W_2)
\subseteq {\cal C}(W_1,W_2)+\nu_n(0,1), 
\eeqno
\end{co}

\newcommand{\Xssp}{
It immediately follows from the above result that
for each pair $(\varepsilon_1,\varepsilon_2)$ $ \in (0,1)^2$ 
satisfying $\varepsilon_1+\varepsilon_2$ $<1$, we have 
\begin{align*}
& {\cal C}_{\rm m, \ABC}(\varepsilon_1,\varepsilon_2|W_1,W_2)
\\
&={\cal C}_{\rm \ABC}(\varepsilon_1,\varepsilon_2|W_1,W_2)
  ={\cal C}_{\rm \ABC}(\varepsilon_1+\varepsilon_2|W_1,W_2)
\\
&={\cal C}_{\rm \ABC}(W_1,W_2)={\cal C}(W_1,W_2).
\end{align*}
In particular, for each $\varepsilon \in (0,1/2)$, we have 
\beqno
&& {\cal C}_{\rm \ABC}(\varepsilon,\varepsilon|W)
  ={\cal C}_{\rm \ABC}(2\varepsilon|W_1,W_2)
={\cal C}(W_1,W_2).
\eeqno
\end{co}
}

Proof of this corollary will be given in 
Section \ref{sec:Secaa}. The above result 
together with
\beq
{\cal C}_{\ABC}(\varepsilon|W_1,W_2)
={\rm cl}\left(
\bigcup_{m\geq 1}
\bigcap_{n \geq m}{\cal C}_{\ABC}(n,\varepsilon | W_1,W_2)
\right).
\label{eqn:Zassd0}
\eeq 
recovers the strong converse theorem by Csisz\'ar and 
K\"orner \cite{ckBook81}. That is, for each fixed 
$\varepsilon$ $ \in (0,1)$, and for any $(W_1,$ $W_2)$, 
we have 
\begin{align*}
&{\cal C}_{\ABC}(\varepsilon|W_1,W_2)
 ={\cal C}_{\ABC}(W_1,W_2)
={\cal C}(W_1,W_2).
\end{align*} 
For each fixed $\varepsilon$ $ \in (0,\frac{1}{2})$, 
and for any $(W_1,W_2)$, we have 
\begin{align*}
& {\cal C}_{\ABC}(\varepsilon,\varepsilon|W_1,W_2)
 ={\cal C}_{\ABC}(W_1,W_2)
\\
&={\cal C}(W_1,W_2).
\end{align*}

\newcommand{\ProofCor}{  
{\it Proof of Corollary \ref{co:StConv}:}
Since $g$ is an inverse function of $\vartheta$, 
the definition (\ref{eqn:zdd}) of $\nu_n$ 
is equivalent to 
\beq
g\left( 
\frac{\nu_{n}}{\rho(W_1,W_2)}\right)
=\sqrt{
\frac{6}{n\rho(W_1,W_2)} \log\left(\frac{6}{1-\varepsilon}\right)}.
\label{eqn:zddQ}
\eeq
By the definition of $n_0=n_0(\varepsilon,\rho(W_1,W_2))$, 
we have that $\nu_n \leq (1/4)\rho(W_1,W_2)$ for $n\geq n_0$.
We assume that for $n\geq n_0$,
$(R_1,R_2)
\in 
{\cal C}_{\ABC}(n,\varepsilon |W_1,W_2).
\label{eqn:Zassd}
$ 
Then there exists a sequence 
$\{(\varphi^{(n)},$ $\psi_1^{(n)},$ $\psi_2^{(n)})\}_{n\geq n_0}$ 
such that for $n\geq n_0$,
\beqno
& &\frac{1}{n}\log |{\cal K}_n| \geq R_1, 
   \frac{1}{n}\log |{\cal L}_n| \geq R_2, 
\\
& &{\rm P}_{\rm e}^{(n)}(\varphi^{(n)}_1,
\varphi^{(n)}_2,\psi^{(n)})
\leq \varepsilon.
\eeqno
Then by Theorem \ref{Th:main} and Property \ref{pr:pro1} part b), 
we have for $n\geq n_0$,
\beqa
1-\varepsilon&\leq& 
{\rm P}_{\rm c}^{(n)}(\varphi^{(n)},\psi_1^{(n)},\psi_2^{(n)})
\nonumber\\
&\leq& 6\exp \left\{-n \tilde{F}(R_1,R_2|W_1,W_2)\right\}.
\label{eqn:Zsddd}
\eeqa
From (\ref{eqn:Zsddd}), we have that for 
$n\geq n_0$,
\beq
\tilde{F}(R_1,R_2|W_1,W_2)\leq \frac{1}{n}
\log \frac{6}{1-\varepsilon}\MEq{a} 
\frac{\rho}{6}g^2\left(\frac{\nu_n}{\rho}\right).
\label{eqn:Zsdddii}
\eeq
Step (a) follows from (\ref{eqn:zddQ}).
Hence, by Property \ref{pr:pro1} part f),  
we have that under $\nu_n \leq (1/4)\rho(W_1,W_2)$,
the bound (\ref{eqn:Zsdddii}) implies 
$(R_1,R_2-\nu_n)\in {\cal C}(W_1,W_2)$ or equivalent to 
\beq
(R_1,R_2)\in {\cal C}(W_1,W_2)+\nu_n(0,1).
\label{eqn:Zsdddij} 
\eeq
Since (\ref{eqn:Zsdddij}) holds for any $n \geq n_0$ 
and any $(R_1,R_2) 
\in {\cal C}_{\rm ABC}($ $n,\varepsilon |W_1,W_2)$, we have
$$
{\cal C}_{\ABC}(n,\varepsilon |W_1,W_2) 
\subseteq {\cal C}(W_1,W_2)+\nu_n(0,1)\mbox{ for }n\geq n_0,
$$ 
completing the proof.
\hfill\IEEEQED
}

\section{Proofs of the Main Results}
\label{sec:Secaa}

We first prove the following lemma. 
\begin{lm}\label{lm:Ohzzz} For any $\eta>0$ and 
for any $(\varphi^{(n)},\psi_1^{(n)},\psi_2^{(n)})$  
satisfying  
$
(1/n)\log |{\cal K}_n| \geq R_1,
(1/n)\log |{\cal L}_n| \geq R_2,
$
we have 
\beqa	
& &{\rm P}_{\rm c}^{(n)}
(\varphi^{(n)},\psi_1^{(n)},\psi_2^{(n)})
\leq p_{L_nX^nY^nZ^n}\hugel
\nonumber\\
& &\:\:\:
0\leq \frac{1}{n}
\log \frac{W_1^n(Y^n|X^n)}{Q^{\One}_{Y^n|X^nZ^nL_n}(Y^n|X^n,Z^n,L_n)}+\eta,
\label{eqn:asXpA}\\
& &\:\:\:
0\leq \frac{1}{n}
\log \frac{W_2^n(Z^n|X^n)}{Q^{\Two}_{Z^n|X^nY^nL_n}(Z^n|X^n,Y^n,L_n)}+\eta,
\label{eqn:asXpB}\\
& &
R_1\leq \frac{1}{n}\log
\frac{W_1^n(Y^n|X^n)}{Q^{\Thr}_{Y^n|L_n}(Y^n|L_n)}+\eta,
\label{eqn:asXpCa}\\
& &
R_2 \leq 
\frac{1}{n}\log\frac{p_{Z^n|L_n}(Z^n|L_n)}{Q^{\Fou}_{Z^n}(Z^n)}+\eta,
\label{eqn:asXpCb}\\
& &
R_1+ R_2 \leq \left.
\frac{1}{n}\log\frac{
W_1^n(Y^n|X^n)
}{Q^{\Fiv}_{Y^n}(Y^n)}+\eta
\right\}
+5{\rm e}^{-n\eta}.
\label{eqn:asXpD}
\eeqa
In (\ref{eqn:asXpA}), we can choose 
any conditional distribution $Q^{\One}_{Y^n|}$ ${}_{X^nZ^nL_n}$ 
on ${\cal Y}^n$ given $({X}^n,Z^n,$ $L_n)$. 
In (\ref{eqn:asXpB}), we can choose 
any conditional distribution $Q^{\Two}_{Z^n|X^nY^nL_n}$ on 
${\cal Z}^n$ given $(X^n,$ $Y^n$,$L_n)$. 
In (\ref{eqn:asXpCa}), we can choose 
any conditional distribution $Q^{\Thr}_{Y^n|L_n}$ on 
${\cal Y}^n$ given $L_n$. 
In (\ref{eqn:asXpCb}), we can choose
any distribution $Q^{\Fou}_{Z^n}$ on ${\cal Z}^n$.
In (\ref{eqn:asXpD}) , we can choose
any distribution $Q^{\Fiv}_{Y^n}$ on ${\cal Y}^n$. 
\end{lm}

Proof of this lemma is given in Appendix \ref{sub:Apda}.
\newcommand{\Apda}{
\subsection{
Proof of Lemma \ref{lm:Ohzzz} 
}\label{sub:Apda}


In this appendix we prove Lemma \ref{lm:Ohzzz}.
For $l\in {\cal L}_n$, set
\begin{align*}
& {\cal A}_1(l) \defeq \left\{(x^n,y^n,z^n):
\frac {p_{Y^n|X^nZ^nL_n}(y^n|x^n,z^n,l)}
      {Q^{\One}_{Y^n|X^nZ^nL_n}(y^n|x^n,z^n,l)}
\right.
\\
& \qquad\quad \left.
= \frac {W_1(y^n|x^n)}
        {Q^{\One}_{Y^n|X^nZ^nL_n}(y^n|x^n,z^n,l)}
        \geq {\rm e}^{-n\eta}
\right\},
\\
& {\cal A}_2(l) \defeq \left\{(x^n,y^n,z^n):
\frac {p_{Z^n|X^nY^nL_n}(z^n|x^n,y^n,l)}
      {Q^{\Two}_{Z^n|X^nY^nL_n}(z^n|x^n,y^n,l)}
\right.
\\
& \qquad\quad \left.
= \frac {W_2^n(z^n|x^n)}
      {Q^{\Two}_{Z^n|X^nY^nL_n}(z^n|x^n,y^n,l)}
       \geq {\rm e}^{-n\eta} 
\right\}.
\end{align*}
Furthermore, for $l\in {\cal L}_n$, set
\beqno
{\cal A}_3(l)
&\defeq&
\{(x^n,y^n,z^n): 
\ba[t]{l}
W_1^n(y^n|x^n)\\
\geq |{\cal K}_n| {\rm e}^{-n\eta}
Q^{\Thr}_{Y^n|L_n}(y^n|l)\},
\ea
\\
{\cal A}_4(l)
&\defeq &
\{(x^n,y^n,z^n): 
\ba[t]{l} p_{Z^n|L_n}(z^n|l)\\
\geq |{\cal L}_n| {\rm e}^{-n\eta}
Q^{\Fou}_{Z^n}(z^n)\},
\ea
\\
{\cal A}_5(l)
&\defeq&
\{(x^n,y^n,z^n): 
\ba[t]{l}
W_1^n(y^n|x^n)\\
\geq |{\cal K}_n| |{\cal L}_n| {\rm e}^{-n\eta}
Q^{\Fiv}_{Y^n}(y^n)\},
\ea
\\
{\cal A}(l)&\defeq& \bigcap_{i=1}^5 {\cal A}_i(l).
\eeqno
Define six quantities $\Delta_i,i=0,1,\cdots,5$ by
\begin{align*}
& \Delta_0 \defeq \Pr\{(X^n,Y^n,Z^n) \in {\cal A}(L_n) \}, 
\\
& \Delta_i \defeq \Pr\{(X^n,Y^n,Z^n)
\notin {\cal A}_i(L_n)\},i=1,2,
\\
& \Delta_3 \defeq 
\Pr \{(X^n,Y^n,Z^n) \notin {\cal A}_3(L_n), Y^n \in {\cal D}_1(K_n,L_n) \},
\\
& \Delta_4 \defeq 
\Pr \{(X^n,Y^n,Z^n) \notin {\cal A}_4(L_n), Z^n \in {\cal D}_2(L_n) \},
\\
& \Delta_5 \defeq 
\Pr \{(X^n,Y^n,Z^n) \notin {\cal A}_5(L_n), Y^n \in {\cal D}_1(K_n,L_n) \}.
\end{align*}

{\it Proof of Lemma \ref{lm:Ohzzz}:} 
On upper bound of ${\rm P}_{\rm c}^{(n)}$, 
we have the following: 
\begin{align*}
&{\rm P}_{\rm c}^{(n)}
=\Pr\{Y^n \in {\cal D}_1(K_n,L_n),Z^n \in {\cal D}_2(L_n)\}
\\
&\leq \Pr\{(X^n,Y^n,Z^n) \in {\cal A}(L_n)\}
\\
&\quad +\Pr\{(X^n,Y^n,Z^n) \notin {\cal A}(L_n), 
\\ 
&\qquad \qquad 
   Y^n \in {\cal D}_1(K_n,L_n),Z^n \in {\cal D}_2(L_n)\}
\\
& \leq \sum_{i=0}^5 \Delta_i.
\end{align*}
By definition we have 
\begin{align}
& \Delta_0
=p_{L_nX^nY^nZ^n}\hugel
\nonumber\\
& \:\:\:
0\leq \frac{1}{n}
\log \frac{W_1^n(Y^n|X^n)}
          {Q^{\One}_{Y^n|X^nZ^nL_n}(Y^n|X^n,Z^n,L_n)}+\eta,
\nonumber\\
& \:\:\:
0\leq \frac{1}{n}
\log \frac{W_2^n(Z^n|X^n)}
          {Q^{\Two}_{Z^n|X^nY^nL_n}(Z^n|X^n,Y^n,L_n)}+\eta,
\nonumber\\
& 
\frac{1}{n}\log |{\cal K}_n| \leq \frac{1}{n}\log
\frac{W_1^n(Y^n|X^n)}{Q^{\Thr}_{Y^n|L_n}(Y^n|L_n)}+\eta,
\nonumber\\
& 
\frac{1}{n}\log |{\cal L}_n| \leq 
\frac{1}{n}\log\frac{p_{Z^n|L_n}(Z^n|L_n)}{Q^{\Fou}_{Z^n}(Z^n)}+\eta
\nonumber\\
&\frac{1}{n}\log(|{\cal K}_n||{\cal L}_n|)
\leq  \left.
\frac{1}{n}\log\frac{W_1^n(Y^n|X^n)}{ Q^{\Fiv}_{Y^n}(Y^n)}+\eta
\right\}.
\label{eqn:azsadaba}
\end{align}
From (\ref{eqn:azsadaba}), it follows that 
if $(\varphi^{(n)},\psi_1^{(n)},\psi_2^{(n)})$  
satisfies   
$(1/n)\log |{\cal K}_n| \geq R_1$ and 
$(1/n)\log |{\cal L}_n| \geq R_2$,
then the quantity $\Delta_0$ is upper bounded by the first term 
in the right members of (\ref{eqn:asXpD}) in Lemma \ref{lm:Ohzzz}.
Hence it suffices to show $\Delta_i\leq {\rm e}^{-n\eta}, i=1,2,\cdots,5$ 
to prove Lemma \ref{lm:Ohzzz}. 
We first prove 
$\Delta_i \leq {\rm e}^{-n\eta}$ for $i=1,2$. 
By a symmetrical structure on ${\cal A}_1(\cdot)$ 
and ${\cal A}_2(\cdot)$, it suffices to prove 
$\Delta_1 \leq {\rm e}^{-n\eta}$. We have the 
following chain of inequalities:  
\begin{align*}
&\Delta_1
=\sum_{l\in {\cal L}_n }
\sum_ {\scs (x^n,y^n,z^n)
          \atop{\scs \notin {\cal A}_1(l) 
          }
   } 
p_{L_nX^nY^nZ^n}(l,x^n,y^n,z^n)
\\
&\leq{\rm e}^{-n\eta}
\sum_{ l\in {\cal L}_n }
\sum_ {\scs (x^n,y^n,z^n)\atop{\scs \notin {\cal A}_1(l)}} 
Q^{\One}_{Y^n|X^nZ^nL_n}(y^n|x^n,z^n,l)
\\
&\quad \times p_{X^nZ^nL_n}(x^n,z^n,l)\leq {\rm e}^{-n\eta}.
\end{align*}
Next, we prove $\Delta_3\leq {\rm e}^{-n\eta}$. 
We have the following chain of inequalities: 
\begin{align*}
&\Delta_3
 = \frac{1}{|{\cal K}_n| |{\cal L}_n|} 
  \sum_{(k,l)\in {\cal K}_n \times {\cal L}_n }
  \sum_{\scs 
        (x^n,y^n,z^n):
         \atop{\scs 
             y^n \in {\cal D}_1(k,l), 
                \atop{\scs 
                    W_1^n(y^n|x^n) <{\rm e}^{-n\eta}
                    \atop{\scs   
                      \times |{\cal K}_n| Q^{\Thr}_{Y^n|L_n}(y^n|l)
                    }
                }
         }
    }1
\\
&\quad \times 
\varphi^{(n)}(x^n|k,l)W_1^n(y^n|x^n)W_2^n(z^n|x^n) 
\\
&\leq 
  \frac{ {\rm e}^{-n\eta} }{|{\cal L}_n|} 
  \sum_{(k,l)\in {\cal K}_n \times {\cal L}_n }
  \sum_{\scs
         (x^n,y^n):
           \atop {\scs 
            y^n \in {\cal D}_1(k,l)
       }
    }
\varphi^{(n)}(x^n|k,l)Q^{\Thr}_{Y^n|L_n}(y^n|l) 
\\
&=
  \frac{ {\rm e}^{-n\eta} }{|{\cal L}_n|} 
  \sum_{ l \in {\cal L}_n }
  \sum_{ k\in {\cal K}_n }
  Q^{\Thr}_{Y^n|L_n}\left(\left.{\cal D}_1(k,l)\right| l \right) 
\\
&=
  \frac{ {\rm e}^{-n\eta} }{|{\cal L}_n|} 
  \sum_{ l\in {\cal L}_n }
  Q^{\Thr}_{Y^n|L_n}
\left(\left. \bigcup_{k\in {\cal K}_n}{\cal D}_1(k,l)\right|l \right) 
\\
&\leq
  \frac{ {\rm e}^{-n\eta} }{|{\cal L}_n|} 
  \sum_{ l\in {\cal L}_n }1= {\rm e}^{-n\eta}.
\end{align*}
We prove $\Delta_4\leq {\rm e}^{-n\eta}$. 
We have the following chain of inequalities:
\begin{align*}
&\Delta_4=  \frac{1}{|{\cal L}_n|} \sum_{l\in {\cal L}_n} 
  \sum_{\scs 
        z^n \in {\cal D}_2(l),
                \atop{\scs 
                    p_{Z^n|L_n}(z^n|l) <{\rm e}^{-n\eta}
                    \atop{\scs   
                      \times |{\cal L}_n| Q^{\Fou}_{Z^n}(z^n)
                    }
                }
    }
\sum_{k\in {\cal K}_n} 
\sum_{(x^n,y^n)\in {\cal X}^n\times {\cal Y}^n }1
\\
& \quad \times 
p_{K_nX^nY^nZ^n|L_n}(k,x^n, y^n, z^n|l) 
\\
&= 
  \frac{1}{|{\cal L}_n|} 
  \sum_{l\in {\cal L}_n} 
  \sum_{\scs 
            z^n \in {\cal D}_2(l),
            \atop{\scs 
                p_{Z^n|L_n}(z^n|l) <{\rm e}^{-n\eta}
                   \atop{\scs   
                        \times |{\cal L}_n| Q^{\Fou}_{Z^n}(z^n)
                    }
                }
        }  
    p_{Z^n|L_n}(z^n|l) 
\\
&\leq  
  {\rm e}^{-n\eta} 
   \sum_{l\in {\cal L}_n }\sum_{z^n \in {\cal D}_2(l)}Q^{\Fou}_{Z^n}(z^n)
= {\rm e}^{-n\eta} 
   \sum_{l\in {\cal L}_n }Q^{\Fou}_{Z^n}\left({\cal D}_2(l)\right)
\\
&={\rm e}^{-n\eta}Q^{\Fou}_{Z^n}
\left(\bigcup_{l \in {\cal L}_n} {\cal D}_2(l) \right)\leq {\rm e}^{-n\eta}. 
\end{align*}
Finally, we prove $\Delta_5\leq {\rm e}^{-n\eta}$.
We have the following chain of inequalities: 
\begin{align*}
&\Delta_5
= 
  \frac{1}{|{\cal K}_n| |{\cal L}_n|} 
  \sum_{(k,l)\in {\cal K}_n \times {\cal L}_n}
  \sum_{\scs (x^n,y^n,z^n):
         \atop{\scs 
             y^n \in {\cal D}_1(k,l), 
                \atop{\scs 
                    W_1^n(y^n|x^n) <{\rm e}^{-n\eta}
                     \atop{\scs   
                      \times |{\cal K}_n ||{\cal L}_n |Q^{\Fiv}_{Y^n}(y^n)
                    }
                }
         }
    }1
\\
&\quad \times 
\varphi^{(n)}(x^n|k,l)W_1^n(y^n|x^n)W_2^n(z^n|x^n) 
\\
&=\frac{1}{|{\cal K}_n| |{\cal L}_n|} 
  \sum_{(k,l)\in {\cal K}_n \times {\cal L}_n }
  \sum_{\scs 
        (x^n,y^n):
         \atop{\scs 
             y^n \in {\cal D}_1(k,l), 
                \atop{\scs 
                    W_1^n(y^n|x^n) <{\rm e}^{-n\eta}
                    \atop{\scs   
                      \times |{\cal K}_n ||{\cal L}_n |Q^{\Fiv}_{Y^n}(y^n)
                    }
                }
         }
    }1
\\
&\quad \times \varphi^{(n)}(x^n|k,l)W_1^n(y^n|x^n)
\\
&\leq
  {\rm e}^{-n\eta}
  \sum_{l \in {\cal L}_n }
  \sum_{k \in {\cal K}_n }
  \sum_{\scs (x^n,y^n):
         \atop{\scs 
             y^n \in {\cal D}_1(k,l), 
                \atop{\scs 
                    W^n(y^n|x^n) <{\rm e}^{-n\eta}
                    \atop{\scs   
                      |{\cal K}_n||{\cal L}_n |Q^{\Fiv}_{Y^n}(y^n)
                    }
                }
         }
    }
\varphi^{(n)}(x^n|k,l)Q^{\Fiv}_{Y^n}(y^n)
\\
&\leq 
  {\rm e}^{-n\eta}
  \sum_{l \in {\cal L}_n }
  \sum_{k \in {\cal K}_n }
  \sum_{
         \atop{\scs 
             y^n \in {\cal D}_1(k,l), 
         }
    }Q^{\Fiv}_{Y^n}(y^n)
\\
&= 
  {\rm e}^{-n\eta}
  \sum_{l \in {\cal L}_n }
  \sum_{k \in {\cal K}_n }
Q^{\Fiv}_{Y^n}\left({\cal D}_1(k,l)\right)
\\
&={\rm e}^{-n\eta}
Q^{\Fiv}_{Y^n}
\left(
\bigcup_{(k,l)\in {\cal K}_n\times{\cal L}_n}{\cal D}_1(k,l)
\right) 
\leq {\rm e}^{-n\eta}.
\end{align*}
Thus Lemma \ref{lm:Ohzzz} is proved. 
\hfill\IEEEQED
}
For $t=1,2,$ $\cdots,n$, set 
\beqno
& & {\cal U}_t\defeq {\cal L}_n \times {\cal Y}^{t-1} 
     \times {\cal Z}_{t+1}^n, 
{U}_t \defeq (L_n,Y^{t-1},Z_{t+1}^n) \in {\cal U}_t,
\\
& & {u}_t \defeq (l,y^{t-1},z_{t+1}^n)\in {\cal U}_t, 
\\
& & {\cal V}_t\defeq {\cal L}_n \times {\cal Z}_{t+1}^{n},
{V}_t \defeq (L_n, Z_{t+1}^n) \in {\cal V}_t. 
\eeqno
For each $t=1,2\cdots,l$, let $\kappa_t$ 
be a natural projection from 
${\cal U}_t$ onto ${\cal V}_t$.  
Using $\kappa_t$, we have $V_t=
\kappa_t(U_t),$ $t=1,2,\cdots,n$.
From Lemma \ref{lm:Ohzzz} we have the following. 
\begin{lm}\label{lm:Ohzzzca} For any $\eta>0$ and 
for any $(\varphi^{(n)},\psi_1^{(n)},\psi_2^{(n)})$  
satisfying
$
(1/n)\log |{\cal K}_n| \geq R_1,
(1/n)\log |{\cal L}_n| \geq R_2,
$
we have 
\beqa
& &{\rm P}_{\rm c}^{(n)}(\varphi^{(n)},\psi_1^{(n)},\psi_2^{(n)})  
\leq 
p_{{L_n}X^nY^nZ^n}\hugel
\nonumber\\
& &0\leq \frac{1}{n}
\sum_{t=1}^n
\log \frac{W_1(Y_t|X_t)}{Q^{\One}_{Y_t|{X_tZ_tU_t}}(Y_t|X_t,Z_t,U_t)}
+\eta,
\nonumber\\
& &0\leq \frac{1}{n}
\sum_{t=1}^n
\log \frac{W_2(Z_t|X_t)}{Q^{\Two}_{Z_t|{X_tU_t}}(Z_t|X_t,U_t)}
+\eta,
\nonumber\\
& &
R_1 \leq \frac{1}{n}\sum_{t=1}^n 
\log \frac{W_1(Y_t|X_t)}{Q^{\Thr}_{Y_t|U_t}(Y_t|U_t)}
\nonumber\\
& &
\qquad\quad
+\frac{1}{n}\sum_{t=1}^n \log \frac{\tilde{Q}^{\Thr}_{Z_t|U_t}(Z_t|U_t) }
{p_{Z_t|V_t}(Z_t|V_t)}
+\eta
\label{eqn:ZZsspp}
\\
& & 
R_2 \leq \frac{1}{n}\sum_{t=1}^n\log 
\frac{p_{Z_t|V_t}(Z_t|V_t)}{Q^{\Fou}_{Z_t}(Z_t)}+\eta
\nonumber\\
& &R_1+R_2\leq \frac{1}{n}\sum_{t=1}^n\log 
\frac{W_1(Y_t|X_t)}{Q^{\Fiv}_{Y_t}(Y_t)} 
+\eta
\huger
+5{\rm e}^{-n\eta},
\label{eqn:Zsdaa}
\eeqa
where 
for each $t=1,2,\cdots,n$, the following probability 
and conditional probability distributions:
\beq
\ba{l}
Q^{\One}_{Y_t|X_tZ_tU_t},
Q^{\Two}_{Z_t|X_tU_t},
Q^{\Thr}_{Y_t|U_t}, 
Q^{\Fou}_{Z_t},
Q^{\Fiv}_{Y_t}
\ea
\eeq
appearing in the first term in the right members of 
(\ref{eqn:Zsdaa}) have a property that we can choose 
their values arbitrary. In (\ref{eqn:ZZsspp}), 
$\tilde{\cal Q}^{\Thr}_{Z_t|U_t}$ can be computed 
from $\{(p_{Z_i|V_i}$, 
${\cal Q}^{\Thr}_{Y_i|U_i}\}_{i=1}^{t-1}$ and $p_{Z_t|V_t}$, 
having the form
\begin{align*}
& \tilde{Q}^{\Thr}_{Z_t|U_t}(z_t|u_t)
\\
&=
\frac{\ds \sum_{\tilde{z}^{t-1}}
\left(\prod_{i=1}^{t-1}\left\{
Q^{\Thr}_{Y_i|U_i}(y_i|\tilde{u}_i)
p_{Z_i|V_i}(\tilde{z}_i|\tilde{v}_i)\right\}
\right)p_{Z_t|V_t}(z_t|v_t)}
{\ds \sum_{\tilde{z}^{t}}
\left(\prod_{i=1}^{t-1}\left\{
Q^{\Thr}_{Y_i|U_i}(y_i|\tilde{u}_i)
p_{Z_i|V_i}(\tilde{z}_i|\tilde{v}_i)\right\}
\right)p_{Z_t|V_t}(\tilde{z}_t|v_t)},
\end{align*}
where for each $i=1,2,\cdots,t-1$, $\tilde{u}_i$ and $\tilde{v}_i$ 
are defined by
\beqno
&&\tilde{u}_i \defeq (y^{i-1},\tilde{z}_{i+1}^t,z_{t+1}^n,l)
\in {\cal Y}^{i-1}\times {\cal Z}_{i+1}^{n}\times {\cal L}_n={\cal U}_i,
\\
&&\tilde{v}_i\defeq (\tilde{z}_{i+1}^t,z_{t+1}^n,l)
\in {\cal Z}_{i+1}^{n}\times {\cal L}_n={\cal V}_i.
\eeqno
\end{lm}

{\it Proof:} 
In (\ref{eqn:asXpA}), we choose $Q^{\One}_{Y^n|X^nZ^nL_n}$ 
so that 
\begin{align}
& Q^{\One}_{Y^n|X^nZ^nL_n}(Y^n|X^n,Z^n,L_n)
\nonumber\\
&=\prod_{t=1}^n 
Q^{\One}_{Y_t|X_tY^{t-1}Z_{t}^nL_n}(Y_t|X_t,Y^{t-1},Z_{t}^n,L_n)
\nonumber\\
&=\prod_{t=1}^n Q^{\One}_{Y_t|X_tU_t}(Y_t|X_t,Z_t,U_t).
\label{eqn:ddsb}
\end{align}
In (\ref{eqn:asXpB}), we choose $Q^{\Two}_{Z^n|X^nY^nL_n}$ so that 
\begin{align}
& Q^{\Two}_{Z^n|X^nY^nL_n}(Z^n|X^n,Y^n,L_n)
\nonumber\\
&=\prod_{t=1}^n 
Q^{\Two}_{Z_t|X_tY^{t-1}Z_{t+1}^nL_n}(Z_t|X_t,Y^{t-1},Z_{t+1}^n,L_n)
\nonumber\\
&= \prod_{t=1}^n 
Q^{\Two}_{Z_t|X_tU_t}(Z_t|X_t,U_t).
\quad \label{eqn:ddsbb}
\end{align}
We define joint conditional distribution 
$\tilde{Q}^{\Thr}_{Y^nZ^n|L_n}(Y^n,Z^n$ $|L_n)$
on $(Y^n,Z^n)$ given $L_n$ by
\begin{align}
& \tilde{Q}^{\Thr}_{Y^nZ^n|L_n}(Y^n,Z^n|L_n)
\nonumber\\
&=\tilde{Q}^{\Thr}_{Y^n|Z^nL_n}(Y^n|Z^n,L_n)
   p_{Z^n|L_n}(Z^n|L_n)
\nonumber\\
&=\left\{\prod_{t=1}^n
Q^{\Thr}_{Y_t|Y^{t-1}Z_{t+1}^nL_n}(Y_t|Y^{t-1},Z_{t+1}^n,L_n)\right\}
\nonumber\\
&\quad \times \left\{\prod_{t=1}^n
p_{Z_t|Z_{t+1}^{n}L_n}(Z_t|Z_{t+1}^n,L_n)\right\}.
\label{eqn:ddsbAzz}
\end{align}
We choose $Q^{\Thr}_{Y^n|L_n}(Y^n|L_n)$ such that it is equal 
to the marginal distribution $\tilde{Q}^{\Thr}_{Y^n|L_n}(Y^n|L_n)$
of $\tilde{Q}^{\Thr}_{Y^nZ^n|L_n}(Y^n,Z^n$ $|L_n)$.
Then in (\ref{eqn:asXpCa}), we have the following chain of equalities:
\begin{align}
& \frac{W_1^n(Y^n|X^n)}{Q^{\Thr}_{Y^n|L_n}(Y^n|L_n) }
=\frac{\tilde{Q}^{\Thr}_{Z^n|L_n}(Z^n|L_n)}
{\tilde{Q}^{\Thr}_{Y^n|L_n}(Y^n|L_n)}
\frac{W_1^n(Y^n|X^n)}{p_{Z^n|L_n}(Z^n|L_n)}
\nonumber\\
&=\prod_{t=1}^n 
\frac{\tilde{Q}^{\Thr}_{Z_t|Y^{t-1}Z_{t+1}^nL_n}(Z_t|Y^{t-1},Z_{t+1}^n,L_n)}
     {\tilde{Q}^{\Thr}_{Y_t|Y^{t-1}Z_{t+1}^nL_n}(Y_t|Y^{t-1},Z_{t+1}^n,L_n)}
\nonumber\\
&\quad \times 
\prod_{t=1}^n \frac{W_1(Y_t|X_t)}{p_{Z_t|Z_{t+1}^nL_n}(Z_t|Z_{t+1}^n,L_n)}
\nonumber\\
&\MEq{a}\prod_{t=1}^n 
\frac{W_1(Y_t|X_t)}{Q^{\Thr}_{Y_t|U_t}(Y_t|U_t)}
\frac{\tilde{Q}^{\Thr}_{Z_t|U_t}(Z_t|U_t)}{p_{Z_t|V_t}(Z_t|V_t)}.
\quad
\label{eqn:ddscb}
\end{align}
Step (a) follows from (\ref{eqn:ddsbAzz}). 
Based on (\ref{eqn:ddsbAzz}), we compute 
${\tilde{Q}^{\Thr}_{Y^{t-1} Z_t|Z_{t+1}^nL_n}(y^{t-1},z_t|z_{t+1}^n,l)}$
to obtain
\begin{align*}
& {\tilde{Q}^{\Thr}_{Y^{t-1}Z_t|Z_{t+1}^nL_n}(y^{t-1},z_t|z_{t+1}^n,l)}
\\
&=\sum_{\tilde{z}^{t-1}}
{\tilde{Q}^{\Thr}_{Y^{t-1}Z^{t-1}Z_t|Z_{t+1}^nl}
(y^{t-1},\tilde{z}^{t-1},z_t|z_{t+1}^n,l)}
\\
&=\sum_{\tilde{z}^{t-1}}
\left(\prod_{i=1}^{t-1}\left\{
Q^{\Thr}_{Y_i|Y^{i-1}Z_{i+1}^{t-1}Z_{t}^nL_n}(y_i|y^{i-1},
\tilde{z}_{i+1}^{t-1},z_{t}^n,l)
\right.
\right.
\nonumber\\
& \quad \left.\times 
p_{Z_i|Z_{i+1}^{t-1}Z_{t}^nL_n}
({z}_i|\tilde{z}_{i+1}^{t-1},z_{t}^n,l)
\right\}\Hugecr
\\
& \quad \times 
p_{Z_t|Z_{t+1}^nL_n}(z_t|z_{t+1}^n,l)
\\
&=\sum_{\tilde{z}^{t-1}}
\left(\prod_{i=1}^{t-1}\left\{
Q^{\Thr}_{Y_i|U_i}(y_i|\tilde{u}_i)
p_{Z_i|V_i}(\tilde{z}_i|\tilde{v}_i)\right\}
\right)p_{Z_t|V_t}(z_t|v_t).
\end{align*}
Hence we have 
\begin{align*}
& \tilde{Q}^{\Thr}_{Z_t|U_t}(z_t|u_t)
 =\tilde{Q}^{\Thr}_{Z_t|Y^{t-1}Z_{t+1}^n L_n}(z_t|y^{t-1},z_{t+1}^n,l)
\\
&=
\frac{\tilde{Q}^{\Thr}_{Y^{t-1}Z_t|Z_{t+1}^nL_n}(y^{t-1},z_t|z_{t+1}^n,l)}
{\tilde{Q}^{\Thr}_{Y^{t-1}|Z_{t+1}^nL_n}(y^{t-1}|z_{t+1}^n,l)}
\\
&=
\frac{\ds \sum_{\tilde{z}^{t-1}}
\left(\prod_{i=1}^{t-1}\left\{
Q^{\Thr}_{Y_i|U_i}(y_i|\tilde{u}_i)
p_{Z_i|V_i}(\tilde{z}_i|\tilde{v}_i)\right\}
\right)p_{Z_t|V_t}(z_t|v_t)}
{\ds \sum_{\tilde{z}^{t}}
\left(\prod_{i=1}^{t-1}\left\{
Q^{\Thr}_{Y_i|U_i}(y_i|\tilde{u}_i)
p_{Z_i|V_i}(\tilde{z}_i|\tilde{v}_i)\right\}
\right)p_{Z_t|V_t}(\tilde{z}_t|v_t)}.
\end{align*}
In (\ref{eqn:asXpCb}), we choose $Q^{\Fou}_{Z^n}$ so that 
\beqa
& &Q^{\Fou}_{Z^n}(Z^n)
=\prod_{t=1}^n Q^{\Fou}_{Z_t}(Z_t).
\label{eqn:ddsc}
\eeqa
In (\ref{eqn:asXpD}), we choose $Q^{\Fiv}_{Y^n}$ so that 
\beqa
& &Q^{\Fiv}_{Y^n}(Y^n)=\prod_{t=1}^n Q^{\Fiv}_{Y_t}(Y_t).
\label{eqn:ddsd}
\eeqa
From Lemma \ref{lm:Ohzzz} and (\ref{eqn:ddsb})-(\ref{eqn:ddsd}), 
we have the bound (\ref{eqn:Zsdaa}) in Lemma \ref{lm:Ohzzzca}.
\hfill \IEEEQED

For each $t=1,2,\cdots,n$, let 
$\underline{\cal Q}_t $ 
be a set of all 
\beqno 
\underline{Q}_t&=&(Q^{\One}_{Y_t|X_tZ_tU_t},
Q^{\Two}_{Z_t|X_tU_t},
Q^{\Thr}_{Y_t|U_t},
Q^{\Fou}_{Z_t},
Q^{\Fiv}_{Y_t}).
\eeqno 
Set 
\beqno
\underline{\cal Q}^n&\defeq& 
\prod_{t=1}^n \underline{\cal Q}_t,
\underline{Q}^n  \defeq  \left\{ \underline{Q}_t \right\}_{t=1}^n 
\in \underline{\cal Q}^n.
\eeqno
To evaluate an upper bound of (\ref{eqn:Zsdaa}) in Lemma \ref{lm:Ohzzzca}.
We use the following lemma, which is well known as 
the Cram\`er's bound in the large deviation principle.
\begin{lm}
\label{lm:Ohzzzb}
For any real valued random variable 
$A$ and any $\theta\geq 0$, we have
$$
\Pr\{A \geq a \}\leq 
\exp
\left[
-\left(
\lambda a -\log {\rm E}[\exp(\theta A)]
\right) 
\right].
$$
\end{lm}

Here we define a quantity which serves as an exponential
upper bound of ${\rm P}_{\rm c}^{(n)}(\varphi^{(n)},$ 
$\psi_1^{(n)},\psi_2^{(n)})$. 
Let ${\cal P}^{(n)}(W_1,W_2)$ be a 
set of all probability distributions 
${p}_{L_nX^nY^nZ^n}$ on 
${\cal L}_n$
$\times {\cal X}^n$
$\times {\cal Y}^n$
$\times {\cal Z}^n$
having the form:
\begin{align*}
& {p}_{L^nX^nY^nZ^n}(l,x^n,y^n,z^n)
\\
&={p}_{L^n}(l)
\prod_{t=1}^n 
{p}_{X_t|L_n X^{t-1}}
(x_t|l,x^{t-1})W_1(y_t|x_t)W_2(z_t|x_t).
\end{align*}
For simplicity of notation we use the notation $p^{(n)}$ 
for $p_{L_nX^nY^nZ^n}$ $\in {\cal P}^{(n)}$
$(W_1,W_2)$. We assume that 
$
p_{U_tX_tY_tZ_t}=p_{L_nX_tY^{t}Z_{t}^n}
$
is a marginal distribution induced by $p^{(n)}$.
For $t=1,2,\cdots, n$, we simply write $p_t=$ $p_{U_tX_tY_tZ_t}$. 
For each $t=1,2,\cdots, n$, let ${\rm Proj}({\cal U}_t \to {\cal V}_t)$ 
be a set of all projection ${\kappa}_t$ from ${\cal U}_t$ onto ${\cal V}_t$.
For $p^{(n)}$ $\in {\cal P}^{(n)}(W_1,W_2)$,
$$
\kappa^n=\{\kappa_t\}_{t=1}^n 
\in \prod_{t=1}^n {\rm Proj}({\cal U}_t \to {\cal V}_t),
$$ 
and $\underline{Q}^n$ $\in \underline{\cal Q}^n$, we define
\begin{align*}
& 
\Omega^{(\mu,\gamma,\pOne,\theta)}(p^{(n)},\kappa^n,\underline{Q}^{n})
\\
&\defeq
\log
{\rm E}_{p^{(n)}}
\left[\left(
\prod_{t=1}^n\left\{
\frac{W_1(Y_t|X_t)}
 {Q^{\One}_{Y_t|X_tZ_tU_t}(Y_t|X_t,Z_t,U_t)}
\right\}^{\theta}\right)\right.
\\
&
\times\left(
\prod_{t=1}^n \left\{
\frac{W_2(Z_t|X_t)}
 {Q^{\Two}_{Z_t|X_tU_t}(Z_t|X_t,U_t)}
\right\}^{\theta}\right)
\\
&
\times 
\left(\prod_{t=1}^n
\left\{
\frac{W_1(Y_t|X_t)\tilde{Q}_{Z_t|U_t}^{\Thr}(Z_t|U_t)}
{Q^{\Thr}_{Y_t|U_t}(Y_t|U_t)p_{Z_t|V_t}(Z_t|V_t)}
\right\}^{\gamma\mu\pOne\theta}\right)
\\
&
\times 
\left(
\prod_{t=1}^n
\left\{
\frac{p_{Z_t|V_t}(Z_t|V_t)}
    {Q_{Z_t}^{\Fou}(Z_t)}
\right\}^{\gamma{\prmtB}\pOne\theta}
\right)
\\
&
\times 
\left(\prod_{t=1}^n
\left\{
\frac{W_1(Y_t|X_t)}{Q^{\Fiv}_{Y_t}(Y_t)}
\right\}^{\bar{\gamma}\pOne\theta}\right)\HUgebr,
\end{align*}
where for each $t=1,2,\cdots,n$, 
the following probability 
and conditional probability distributions:
\beq
\ba{l}
Q^{\One}_{Y_t|X_tZ_tU_t},
Q^{\Two}_{Z_t|X_tU_t},
Q^{\Thr}_{Y_t|U_t}, 
Q^{\Fou}_{Z_t},
Q^{\Fiv}_{Y_t}
\ea
\eeq
appearing in the definition of 
$
\Omega^{(\mu,\gamma,\pOne,\theta)}(p^{(n)}, \kappa^n, \underline{Q}^{n})
$
can be chosen arbitrary.
\newcommand{\XzQQ}{
Here we give a remark on an essential difference 
between $p^{(n)}$ $\in {\cal P}^{(n)}(W_1,W_2)$ 
and $q^n$ $\in {\cal Q}^n$. 
For the former the $n$ probability distributions 
$p_t,$ $t=1,2,\cdots, n,$ are consistent with $p^{(n)}$, 
since all of them are marginal distributions 
of $p^{(n)}$. On the other hand, for the latter, $q^{n}$ 
is just {\it a sequence} of $n$ 
probability distributions. Hence, we may not have the 
consistency between the $n$ elements $q_t$, $t=1,2,\cdots,n,$ 
of $q^n$. 
}

By Lemmas \ref{lm:Ohzzzca} and \ref{lm:Ohzzzb}, we have 
the following proposition. 
\begin{pro} \label{pro:abcOhzzp}
For any $(\mu,\gamma) \in[0,1]^2$, any $\pOne,$ 
$\theta \geq 0$, any $\underline{Q}^n \in \underline{\cal Q}^n$, and 
any $(\varphi^{(n)},\psi_1^{(n)},\psi_2^{(n)})$  
satisfying  
$(1/n)\log |{\cal K}_n|$ $\geq R_1,$
$(1/n)\log |{\cal L}_n|$$\geq R_2,$
we have 
\begin{align*}	
& {\rm P}_{\rm c}^{(n)}(\varphi^{(n)},\psi_1^{(n)},\psi_2^{(n)})
\leq 6 \exp \biggl[-n
[1+\theta(2+\pOne(\gamma\mu+1)]^{-1}
\\
& \times 
\left\{\theta \alpha[(\bar{\gamma}+\gamma\mu)R_1 + R_2] 
-\frac{1}{n}
\Omega^{(\mu,\gamma,\pOne,\theta)}(p^{(n)}, \kappa^n, \underline{Q}^{n})
\right\}\biggr].
\end{align*}

\end{pro}

{\it Proof:} 
When 
$$
n \theta[(\bar{\gamma}+\gamma\mu)R_1+{\prmtB} R_2] \leq 
\Omega^{(\mu,\gamma,\pOne, \theta)}(p^{(n)},
{\kappa}^n,\underline{Q}^n),
$$ 
the bound we wish to prove is obvious.
In the following argument we assume that
$$
n\theta[(\mu \bar{\gamma}+\gamma\mu) R_1+{\prmtB} R_2]
>\Omega^{(\mu,\theta)}(p^{(n)},{\kappa}^n,\underline{Q}^n).
$$
We define six random variables $A_i,$$i=1,2,\cdots,5$ by
\beqno
& & A_1=\frac{1}{n}\sum_{t=1}^n
\log \frac{W_1(Y_t|X_t)}{Q^{\One}_{Y_t|{X_tZ_tU_t}}(Y_t|X_t,Z_t,U_t)},
\\
& & A_2=\frac{1}{n}\sum_{t=1}^n
\log \frac{W_2(Z_t|X_t)}{Q^{\Two}_{Z_t|{X_tU_t}}(Z_t|X_t,U_t)},
\\
& & A_3=\frac{1}{n}\sum_{t=1}^n
\log \frac{W_1(Y_t|X_t)\tilde{Q}_{Z_t|U_t}^{\Thr}(Z_t|U_t)} 
          {Q_{Y_t|U_t}^{\Thr}(Y_t|U_t)p_{Z_t|V_t}(Z_t|V_t)},
\\
& & A_4=\frac{1}{n}\sum_{t=1}^n
  \log \frac{p_{Z_t|V_t}(Z_t|V_t)}{Q_{Z_t}^{\Fiv}(Z_t)},
A_5=
\frac{1}{n}\sum_{t=1}^n\log
\frac{W_1(Y_t|X_t)}{Q_{Y_t}^{\Fiv}(Y_t)}.
\eeqno
Then by Lemma \ref{lm:Ohzzzca}, 
for any $(\varphi^{(n)},\psi_1^{(n)},\psi_2^{(n)})$  
satisfying  
$
\frac{1}{n}\log |{\cal K}_n| \geq R_1,
$
$
\frac{1}{n}\log |{\cal L}_n| \geq R_2,
$
we have
\begin{align}
& {\rm P}_{\rm c}^{(n)}(\varphi^{(n)},\psi_1^{(n)},\psi_2^{(n)})
\leq p_{{L_n}X^nY^nZ^n}\{
A_i\geq -\eta\mbox{ for }i=1,2,
\nonumber\\
& A_3\geq R_1-\eta, 
  A_4\geq R_2-\eta, A_5\geq R_1+R_2-\eta\}+5{\rm e}^{-n\eta}
\nonumber\\
&\leq p_{{L_n}X^nY^nZ^n}\{
A_1+A_2+\pOne(\gamma[\mu A_3+ A_4]+\bar{\gamma} A_5])
\nonumber\\
&\quad  \geq \pOne([\gamma\mu+\bar{\gamma}] R_1 + R_2 
 -\eta [2+\pOne(\gamma\mu+1)]\}+5{\rm e}^{-n\eta}
\nonumber\\
&= p_{{L_n}X^nY^nZ^n}\{A\geq a\}+5{\rm e}^{-n\eta},
\label{eqn:abcawxx}
\end{align}
where we set 
\beqno
A& \defeq &A_1+A_2+\pOne(\gamma[\mu  A_3+ {\prmtB} A_4]
     +\bar{\gamma}A_5),
\\
a& \defeq &\pOne([\gamma\mu+\bar{\gamma}] R_1+R_2) 
 -\eta [2+\pOne(\gamma\mu+1)]. 
\eeqno
Applying Lemma \ref{lm:Ohzzzb} to the first term in the right member 
of (\ref{eqn:abcawxx}), we have  
\begin{align}
& 
{\rm P}_{\rm c}^{(n)}(\varphi^{(n)},\psi_1^{(n)},\psi_2^{(n)})
\nonumber\\
&\leq
\exp \left[
-\left(\theta a -\log {\rm E}_{p^{(n)}}[\exp(\theta A)]\right) 
\right]+5{\rm e}^{-n\eta}
\nonumber\\
&=
\exp\biggl[n\biggl\{
\theta[2+\pOne(\gamma\mu+1)]\eta
-\theta\pOne
\bigl[({\bar{\gamma}+\gamma\mu})R_1+R_2 \bigr]
\nonumber\\
& \qquad\left.\left.
 +\frac{1}{n}
\Omega^{(\mu,\gamma,\pOne,\theta)}(p^{(n)}, \kappa^n, \underline{Q}^{n})
\right\}\right]
  +5{\rm e}^{-n\eta}.\qquad 
\label{eqn:aaabv}
\end{align}
We choose $\eta$ so that 
\beqa
-\eta&=&
\theta\{2+\pOne(\gamma\mu+1)\}\eta
-\theta\pOne
\bigl[({\bar{\gamma}+\gamma\mu})R_1+R_2 \bigr]
\nonumber\\ 
    & &+\frac{1}{n}
\Omega^{(\mu,\gamma,\pOne,\theta)}(p^{(n)}, \kappa^n, \underline{Q}^{n}).
\label{eqn:aaappp}
\eeqa
Solving (\ref{eqn:aaappp}) with respect to $\eta$, we have 
\beqno
\eta&=&
\left[1+\theta\{2+\pOne(\gamma\mu+1)\}\right]^{-1}
\biggl\{\theta\pOne\bigl[({\bar{\gamma}+\gamma\mu})R_1
+R_2\bigr]
\\
& &
\left. \left. 
\quad 
-\frac{1}{n}
\Omega^{(\mu,\gamma,\theta)}(p^{(n)}, \kappa^n, \underline{Q}^{n})
\right.\right\}.
\eeqno
For this choice of $\eta$ and (\ref{eqn:aaabv}), we have
\begin{align*}
&{\rm P}_{\rm c}^{(n)}(\varphi^{(n)},\psi_1^{(n)},\psi_2^{(n)})
\leq 6{\rm e}^{-n\eta}
\\
&=
6\exp
\biggl[
-n\left[1+\theta\{2+\pOne(\gamma\mu+1)\}\right]^{-1}
\\
&\times 
\biggl\{\theta\pOne\bigl[({\bar{\gamma}+\gamma\mu})R_1+R_2\bigr]
-\frac{1}{n}
\Omega^{(\mu,\gamma,\pOne,\theta)}(p^{(n)}, \kappa^n, \underline{Q}^{n})
\biggr\}\biggr],
\end{align*}
completing the proof. 
\hfill \IEEEQED

Set 
\begin{align*}
& \overline{\Omega}^{(\mu,\gamma,\pOne,\theta)}(W_1,W_2)
\\
&\defeq  
\sup_{n\geq 1}
\max_{\scs 
{p}^{(n)} \in
\scs
     {\cal P}^{(n)}(W_1,W_2),
      \atop{\scs 
        \kappa^n \in
         \prod_{t=1}^n {\rm Proj}({\cal U}_t \to {\cal V}_t) 
      }
}
\min_{\scs \underline{Q}^n \in \underline{\cal Q}^n}1
\\
&\qquad \times
\frac{1}{n}\Omega^{(\mu,\gamma,\pOne,\theta)}
(p^{(n)},\kappa^n,\underline{Q}^{n}).
\end{align*}
Then we have the following corollary from Proposition \ref{pro:abcOhzzp}.
\begin{co}\label{co:ProA}
For any $R_1,R_2\geq 0$, any $(W_1,W_2)$, 
any $\theta \geq 0,\mu\in [0,1]$, and 
for any $(\varphi^{(n)},$ $\psi_1^{(n)},$ 
$\psi_2^{(n)})$ satisfying 
$
(1/n)\log |{\cal K}_n| \geq R_1,  
(1/n)\log |{\cal L}_n| \geq R_2,  
$
we have 
\begin{align*}	
&  {\rm P}_{\rm c}^{(n)}(\varphi^{(n)},\psi_1^{(n)},\psi_2^{(n)})
\\
& 
\leq 6 \exp \biggl[ -n
[1+\theta\{2+\pOne(1+\gamma\mu)\} ]^{-1}
\\
&\quad \times 
\left\{ \theta \alpha[({\bar{\gamma}+\gamma\mu})R_1+R_2] 
-\overline{\Omega}^{(\mu,\gamma,\pOne, \theta)}(W_1,W_2)
\right\}\biggr].
\end{align*}
\end{co}

We shall call $\overline{\Omega}^{(\mu,\gamma,\pOne,\theta)}(W_1,W_2)$ 
the communication potential. The above corollary implies that the 
analysis of $\overline{\Omega}^{(\mu,\gamma,\pOne,\theta)}($$W_1,W_2)$ 
leads to an establishment of a strong converse theorem for the ABC. 

\newcommand{\Ft}{ {\cal F}^t } 
\newcommand{\Fi}{ {\cal F}^i } 
In the following argument we drive an explicit upper bound of 
$\overline{\Omega}^{(\mu,\gamma,\pOne,\theta)}$ $(W_1,W_2)$. To 
this end we use a new novel technique called {\it the recursive method}. 
This method is a powerful tool to drive a single letterized exponent 
function for rates below the rate distortion function. This method is 
also applicable to prove the exponential strong converse theorems for 
other network information theory problems \cite{OhIsit15AKWStConv}, 
\cite{OhIsit15DBCStConv}, \cite{OhIsit15DBCFBStConv}.
Set
$
{\cal F}_t\defeq (p_{Z_t|V_t},\kappa_t,\underline{Q}_t),
{\cal F}^t \defeq \{{\cal F}_i\}_{i=1}^{t}.
$
For each $t=1,2,\cdots,n$, define a function of 
$(u_t,x_t,y_t,z_t)$
$\in {\cal U}_t$
$\times {\cal X}$
$\times {\cal Y}$
$\times {\cal Z}$ 
by 
\begin{align*}
& f_{\Ft}^{(\mu,\gamma,\pOne,\theta)}
(x_t,y_t,z_t|u_t)
\nonumber\\
&\defeq  
\left\{
\frac{W_1(y_t|x_t)W_2(z_t|x_t)}
{Q_{Y_t|X_tZ_tU_t}^{\One}(y_t|x_t,z_t,u_t)
Q_{Z_t|X_tU_t}^{\Two}(z_t|x_t,u_t)}
\right\}^{\theta}
\\
&\quad \times 
\left\{
\frac{W_1(y_t|x_t)}{Q^{\Thr}_{Y_t|U_t}(y_t|u_t)}
\right\}^{\mu\gamma\pOne\theta}
\left\{\frac{\tilde{Q}_{Z_t|U_t}^{\Thr}(z_t|u_t)}
{p_{Z_t|V_t}(z_t|v_t)}
\right\}^{\mu\gamma\pOne\theta}
\\
&\quad \times
\left\{
\frac{p_{Z_t|V_t}(z_t|v_t)}
    {Q_{Z_t}^{\Fou}(z_t)}
\right\}^{\gamma\pOne\theta}
\left\{
\frac{W_1(y_t|x_t)}{Q^{\Fiv}_{Y_t}(y_t)}
\right\}^{\bar{\gamma}\pOne\theta}.
\end{align*}
Here we note that $\tilde{Q}^{\Thr}_{Z_t|U_t}$ is 
uniquely determined by the component 
$\{p_{Z_i|V_i},Q^{\Thr}_{Y_i|U_i}\}_{i=1}^{t-1}$ of ${\cal F}^{t-1}$ and 
$p_{Z_t|V_t}$, that is,
$$
 \tilde{Q}^{\Thr}_{Z_t|U_t}
=\tilde{Q}^{\Thr}_{Z_t|U_t;({\cal F}^{t-1},p_{Z_t|V_t})}.
$$
For each $t=1,2,\cdots,n$, we define a conditional probability 
distribution of $(X^t,Y^t)$ given $(L_n,Z^n)$ by
\begin{align*}
& {p}_{X^tY^{t}|L_nZ^n;{\cal F}^t}^{(\mu,\gamma,\pOne,\theta)}
\defeq \biggl\{
p_{X^tY^{t}|L_nZ^n;{\cal F}^t}^{(\mu,\gamma,\pOne,\theta)}
(x^t,y^t|l,z^n)
\\
&\qquad \qquad\qquad\qquad\qquad
  \biggr\}_{(x^t,y^t,l,z^n)
  \in {\cal X}^t \times {\cal Y}^t
\times {\cal L}_n \times {\cal Z}^n},
\\
& p_{X^tY^{t}|L_n{{Z}}^n;{\cal F}^t}^{(\mu,\gamma,\pOne,\theta)}
(x^t,y^t|l,{z}^n)\defeq C_t^{-1}(l,{z}^n)
\\ 
&\quad \times 
p_{X^tY^{t}|L_n{{Z}}^n}(x^t,y^t|l,{z}^n)
\prod_{i=1}^t
f_{\Fi}^{(\mu,\gamma,\pOne,\theta)}(x_i,y_i,z_i|u_i),
\end{align*} 
where
\beqno
C_t(l,{z}^n)
&\defeq & 
\ba[t]{l}
\ds \sum_{x^t,y^t}p_{X^tY^{t}|L_n{{Z}}^n}(x^t,y^t|l,{z}^n)
\vspace*{-2mm}\\
\ds \qquad \times \prod_{i=1}^t 
f_{\Fi}^{(\mu,\gamma,\pOne,\theta)}(x_i,y_i,z_i|u_i)
\ea
\eeqno
are constants for normalization. For $t=1,2,\cdots,n$, 
define 
\beq
\Phi_{t,{\cal F}^t}^{(\mu,\gamma,\pOne,\theta)}(l,{z}^n)
\defeq C_t(l,z^n)C_{t-1}^{-1}(l,z^n),
\label{eqn:defa}
\eeq
where we define $C_{0}(l,z^n)=1$ for 
$(l,z^n)\in {\cal L}_n $ $\times {\cal Z}^n.$
Then we have the following lemma.
\begin{lm}\label{lm:aaa}
For each $t=1,2,\cdots,n$, and for any 
$(l,$ $z^n$ $x^t, y^t)\in {\cal L}_n$
$\times {\cal Z}^n$
$\times {\cal X}^t$
$\times {\cal Y}^t$,
we have
\begin{align*}
& {p}_{X^tY^t|L_nZ^n;{\cal F}^t}^{(\mu,\gamma,\pOne,\theta{})}
(x^t,y^t|l,z^n)
=(\Phi_{t,{\cal F}^t}^{(\mu,\gamma,\pOne,\theta)}(l,{z}^n))^{-1}
\\
& \quad \times
 p_{X^{t-1}Y^{t-1}|L_n{{Z}}^n;{\cal F}^{t-1}}^{(\mu,\gamma,\pOne,\theta)}
(x^{t-1},y^{t-1}|l,{z}^n)
\\
& \quad \times p_{X_tY_t|L_nX^{t-1}Z^n}
(x_t,y_t|l,x^{t-1},y^{t-1},z^n)
\nonumber\\
& \quad \times 
f_{\Ft{}}
^{(\mu,\gamma,\pOne,\theta)}(x_t,y_t,z_t|u_t).
\end{align*}
Furthermore, we have  
\begin{align}
& \Phi_{t,{\cal F}^t}^{(\mu,\gamma,\pOne,\theta)}(l,z^n)
\nonumber\\
& = \sum_{x^t,y^t} 
p_{X^{t-1}Y^{t-1}|L_n{{Z}}^n;{\cal F}^{t-1}}^{(\mu,\gamma,\pOne,\theta)}
(x^{t-1},y^{t-1}|l,{z}^n)
\nonumber\\
& \quad \times p_{X_tY_t|L_nX^{t-1}Z^n}
(x_t,y_t|l,x^{t-1},y^{t-1},z^n)
\nonumber\\
& \quad\times 
f_{\Ft{}}
^{(\mu,\gamma,\pOne,\theta)}
(x_t,y_t,z_t|u_t).
\label{eqn:DfaaaK}
\end{align}
\end{lm}

Proof of this lemma is given in Appendix \ref{sub:sdfa}.
\newcommand{\Apdc}{
\subsection{Proof of Lemma \ref{lm:aaa}}\label{sub:sdfa}
In this appendix we prove Lemma \ref{lm:aaa}.  

{\it Proof of  Lemma \ref{lm:aaa}:} By the definition 
of ${p}_{X^tY^{t}|L_nZ^n;{\cal F}^t}^{(\mu,\gamma,\pOne,\theta{})}$ 
$(x^t,y^t|l,{z}^n)$, for $t=1,2,\cdots,n$, we have 
\begin{align}
& p_{X^tY^{t}|L_nZ^n;{\cal F}^t}^{(\mu,\gamma,\pOne,\theta{})}(x^t,y^t|l,{z}^n)
\nonumber\\
&=C_t^{-1}(l,z^n)
p_{X^tY^t|L_nZ^n;{\cal F}^t}(x^t,y^t|l,z^n) 
\nonumber\\
&\quad \times
\prod_{i=1}^t
f_{\Fi}^{(\mu,\gamma,\pOne,\theta)}
(x_i,y_i,z_i|u_i).
\label{eqn:azaq}
\end{align} 
Then we have the following chain of equalities:
\begin{align} 
& p_{X^tY^{t}|L_n{{Z}}^n;{\cal F}^t}^{(\mu,\gamma,\pOne,\theta{})}
(x^t,y^{t}|l,{z}^n)
\nonumber\\
&\MEq{a}
C_t^{-1}(l,z^n)p_{X^tY^{t}|L_nZ^n}(x^t,y^t|l,z^n) 
\nonumber\\
&\quad \times \prod_{i=1}^t 
f_{\Fi}^{(\mu,\gamma,\pOne,\theta)}
(x_i,y_i,z_i|u_i)
\nonumber\\
&=C_t^{-1}(l,{z}^n)p_{X^{t-1}Y^{t-1}|L_nZ^n}(x^{t-1},y^{t-1}|l,z^n)
\nonumber\\
&\quad\times \prod_{i=1}^{t-1}
f_{\Fi}^{(\mu,\gamma,\pOne,\theta)}(x_i,y_i,z_i|u_i)
\nonumber\\
&\quad \times p_{ X_tY_t|X^{t-1}Y^{t-1}L_nZ^n}(x_t,y_t|x^{t-1},y^{t-1},l,z^n) 
\nonumber\\
&\quad \times f_{\Ft{}}^{(\mu,\gamma,\pOne,\theta)}(x_t,y_t|u_t)
\nonumber\\
&\MEq{b}
\frac{C_{t-1}(l,{z}^n)}{C_t(l,{z}^n)}
p_{X^{t-1}Y^{t-1}|L_n{{Z}}^n;{\cal F}^{t-1}}^{(\mu,\gamma,\pOne,\theta)}
(x^{t-1},y^{t-1}|l,{z}^n)
\nonumber\\
&\quad 
\times p_{X_t|Y_t|X^{t-1}Y^{t-1}L_n{{Z}}^n}(x_t,y_t|x^{t-1},y^{t-1},l,{z}^n) 
\nonumber\\
& \quad \times f_{\Ft{}}^{(\mu,\gamma,\pOne,\theta)}(x_t,y_t,z_t|u_t)
\nonumber\\
&=(\Phi_{t,{\cal F}^t}^{(\mu,\gamma,\pOne,\theta)}(l,{z}^n))^{-1}
\nonumber\\
&\quad \times p_{X^{t-1}Y^{t-1}|L_nZ^n;{\cal F}^{t-1}}^{(\mu,\gamma,\pOne,\theta)}
(x^{t-1},y^{t-1}|l,z^n)
\nonumber\\
&\quad  \times p_{X_tY_t|X^{t-1}Y^{t-1}L_n{{Z}}^n}(x_t,y_t|x^{t-1},z^{t-1},l,{z}^n) 
\nonumber\\
& \quad \times f_{\Ft{}}^{(\mu,\gamma,\pOne,\theta)}(x_t,y_t, z_t|u_t).
\label{eqn:daaaq}
\end{align}
Steps (a) and (b) follow from (\ref{eqn:azaq}). 
From (\ref{eqn:daaaq}), we have 
\begin{align} 
& \Phi_{t,{\cal F}^t}^{(\mu,\gamma,\pOne,\theta)}(l,z^n)
p_{X^tY^{t}|L_nZ^n;{\cal F}^t}^{(\mu,\gamma,\pOne,\theta{})}
(x^t,y^t|l,{z}^n)
\label{eqn:daxx}
\\
&=p_{X^{t-1}Y^{t-1}|L_nZ^n;{\cal F}^{t-1}}
^{(\mu,\gamma,\pOne,\theta)}
(x^{t-1},y^{t-1}|l,z^n)
\nonumber\\
&\quad \times
p_{X_tY_t|X^{t-1}Y^{t-1}L_nZ^n}(x_t,y_t|x^{t-1},y^{t-1},l,z^n)
\nonumber\\
&\quad \times
f_{\Ft{}}^{(\mu,\gamma,\pOne,\theta)}(x_t,y_t,z_t|u_t).
\label{eqn:daaxx}
\end{align}
Taking summations of (\ref{eqn:daxx}) and 
(\ref{eqn:daaxx}) with respect to $x^t,y^t$, 
we obtain 
\begin{align*} 
& \Phi_{t,{\cal F}^t}^{(\mu,\gamma,\pOne,\theta)}(l,z^n)
\\
&=\sum_{x^t,y^t}
p_{X^{t-1}Y^{t-1}|L_nZ^n;{\cal F}^{t-1}}
^{(\mu,\gamma,\pOne,\theta)}
(x^{t-1},y^{t-1}|l,z^n)
\nonumber\\
&\quad \times
p_{X_tY_t|X^{t-1}Y^{t-1}L_nZ^n}(x_t,y_t|x^{t-1},y^{t-1},l,z^n)
\nonumber\\
&\quad \times
f_{\Ft{}}^{(\mu,\gamma,\pOne,\theta)}
  (x_t,y_t,z_t|u_t),
\end{align*}
completing the proof.
\hfill \IEEEQED
}
Next we define the probability distribution 
$$
p_{L_nZ^n;{\cal F}^t}^{(\mu,\gamma,\pOne,\theta{})}
=\left\{
p_{L_nZ^{{n}};{\cal F}^t}^{(\mu,\gamma,\pOne,\theta{})}(l,z^n)
\right\}_{ (l,z^n)\in  {\cal L}_n\times {\cal Z}^n }
$$
of the random variable $(L_n,Z^n)$ 
taking values in ${\cal L}_n$ $\times {\cal Z}^n$ by 
\begin{align}
& p_{L_nZ^{{n}};{\cal F}^t}^{(\mu,\gamma,\pOne,\theta{})}(l,z^n)
\nonumber\\
&=\tilde{C}_t^{-1} 
p_{L_nZ^{{n}}}(l,z^n)
\prod_{i=1}^t 
\Phi_{i,{\cal F}^i}^{(\mu,\gamma,\pOne,\theta)}(l,z^n),
\label{eqn:defzz}
\end{align}
where $\tilde{C}_t$ is a constant for normalization given by 
$$
\tilde{C}_t=\sum_{l,z^n}p_{L_nZ^{{n}}}(l,z^n)
\prod_{i=1}^t 
\Phi_{i,{\cal F}^i}^{(\mu,\gamma,\pOne,\theta)}(l,z^n).
$$
By the above definition, we have
\begin{align}
\tilde{C}_n=\exp\left\{
\Omega^{(\mu,\gamma,\pOne,\theta)}
(p^{(n)},\kappa^n,\underline{Q}^{n})
\right\}.
\label{eqn:Kds}
\end{align}
For $t=1,2,\cdots,n$, define 
\beq
\Lambda_{t,{\cal F}^t}^{(\mu,\gamma,\pOne,\theta)}
\defeq \tilde{C_t}\tilde{C}_{t-1}^{-1},
\label{eqn:passdf}
\eeq
where we define $\tilde{C}_0=1$. Furthermore, define
\begin{align}
& p_{L_n X_tY^tZ_t^n;{\cal F}^{t-1}}^{(\mu,\gamma,\pOne,\theta)}
(l,x_t,y^t,z_t^n)
\nonumber\\
&
= p_{U_tX_tY_tZ_t;{\cal F}^{t-1}}^{(\mu,\gamma,\pOne,\theta)}
(u_t,x_t,y_t,z_t)
\nonumber\\
& \defeq  \sum_{x^{t-1},z^{t-1}}
p_{L_nZ^n;{\cal F}^{t-1}}^{(\mu,\gamma,\pOne,\theta )}(l,z^n)
\nonumber\\
& \quad \times
p_{X^{t-1}Y^{t-1}|L_n Z^n;{\cal F}^{t-1}}^{(\mu,\gamma,\pOne,\theta)}
(x^{t-1},y^{t-1}|l,z^n)
\nonumber\\
& \quad \times 
p_{X_tY_t|X^{t-1}Y^{t-1}L_nZ^n}(x_t,y_t|x^{t-1},y^{t-1},l,z^n). 
\end{align}
Then, we have the following lemma, which is a key result 
to derive a single-letterized upper bound of 
$\overline{\Omega}^{(\mu,\gamma,\pOne,\theta)}(W_1,W_2)$. 
\begin{lm}\label{lm:keylm}For any $\pOne,\theta \geq 0$, 
any $p^{(n)}\in {\cal P}^{(n)}$,    
any $\kappa^n \in \prod_{t=1}^n
               {\rm Proj}({\cal U}_t \to {\cal V}_t)$, 
and any $\underline{Q}^n \in \underline{\cal Q}^n$, we have 
\begin{align} 
& \Omega^{(\mu,\gamma,\pOne,\theta)}
(p^{(n)},\kappa^n, \underline{Q}^n)
=\sum_{t=1}^n 
\log \Lambda_{t,{\cal F}^t}^{(\mu,\gamma,\pOne,\theta)}
\quad\label{eqn:uzapa}
\\
& 
\Lambda_{t,{\cal F}^t}^{(\mu,\gamma,\pOne,\theta)}
= \sum_{u_t,x_t,y_t,z_t }
p_{U_tX_tY_tZ_t;{\cal F}^{t-1}}^{(\mu,\gamma,\pOne,\theta)}
(u_t,x_t,y_t,z_t) 
\nonumber\\
&\quad \qquad\qquad \times
f_{\Ft}^{(\mu,\gamma,\pOne,\theta)}(x_t,y_t,z_t|u_t). 
\label{eqn:uzapaPPP}
\end{align}
\end{lm}

{\it Proof}: We first prove (\ref{eqn:uzapa}). We have the following: 
\begin{align}
& \exp\left\{
\Omega^{(\mu,\gamma,\pOne,\theta)}
(p^{(n)},\kappa^n, \underline{Q}^n)
\right\}
\nonumber\\
&=\tilde{C}_n=\prod_{t=1}^n\tilde{C}_t\tilde{C}_{t-1}^{-1}
\MEq{a}\prod_{t=1}^n
\Lambda_{t,{\cal F}^t}^{(\mu,\gamma,\pOne,\theta)}. 
\label{eqn:zza}
\end{align}
Step (a) follows from (\ref{eqn:Kds}).
Step (b) follows from the definition (\ref{eqn:passdf}) of 
$\Lambda_{t,q^t}^{(\mu,\gamma,\pOne,\theta)}.$ 
From (\ref{eqn:zza}), we have (\ref{eqn:uzapa}) 
in Lemma \ref{lm:keylm}. We next prove (\ref{eqn:uzapaPPP}). 
Multiplying 
$\Lambda_{t,{\cal F}^t}^{(\mu,\gamma,\pOne,\theta)}
=\tilde{C}_{t}/\tilde{C}_{t-1}$ to both sides of (\ref{eqn:defzz}),
we have 
\begin{align}
& \Lambda_{t,{\cal F}^t}^{(\mu,\gamma,\pOne,\theta)}
 p_{L_nZ^n;{\cal F}^t}^{(\mu,\gamma,\pOne,\theta{})}(l,z^n)
\label{eqn:aadff}
\\
&=\tilde{C}_{t-1}^{-1}
p_{L_nZ^{{n}}}(l,z^n)
\prod_{i=1}^t 
\Phi_{i,{\cal F}^i}^{(\mu,\gamma,\pOne,\theta)}(l,z^n),
\nonumber\\
&=
p_{L_nZ^n;{\cal F}^{t-1}}^{(\mu,\gamma,\pOne,\theta)}(l,z^n)
\Phi_{t,{\cal F}^t}^{(\mu,\gamma,\pOne,\theta)}(l,z^n).
\label{eqn:aadffsss}
\end{align}
Taking summations of (\ref{eqn:aadff}) and (\ref{eqn:aadffsss}) 
with respect to $(l,z^n)$, we have 
\begin{align}
& \Lambda_{t,{\cal F}^t}^{(\mu,\gamma,\pOne,\theta)}
=\sum_{l,z^n}
p_{L_nZ^n;{\cal F}^{t-1}}^{(\mu,\gamma,\pOne,\theta)}(l,z^n)
\Phi_{t,{\cal F}^t}^{(\mu,\gamma,\pOne,\theta)}(l,z^n)
\nonumber\\
&\MEq{a}
\sum_{l,z^n}\sum_{x^t,y^t} 
p_{L_nZ^n;{\cal F}^{t-1}}^{(\mu,\gamma,\pOne,\theta)}(l,z^n)
\nonumber\\
& \quad \times
p_{X^{t-1}Y^{t-1}|L_nZ^n;{\cal F}^{t-1}}^{(\mu,\gamma,\pOne,\theta)}
(x^{t-1},y^{t-1}|l,z^n)
\nonumber\\
& \quad \times 
p_{X_tY_t|X^{t-1}Y^{t-1}L_nZ^n}(x_t,y_t|x^{t-1},y^{t-1},l,z^n) 
\nonumber\\
& \quad \times 
f_{\Ft{}}^{(\mu,\gamma,\pOne,\theta)}(x_t,y_t,z_t|u_t). 
\label{eqn:uzapazz}
\end{align}
Step (a) follows from (\ref{eqn:DfaaaK}) in Lemma \ref{lm:aaa}.
From (\ref{eqn:uzapazz}) and the definition of 
$p_{U_tX_tY_tZ_t;{\cal F}^{t-1}}^{(\mu,\gamma,\pOne,\theta)}$, 
we have (\ref{eqn:uzapaPPP}) in Lemma \ref{lm:keylm}. 
\hfill \IEEEQED

The following proposition is a mathematical core 
to prove our main result.
\begin{pro}
\label{pro:mainpro}
For $\gamma \in [0,1]$, $\pOne\geq 0$, and $\pTwo\geq 0$, set 
\beq
\theta \defeq \frac{\pTwo}{1+\gamma\pOne\pTwo}.
\label{eqn:Dsxxx}
\eeq
Then, for any 
$(\mu, \gamma)\in$ $[0,1]^2$, $\pOne,\pTwo \geq 0$, 
we have 
\beqno
\overline{\Omega}^{(\mu,\gamma,\pOne,\theta)}(W_1,W_2)
&\leq &\frac{\Omega^{(\mu,\gamma, \pOne, \pTwo)}(W_1,W_2)}
{1+\gamma \pOne \pTwo}.
\eeqno
\end{pro}

{\it Proof:} 
Set
\beqno
& &\hat{\cal Q}_n
\defeq 
\{q=q_{UXYZ}: 
\pa {\cal U} \pa \leq 
\pa {\cal L}_n \pa \pa {\cal Y}^{n-1}\pa \pa {\cal Z}^{n-1}\pa\},
\\
& &\hat{\Omega}_n^{(\mu,\gamma,\pOne,\pTwo)}(W_1,W_2)
\defeq
\min_{\scs 
     \atop{\scs 
     q\in \hat{\cal Q}_n
     }
}
\Omega^{(\mu,\gamma,\pOne,\pTwo)}(q |W_1,W_2).
\eeqno
We recursively determine 
the sequence $\{{\cal F}^t\}_{t=1}^n$. Note 
that the component $\{p_{Y_t|U_t},\kappa_t\}_{t=1}^n$ 
of $\{{\cal F}^t\}_{t=1}^n$ is given. 
Hence we determine the remaining component 
$\{ \underline{Q}_t \}_{t=1}^n$.
For given ${\cal F}^{t-1}$, we choose 
$q_t=q_{U_tX_tY_tZ_t}$ so that 
\begin{align}
&  q_{U_tX_tY_tZ_t}
=p_{U_tX_tY_tZ_t;{\cal F}^{t-1}}^{(\mu,\gamma,\pOne,\theta)}
\label{eqn:Sdqw}
\end{align}
and choose the components 
$$
\ba{l} 
Q^{\One}_{Y_t|X_tZ_tU_t},
Q^{\Two}_{Z_t|X_tU_t},
Q^{\Thr}_{Y_t|U_t},
Q^{\Fou}_{Z_t},
Q^{\Fiv}_{Y_t}
\ea
$$
of $\underline{Q}_t$ such that they are the distributions 
induced by $q_{U_t}$${}_{X_tY_tZ_t}$. Note that 
$\tilde{Q}^{\Thr}_{Z_t|U_t}$ is uniquely determined 
by $({\cal F}^{t-1},$ $p_{Z_t|V_t})$. We denote 
it by $\tilde{q}_{Z_t|U_t}$. 
Note that since $\tilde{q}_{Z_t|U_t}$ is uniquely 
determined by $({\cal F}^{t-1}, p_{Z_t|V_t})$, 
it may not be consistent with $q_t$.
Then, for each $t=1,2,\cdots,n$, 
we have      
\begin{align}
& f_{\Ft}^{(\mu,\gamma,\pOne,\theta)}
(x_t,y_t,z_t|u_t)
\nonumber\\
&= 
\left\{
\frac{W_1(y_t|x_t)W_2(z_t|x_t)}
{q_{Y_t|X_tZ_tU_t}(y_t|x_t,z_t,u_t)
 q_{Z_t|X_tU_t}(z_t|x_t,u_t)}
\right\}^{\theta}
\notag\\
&\quad \times 
\left\{
\frac{W_1(y_t|x_t)}{q_{Y_t|U_t}(y_t|u_t)}
\right\}^{\mu\gamma\pOne\theta}
\left\{\frac{\tilde{q}_{Z_t|U_t}(z_t|u_t)}
{p_{Z_t|V_t}(z_t|v_t)}
\right\}^{\mu\gamma\pOne\theta}
\notag\\
&\quad \times 
\left\{
\frac{p_{Z_t|V_t}(z_t|v_t)}
     {q_{Z_t}(z_t)}
\right\}^{{\prmtB}\gamma\pOne\theta}
\left\{
\frac{W_1(y_t|x_t)}{q_{Y_t}(y_t)}
\right\}^{\bar{\gamma}\pOne\theta}.
\label{eqn:SdqwB}
\end{align}
We first consider the case where 
$(\mu,\gamma)\in [0,1]^2$ and $\gamma\pOne>0$.
In this case for $\theta \in (0,[\gamma\pOne]^{-1})$, 
we have the following: 
\beq
\pTwo=\frac{\theta}{1-\gamma\pOne\theta} 
\Leftrightarrow  
\theta=\frac{\pTwo}{1+\gamma\pOne\pTwo}. 
\label{eqn:abaddd}
\eeq 
On upper bounds of 
$\Lambda_{t,{\cal F}^t}^{(\mu,\gamma,\pOne,\theta)}$
for $t=1,2,$ $\cdots, n$, we have the following 
chains of inequalities:
\begin{align}
& \Lambda_{t,{\cal F}^t}^{(\mu,\gamma,\pOne,\theta)}
\MEq{a}{\rm E}_{q_t}
\left[
\left\{
\frac
{W_1^{\theta}(Y_t|X_t)W_2^{\theta}(Z_t|X_t)}
 {q^{\theta}_{Y_tZ_t|X_tU_t}(Y_t,Z_t|X_t,U_t)}
\right.\right.
\nonumber\\
& \quad \times
\left.
\frac
{W_1^{\mu\gamma \pOne \theta}(Y_t|X_t)}
{q^{\mu \gamma \pOne \theta}_{Y_t|U_t}(Y_t|U_t)}
\frac
{q_{Z_t|U_t}^{{\prmtB}\gamma\pOne\theta }(Z_t|U_t)}
{q^{ {\prmtB}\gamma\pOne\theta}_{Z_t}(Z_t)}
\frac
{W_1^{\bar{\gamma}\pOne \theta}(Y_t|X_t)}
{q^{\bar{\gamma}\pOne \theta}_{Y_t}(Y_t)}
\right\}
\nonumber\\
& \quad \times
\left.\left\{
\frac
{\tilde{q}^{{\mu}\gamma\pOne\theta}_{Z_t|U_t}(Z_t|U_t)}
{q^{{\mu}\gamma\pOne\theta}_{Z_t|U_t}(Z_t|U_t)}
\right\}
\left\{
\frac 
{p^{\bar{\mu}{\gamma} \pOne \theta}_{Z_t|V_t}(Z_t|V_t)}
{q^{\bar{\mu}{\gamma} \pOne \theta}_{Z_t|U_t}(Z_t|U_t)}
\right\}\right]
\label{eqn:AsssRR}\\
&\MLeq{b}
\left({\rm E}_{q_t}
\left[
\left\{
\frac
{W_1^{\theta}(Y_t|X_t)W_2^{\theta}(Z_t|X_t)}
 {q^{ \theta}_{Y_tZ_t|X_tU_t}(Y_t,Z_t|X_t,U_t)}
\frac 
{W_1^{\mu\gamma \pOne \theta}(Y_t|X_t)}
{q^{\mu\gamma \pOne\theta}_{Y_t|U_t}(Y_t|U_t)}
\right.\right.\right.
\nonumber\\
& \quad \times \left.\left.\left.
\frac 
{q_{Z_t|U_t}^{{\prmtB}\gamma\pOne\theta }(Z_t|U_t)}
{q^{ \gamma{\prmtB}\theta}_{Z_t}(Z_t)}
\frac
{W_1^{\bar{\gamma}\pOne\theta}(Y_t|X_t)}
{q^{\bar{\gamma}\theta}_{Y_t}(Y_t)}
\right\}^{\!\! \frac{1}{1-\gamma\pOne\theta}}
\right] \right)^{\!\!\! 1-\gamma\pOne\theta}
\nonumber\\
& \quad
\times\left\{ {\rm E}_{q_t}\left[ 
\frac 
{\tilde{q}_{Z_t|U_t}(Z_t|U_t)}
{q_{Z_t|U_t}(Z_t|U_t)}
\right]
\right\}^{\gamma \mu \pOne\theta}
\nonumber\\
& \quad \times
\left\{
{\rm E}_{q_t}\left[ 
\frac 
{p_{Z_t|V_t}(Z_t|V_t)}
{q_{Z_t|U_t}(Z_t|U_t)}
\right]
\right\}^{\gamma \bar{\mu} \pOne \theta}
\nonumber\\
&=
\exp\biggl\{\left[1-{\gamma}\pOne \theta \right]
\Omega^{(\mu,\gamma,\pOne,\frac{\theta}
{1-\gamma\pOne\theta})}(q_t|W_1,W_2)
\biggr\}
\nonumber\\
&\MEq{c}
\exp\left\{\frac{\Omega^{(\mu,\gamma,\pOne,\pTwo)}(q_t|W_1,W_2)
}{1+{\gamma}\pOne\pTwo}
\right\}.
\label{eqn:azxxaa}
\end{align}
Step (a) follows from (\ref{eqn:Sdqw}), 
                      (\ref{eqn:SdqwB}), 
and Lemma \ref{lm:keylm}. 
Step (b) follows from H\"older's inequality. 
Step (c) follows from (\ref{eqn:abaddd}). 
We next consider the case where  
$(\mu, \gamma)\in$ $[0,1]^2$, $\pOne\geq 0$, and $\gamma\pOne=0$. 
In this case we have the following equalities 
on $\Lambda_{t,{\cal F}^t}^{(\mu,\gamma,\pOne,\theta)}$
for $t=1,2,$ $\cdots, n$: 
\begin{align}
&  \Lambda_{t,{\cal F}^t}^{(\mu,\gamma,\pOne,\theta)}
\MEq{a}
{\rm E}_{q_t}
\left[
\left\{
\frac
{W_1^{\theta}(Y_t|X_t)W_2^{\theta}(Z_t|X_t)}
 {q^{\theta}_{Y_tZ_t|X_tU_t}(Y_t,Z_t|X_t,U_t)}
\right.\right.
\nonumber\\
& \quad \left.\left.\times 
\frac 
{W_1^{\mu \gamma \pOne \theta}(Y_t|X_t)}
{q^{\mu \gamma \pOne \theta}_{Y_t|U_t}(Y_t|U_t)}
\frac
{q_{Z_t|U_t}^{{\prmtB}\gamma\pOne\theta }(Z_t|U_t)}
{q^{ {\prmtB}\gamma\pOne\theta}_{Z_t}(Z_t)}
\frac
{W_1^{\bar{\gamma}\pOne \theta}(Y_t|X_t)}
{q^{\bar{\gamma}\pOne \theta}_{Y_t}(Y_t)}
\right\}
\right]
\nonumber\\
&=\exp\biggl\{
\Omega^{(\mu,\gamma,\pOne,\theta)}(q_t|W_1,W_2)
\biggr\}
\nonumber\\
&=\exp\left\{\frac{\Omega^{(\mu,\gamma,\pOne,\pTwo)}(q_t|W_1,W_2)
}{1+{\gamma}\pOne\pTwo}
\right\}.
\label{eqn:azxbaa}
\end{align}
Step (a) follows from (\ref{eqn:AsssRR}) and $\gamma\pOne=0$.
Hence for any $(\mu,\gamma)\in [0,1]^2$, $\pOne, \pTwo\geq 0$, 
and any $\theta \geq 0$ satisfying 
(\ref{eqn:Dsxxx}), we have the following upper bounds of 
$\Lambda_{t,{\cal F}^t}^{(\mu,\gamma,\pOne,\theta)}$ for 
$t=1,2,\cdots,n$:   
\beqa
\Lambda_{t,{\cal F}^t}^{(\mu,\gamma,\pOne,\theta)}
&\MLeq{a}&
\exp\left\{\frac{\Omega^{(\mu,\gamma,\pOne,\pTwo)}(q_t|W_1,W_2)
}{1+{\gamma}\pOne\pTwo}
\right\}
\nonumber\\
&\MLeq{b}&
\exp \left\{\frac{\hat{\Omega}_n^{(\mu,\gamma,\pOne,\pTwo)}(W_1,W_2)}
{1+{\gamma}\pOne\pTwo}
\right\}
\nonumber\\
&\MEq{c}&
\exp\left\{
\frac{{\Omega}^{(\mu,\gamma,\pOne,\pTwo)}(W_1,W_2)}
{1+\gamma\pOne\pTwo}
\right\}.\:\quad\label{eqn:ssstoZ}
\eeqa
Step (a) follows from 
(\ref{eqn:azxxaa}) and (\ref{eqn:azxbaa}).
Step (b) follows from 
$q_t \in \hat{\cal P}_n(W_1,W_2)$ and 
the definition of 
$\hat{\Omega}_n^{(\mu,\gamma,\pOne,\pTwo)}$ 
$(W_1,W_2)$. Step (c) follows from Property 
\ref{pr:pro1} part a), the proof 
of which is in Appendix \ref{sub:ApdaAAA}. 
To prove this lemma we bound the cardinality $|{\cal U}|$ 
appearing in the definition of 
$\hat{\Omega}_n^{(\mu, \gamma,\pOne,\pTwo)}(W_1,W_2)$ 
to show that the bound 
$|{\cal U}|\leq $$|{\cal Y}|+$$|{\cal Z}|-1$
is sufficient to describe 
$\hat{\Omega}_n^{(\mu,\gamma,\pOne,\pTwo)}(W_1,W_2)$.
Hence for any $R_1,R_2\geq 0$ and 
for any $(\varphi^{(n)},$ $\psi_1^{(n)},$ 
$\psi_2^{(n)})$ satisfying 
$(1/n)\log |{\cal K}_n| \geq R_1,$
$(1/n)\log |{\cal L}_n| \geq R_2,$  
we have the following:
\begin{align}
& \min_{\scs q^{n}\in {\cal Q}^n}
\frac{1}{n}\Omega^{(\mu,\gamma,\pOne,\theta)}
(p^{(n)}, \kappa^n, \underline{Q}^n)
\nonumber\\
&\leq \frac{1}{n}
\Omega^{(\mu,\gamma, \pOne,\theta)}
(p^{(n)}, \kappa^n, \underline{q}^n)
\MEq{a}
\frac{1}{n}\sum_{t=1}^n \log 
\Lambda_{t,{\cal F}^t}^{(\mu,\gamma,\pOne,\theta)}
\nonumber\\
&\MLeq{b}
\frac{{\Omega}^{(\mu,\gamma,\pOne,\pTwo)}(W_1,W_2)}
{1+\gamma \pOne\pTwo}.
\qquad \label{eqn:aQ1}
\end{align}
Step (a) follows from (\ref{eqn:uzapa}) in Lemma \ref{lm:keylm}.
Step (b) follows from (\ref{eqn:ssstoZ}). 
Since (\ref{eqn:aQ1}) holds for any ${n\geq 1}$ 
and any $p^{(n)}\in {\cal P}^{(n)}$ $(W_1,W_2)$, we have  
$$
\overline{\Omega}^{(\mu,\gamma,\pOne,\theta)}(W_1,W_2)
\leq 
\frac{{\Omega}^{(\mu,\gamma,\pOne,\pTwo)}(W_1,W_2)}
{1+\gamma\pOne\pTwo},
$$
completing the proof. 
\hfill \IEEEQED

{\it Proof of Theorem \ref{Th:main}: }
We fix $\lambda\geq 0, (\mu,\gamma)\in [0,1]^2$, $\pOne \geq 0$ 
arbitrary. For $\pTwo\geq 0$, set 
\beq
\theta=\frac{\pTwo}{1+\gamma\pOne\pTwo}. 
\label{eqn:abadd}
\eeq
Then for any $R_1,R_2\geq 0$ and 
for any $(\varphi^{(n)},$ $\psi_1^{(n)},$ 
$\psi_2^{(n)})$ satisfying 
$(1/n)\log |{\cal K}_n| \geq R_1$ and 
$(1/n)\log |{\cal L}_n| \geq R_2$,
we have the following:
\begin{align}
& \frac{1}{n}\log
\left\{
\frac{6}{{\rm P}_{\rm c}^{(n)}
(\varphi^{(n)},\psi_1^{(n)},\psi_2^{(n)})}
\right\}
\nonumber\\
&\MGeq{a} 
\frac{
\theta \pOne [(\gamma\mu +\bar{\gamma})R_1+R_2]
-\overline{\Omega}^{(\mu,\gamma,\pOne,\theta)}(W_1,W_2)
}
{1+\theta[2+\pOne(\gamma\mu+1)] }
\nonumber\\
&\MGeq{b} 
\frac{\ds \frac{
        \pTwo\pOne[
        (\gamma\mu +\bar{\gamma})R_1+R_2
         ]-\Omega^{(\mu,\gamma,\pOne,\pTwo)}(W_1,W_2)}
          {1+\gamma \pOne \pTwo}
     }{\ds 1+\frac{\pTwo[2+\pOne(\gamma\mu+1)]}
      {1+\gamma\pOne\pTwo}}
\nonumber\\
&=
\frac{\pTwo \pOne [(\gamma\mu+\bar{\gamma})R_1+R_2]
-\Omega^{(\mu,\gamma,\pOne,\pTwo)}(W_1,W_2)}
{1+\pTwo[2+\pOne(1+\gamma+\gamma\mu)] }
\nonumber\\
&=F^{(\mu,\gamma,\pOne,\pTwo)}(R_1,R_2|W_1,W_2). 
\label{eqn:asff}
\end{align}
Step (a) follows from Corollary \ref{co:ProA}. Step (b) follows from 
Proposition \ref{pro:mainpro} and (\ref{eqn:abadd}). 
Since (\ref{eqn:asff}) holds for any 
$(\mu, \gamma)$  $\in [0,1]^2$, $\pOne\geq 0$, and $\pTwo \geq 0$,
we have (\ref{eqn:mainIeq}) in Theorem \ref{Th:main}. 
\hfill\IEEEQED

\ProofCor

\newcommand{\ApdaAAAb}{

Next we prove the following lemma.
\begin{lm} \label{lm:CardLmd} 
For each integer $n \geq 2$, we define 
\beqno
&      &\hat{\Omega}_n^{(\pTwo,\pOne,\gamma,\mu{ },\lambda)}(W_1,W_2)
\\
&\defeq& \max_{\scs q=q_{UXYZ}: 
               \atop{\scs 
                   |{\cal U}|\leq |{\cal L}_n|
                   |{\cal Y}|^{n-1}|{\cal Z}|^{n-1}  
              }
          }
\Omega^{(\pTwo,\pOne,\gamma,\mu{ },\lambda)}(q|W_1,W_2),
\\
& &\Omega^{(\pTwo,\pOne,\gamma,\mu{ },\lambda)}(W_1,W_2)
\\
&\defeq&
\max_{\scs q=q_{UXYZ}: 
        \atop{\scs
           \atop{\scs
             \atop{\scs  
            |{\cal U}|\leq |{\cal Y}|+|{\cal Z}|-1
     }}}}
\Omega^{(\pTwo,\pOne,\gamma,\mu{ },\lambda)}(q|W_1,W_2).
\eeqno
Then we have 
\beqno
& &\hat{\Omega}^{(\pTwo,\pOne,\gamma,\mu{ },\lambda)}(W_1,W_2)
       =\Omega^{ (\pTwo,\pOne,\gamma,\mu{ },\lambda)}(W_1,W_2).
\eeqno

\end{lm}

{\it Proof:} We bound  the cardinality $|{\cal U}|$ of ${U}$ 
to show that the bound 
$|{\cal U}| \leq |{\cal Y}|$$+|{\cal Z}|-1$
is sufficient to describe 
$\hat{\Omega}_n^{(\pTwo,\pOne,\gamma,\mu{ },\lambda)}$ 
$(W_1,W_2)$. Observe that 
\beqa
& &
\left.
\ba{l} 
\ds
q_{{Y}}(y)
=\sum_{u\in {\cal U}}q_U(u)
q_{Y|U}(y|u),
\vspace*{1mm}\\
\ds q_{Z}(z)
=\sum_{u\in {\cal U}}q_U(u)
q_{Z|U}(z|u),
\ea
\right\}
\label{eqn:asdf}
\\
& &\Lambda^{(\pTwo,\pOne,\gamma,\mu{ },\lambda)}(q|W_1,W_2)
\nonumber\\
&=&\sum_{u\in {\cal U}}q_U(u)
\zeta^{(\pTwo,\pOne,\gamma,\mu{ },\lambda)}(q_{XYZ|U}(\cdot|u)),
\label{eqn:aqqqa}
\eeqa
where we set
\beqno
& &\zeta^{(\pTwo,\pOne,\gamma,\mu{ },\lambda)}(q_{XYZ|U}(\cdot,\cdot,\cdot|u))
\\
&\defeq & \sum_{(x,y,z)\in{\cal X}\times{\cal Y}\times{\cal Z}}
q_{{XYZ}|U}(x,y,z|u)
\\
& &\times \exp\left\{\lambda \omega^{
  (\pTwo,\pOne,\gamma,\mu{ })}_{q}(x,y,z|u)\right\}.
\eeqno
For the quantities $q_{Y}(\cdot)$ and $q_{Z}(\cdot)$ 
contained in the forms of 
$\zeta^{(\pTwo,\pOne,\gamma,\mu{ },\lambda)}$ 
$(q_{XYZ|U}(\cdot|u)),$ $u\in {\cal U}$, we regard them as 
constants under (\ref{eqn:asdf}). For each $u \in{\cal U}$, 
$
\zeta^{(\pTwo,\pOne,\gamma,\mu{ },\lambda)}$ $(q_{XYZ|U}(\cdot|u))
$
is a continuous function of $q_{XYZ|U}(\cdot,\cdot,\cdot|u)$.
Then by the support lemma, 
$$
|{\cal U}| \leq |{\cal Y}|+|{\cal Z}|-2+1
=|{\cal Y}|+|{\cal Z}|-1 
$$
is sufficient to express $|{\cal Y}|+|{\cal Z}|-2$ 
values of (\ref{eqn:asdf}) 
and one value of (\ref{eqn:aqqqa}). 
\hfill \IEEEQED
}
%
%
%
%

\section*{\empty}
\appendix

\ApdaAAA
\ApdaAACaa
%
%
\ApdaAAB
%
\ApdaAACa
%
\ApdaAACb
%
\ApdaAACc 
%
\Apda
%
%
\Apdc

\vspace*{3mm}

\noindent
{\bf Acknowledgements}

I am very grateful to Dr. Shun Watanabe for his 
helpful comments.

\end{document}